\documentstyle[12pt,epsf]{article}
\newcommand{\bfig}{\begin{figure}}
\newcommand{\efig}{\end{figure}}

\newcommand{\maru}{\mbox{\tiny$\stackrel{\circ}{\scriptstyle\circ}$}}

\newcommand{\pdot}{{\displaystyle{\raisebox{-1.5ex}[0.25ex]{$\cdot$}
     \atop\raisebox{0.6ex}[0.25ex]{$\scriptstyle (p)$}}}}
\newcommand{\prdot}{{\displaystyle{\raisebox{-1.5ex}[0.25ex]{$\cdot$}
     \atop\raisebox{0.6ex}[0.25ex]{$\scriptstyle (p^{\prime})$}}}}
\newcommand{\ppdot}{{\displaystyle{\raisebox{-1.5ex}[0.25ex]{$\cdot$}
     \atop\raisebox{0.6ex}[0.25ex]{$\scriptstyle (p,p^{\prime})$}}}}
\newcommand{\kuroten}{{\displaystyle{\raisebox{-1.5ex}{$\odot$}
     \atop\raisebox{0.3ex}[1.25ex]{$\scriptstyle(p,p^{\prime})$}}}}
\newcommand{\peke}{{\displaystyle{\raisebox{-1.5ex}{$\times$}
      \atop\raisebox{0.3ex}[1.25ex]{$\scriptstyle (p,p^{\prime})$}}}}

\def\12{\frac{1}{2}}
\def\bea{\begin{eqnarray}}
\def\eea{\end{eqnarray}}
\def\ba{\begin{array}}
\def\ea{\end{array}}

\def\one-loop{\mbox{\scriptsize one-loop}}

\def\G{\Gamma}

\catcode`\@=11

\@addtoreset{equation}{section}
\def\theequation{\arabic{section}.\arabic{equation}}

\def\@normalsize{\@setsize\normalsize{15pt}\xiipt\@xiipt
\abovedisplayskip 14pt plus3pt minus3pt%
\belowdisplayskip \abovedisplayskip
\abovedisplayshortskip  \z@ plus3pt%
\belowdisplayshortskip  7pt plus3.5pt minus0pt}

\def\small{\@setsize\small{13.6pt}\xipt\@xipt
\abovedisplayskip 13pt plus3pt minus3pt%
\belowdisplayskip \abovedisplayskip
\abovedisplayshortskip  \z@ plus3pt%
\belowdisplayshortskip  7pt plus3.5pt minus0pt
\def\@listi{\parsep 4.5pt plus 2pt minus 1pt
            \itemsep \parsep
            \topsep 9pt plus 3pt minus 3pt}}

\def\underline#1{\relax\ifmmode\@@underline#1\else
        $\@@underline{\hbox{#1}}$\relax\fi}
\@twosidetrue
\relax

\catcode`@=12

\catcode`\@=11

\def\section{\@startsection{section}{1}{\z@}{3.5ex plus 1ex minus
   .2ex}{2.3ex plus .2ex}{\large\bf}}

\def\thesection{\Roman{section}.}

\def\appendix{\setcounter{section}{0}
        \def\thesection{Appendix }
        \def\theequation{\Alph{section}.\arabic{equation}}}
\def\ps@headings{\def\@oddfoot{}\def\@evenfoot{}
\def\@oddhead{\hbox{}\hfill
        \makebox[.5\textwidth]{\raggedright\ignorespaces --\thepage{}--
        \hfill {}}}
\def\@oddhead{\hbox{}\hfill --\thepage{}-- \hfill
        {}}
\def\@evenhead{\@oddhead}
\def\subsectionmark##1{\markboth{##1}{}}
}

\ps@headings

\catcode`\@=12

\relax

\def\figcap{\section*{Figure Captions\markboth
        {FIGURECAPTIONS}{FIGURECAPTIONS}}\list
        {Fig. \arabic{enumi}:\hfill}{\settowidth\labelwidth{Fig. 999:}
        \leftmargin\labelwidth
        \advance\leftmargin\labelsep\usecounter{enumi}}}
 \relax
\def\tablecap{\section*{Table Captions\markboth
        {TABLECAPTIONS}{TABLECAPTIONS}}\list
        {Table \arabic{enumi}:\hfill}{\settowidth\labelwidth{Table 999:}
        \leftmargin\labelwidth
        \advance\leftmargin\labelsep\usecounter{enumi}}}
 \relax
\def\reflist{\section*{References\markboth
        {REFLIST}{REFLIST}}\list
        {[\arabic{enumi}]\hfill}{\settowidth\labelwidth{[999]}
        \leftmargin\labelwidth
        \advance\leftmargin\labelsep\usecounter{enumi}}}
 \relax

\catcode`\@=11

\def\ps@headings{\def\@oddfoot{}\def\@evenfoot{}
\def\@oddhead{\hbox{}\hfill
        \makebox[.5\textwidth]{\raggedright\ignorespaces --\thepage{}--
        \hfill {}}}
\def\@evenhead{\@oddhead}
\def\subsectionmark##1{\markboth{##1}{}}
}

\ps@headings

\relax

\newskip\humongous \humongous=0pt plus 1000pt minus 1000pt

\newif\ifdtup

\def\beq{\begin{equation}}
\def\eeq{\end{equation}}

\def\beqn{\begin{eqnarray}}
\def\eeqn{\end{eqnarray}}
\relax

\def\G2{{\; \rm GeV/}c^2}
\def\G{\; \rm GeV}

\def\dotx{\dotx{\dot\overline{x}}}

\relax

\begin{document}

%%%%%%%%%%%%%%%%%%%%%%%%%%%%%%%%%%%%%%
%
%        title page
%
%%%%%%%%%%%%%%%%%%%%%%%%%%%%%%%%%%%%%%
\begin{titlepage}

\renewcommand{\thefootnote}{\fnsymbol{footnote}}

\begin{flushright}
      \normalsize
         OU-HET 345,  May, 2000 \\ 
         hep-th/0005283  \\
\end{flushright}

%\vfill
%
\begin{center}
  {\large\bf Worldsheet and Spacetime Properties of  \\
 $p-p^\prime$ System with  $B$ Field  \\ and Noncommutative Geometry }%
\footnote{This work is supported in part
 by the Grant-in-Aid  for Scientific Research
(10640268, 12640272, 12014210) from the Ministry of Education,
Science and Culture, Japan,
and in part by
the Japan Society for the Promotion of Science for Young
Scientists.}
\end{center}

\vfill

\begin{center}
    { {  B. Chen}\footnote{e-mail address:
                                 chenb@funpth.phys.sci.osaka-u.ac.jp},
      { H. Itoyama}\footnote{e-mail address:
                               itoyama@funpth.phys.sci.osaka-u.ac.jp},
      { T. Matsuo}\footnote{e-mail address:
                                matsuo@funpth.phys.sci.osaka-u.ac.jp}
      and { K. Murakami}\footnote{e-mail address:
                              murakami@funpth.phys.sci.osaka-u.ac.jp}}\\
\end{center}

\vfill

\begin{center}
      \it  Department of Physics,
        Graduate School of Science, Osaka University,\\
        Toyonaka, Osaka 560-0043, Japan
\end{center}

\vfill

%%%%%%%%   abstract  %%%%%%%%%%%%%%%%%%%%%

\begin{abstract}

We study worldsheet and spacetime properties of
the $p$-$p^\prime$ $(p < p^\prime)$ open string system
with constant $B_{ij}$ field viewed from the D$p^\prime$-brane.
The description of this system in terms of the CFT with spin and twist
fields leads us to consider the renormal ordering procedure from
the $SL(2,{\bf R})$ invariant vacuum to the oscillator vacuum.
We compute the attendant two distinct superspace two-point functions
as well as their difference (the subtracted two-point function).
These bring us an integral (Koba-Nielsen) representation
for the multiparticle tree scattering amplitudes consisting of
$N-2$ vectors and two tachyons.
We evaluate them explicitly for the $N=3,4$ cases.
Several novel features are observed which include a momentum dependent
multiplicative factor to each external vector leg and the emergence of
a symplectic tensor multiplying the polarization vectors.
In the zero slope limit, the principal parts of the amplitudes
translate into a noncommutative field theory in $p^{\prime}+1$
dimensions in which a scalar field decaying exponentially
in $(p^{\prime}-p)$ dimensions and a noncommutative $U(1)$ gauge field
interact via the minimal coupling and a new interaction.
A large number of nearly massless states noted before are shown
to propagate in the $t$-channel.

\end{abstract}

\vfill

\setcounter{footnote}{0}
\renewcommand{\thefootnote}{\arabic{footnote}}

\end{titlepage}

%%%%%%%%%%%%%%%%%%%%%%%%%%%%%%%%%%%%%%%%%%%%%%%%%%%%%%%%%%%%%%%%%%%%%%

\section{Introduction}

After decades of investigations, string perturbation theory has
now become a well-established old subject.
It is yet a nontrivial task to uncover spacetime properties,
given a first quantized worldsheet theory.
Some of our previous endeavors on strings are regarded as a search for
a formulation in which these spacetime properties come out
in a more transparent way.
String theory with constant $B_{ij}$ background offers an intriguing
situation in which  the emerging spacetime picture
is given in terms of noncommutative geometry
and is a focus of the recent intensive
studies \cite{CDS}\cite{DH}\cite{CH}\cite{Jabba}\cite{SJ}.

Several important steps have been taken in \cite{SW}.
In particular, the proper spacetime metric on the D$p$-brane
(the open string metric) has been extracted from the worldsheet theory
of an open string with its both ends on a D$p$-brane,
(the $p$-$p$ open string system): the distances measured
with respect to this metric are kept finite at all scales.
The attendant noncommutative field theory in the zero slope limit
lives on this metric together with the parameter representing
noncommutativity of spacetime.

In the previous paper \cite{CIMM}, we have examined the more complex
$p$-$p^\prime$ $(p < p^\prime)$ open string system where the both ends
of the open string are on a D$p$-brane and on a D$p^{\prime}$-brane
respectively.
We have obtained the open string metric and the noncommutativity
parameter on the D$p^\prime$-brane from the worldsheet two-point
function.
They in fact agree with those of the $p$-$p$ system.
We have computed the spectrum in each case of $(p,p^\prime)$,
uncovered the emergence of a large number of nearly massless states
in some cases and clarified the connections among the GSO projection,
branes at angles and supersymmetry.
Yet a number of other properties, in particular, spacetime properties
on the D$p^\prime$-brane  with D$p$-brane inside have remained elusive.
Elucidating upon these is a major goal of the present paper.

It should be mentioned that,
in the vanishing $B_{ij}$ background, several properties
of the $p$-$p^{\prime}$ open string system have been studied.
For example,
amplitudes of some scattering processes taking place
on the D$p$-brane worldvolume (i.e.\ that of
the lower dimensional D-brane) has been evaluated in \cite{AH}
and 
the conformal field theory correlation functions have
been studied in \cite{FGRS}.
In the presence of $B_{ij}$ background
the four point
tachyon amplitudes have been given in \cite{GNS}
for the $p^{\prime}=p+2$ case.

In \cite{SW}
several properties of the $0$-$4$ system have been derived.
The description of the system from the D$0$-brane
has offered a new perspective to the moduli space of
noncommutative instantons \cite{Nek}  where the noncommutativity
of the system is measured by  the presence
of the Fayet-Iliopoulos $D$ term.
In contrast to this moduli space point of view,
the thrust of the present paper is to uncover the spacetime properties
of the system viewed from the higher dimensional D-brane,
namely the D$p^\prime$-brane.
This can be accomplished by placing vertex operators
on the worldvolume of the  D$p^\prime$-brane and by considering
the scattering processes.
This is an extension of the computation of  scattering
amplitudes in the $p$-$p$ open string system
with and without $B_{ij}$ background
\cite{GM}\cite{HK}\cite{SJ}\cite{AIKT}.
This line of reasonings has led us to carry out a systematic
study which begins with evaluating superspace two-point functions
in the relevant CFT with the spin and twist fields,
proceeds to the computation of scattering amplitudes on
the D$p^\prime$-brane and ends with identifying  a proper low energy
noncommutative field theory in the zero slope limit.

In the next section,  we  begin with quantizing an open string
ending on D$p$ and D$p^\prime$  $(p < p^\prime)$ branes.
We exploit superspace formulation, which we find extremely efficient
in the calculation pursued in the subsequent sections. 
We evaluate two distinct two-point functions on superspace
and observe the importance of the renormal ordering procedure
{} from the $SL(2, {\bf R})$ invariant vacuum to the oscillator vacuum. 
This procedure leads us to consider the difference of these 
two two-point functions as well. This third quantity
plays an important role in  section $4$  and will be referred to as
subtracted two-point function.  The conformal weights of the twist and
spin fields  are readily computed from  the renormal ordering procedure.

In section $3$, we examine the tachyon vertex operator of
the $p$-$p^\prime$ open string and the vector vertex operator of the
$p^\prime$-$p^\prime$ open string. We derive the on-shell conditions
in terms of the open string metric, $p+1$ dimensional momenta
of the tachyon, $p^{\prime}+1$ dimensional momenta and polarizations
of the massless vector and mass of the tachyon.

In section $4$, we consider the multiparticle tree scattering amplitudes
consisting of external states of $N-2$ vectors and  two tachyons.
We are able to derive an integral (Koba-Nielsen) representation of
these quantities as integrals over $N-3$ locations of the vertex
operators as well as the $N-2$ Grassmann counterparts
and the $N-2$ Grassmann sources conjugate to the polarization vectors.
(See \cite{IM} for the case of the $p$-$p$ string with vanishing $B$
field.)
Several striking properties emerge from this representation. 
Among other things,
we find  a momentum dependent  exponential factor
to each external vector leg, which
the subtracted  two-point function is responsible for,  as well as
a new symplectic tensor multiplying the polarizations
of the massless vector.
We observe that some parts of the amplitudes are expressible
in terms of the inner products of polarizations, momenta
and the symplectic tensor with respect to the open string metric,
while there are a host of other parts which do not permit such
generic description by the inner product.
We evaluate the amplitudes for the $N=3,4$ cases explicitly.

In section $5$,  we examine the zero slope limit of the system.
We find that the parts of the amplitudes expressible in terms of the
inner product in this limit (the principal parts)
can be summarized as a noncommutative field theory
of a scalar field and a noncommutative $U(1)$ gauge
field in
$p^{\prime}+1$ dimensions in which the scalar field
decays exponentially
in the $x^{p+1},\ldots,x^{p^{\prime}}$-directions.
They interact via the minimal coupling and a new interaction
which consists of the field strength and the scalar bilinear
contracted with the symplectic tensor.
The contributions from the residual parts are consistent with
the propagations in the $t$-channel of a large number of
nearly massless states found in \cite{CIMM}.

%%%%%%%%%%%%%%%%%%%%%%%%%%%%%%%%%%%%%%%%%%%%%%%%%%%%%%%%%%%%%%%%%%%%
\section{Basic Properties of $p$-$p^{\prime}$ System with $B_{ij}$
         Field \label{sect:pprime}}
%%%%%%%%%%%%%%%%%%%%%%%%%%%%%%%%%%%%%%%%%%%%%%%%%%%%%%%%%%%%%%%%%%%%%
In this section,  we will provide the two-point functions, and 
the twist and the spin fields  
for a $p$-$p^{\prime}$ open string in  constant $B$ field background.
This will also help us to establish our notations.
We introduce two types of normal ordering:
the one is taken with respect to the $SL(2,{\bf R})$ invariant
vacuum and the other is with respect to the oscillator vacuum.
We will establish the relationship between these two,
which will be important for our calculation in the subsequent sections.

%%%%%%%%%%%%%%%%%%%%%%%%%%%%%%%%%%%%%%%%%%%%%%%%%%%%%%%
\subsection{Action and boundary condition}

The action of the NSR superstring in the constant $B$ 
background takes the form of
\begin{equation}
 S=\frac{1}{2\pi}\int d^{2}{\xi}
   \int d\theta d\overline{\theta} \left(
     g_{\mu\nu}+2\pi\alpha^{\prime}B_{\mu\nu}
   \right)\overline{D} {\bf X}^{\mu}({\bf z},\overline{\bf z})
           D {\bf X}^{\nu}({\bf z},\overline{\bf z})~,
\label{eq:action-1}
\end{equation}
where ${\bf z}=(z,\theta)$ and
$\overline{\bf z}=(\overline{z},\overline{\theta})$
are the superspace coordinates on the worldsheet,
$ D=\frac{\partial}{\partial \theta}
+ \theta \frac{\partial}{\partial z}$
and
$ \overline{D}=\frac{\partial}{\partial \overline{\theta}}
  +\overline{\theta}\frac{\partial}{\partial \overline{z}}$
are the superspace covariant derivatives
and $g_{\mu\nu}$ denotes the space-time metric which is taken to be flat.
$z=\xi^{1}+i\xi^{2}$ and
$\overline{z}=\xi^{1}-i\xi^{2}$ are complex coordinates on the plane
which are related to the strip coordinates $(\tau,\sigma)$
by $z=e^{\tau+i\sigma}$ and $\overline{z}=e^{\tau-i\sigma}$
respectively.
The superfield ${\bf X}^{\mu}({\bf z},\overline{\bf z})$
is the string coordinate which is expressed 
in terms of component fields as
\begin{equation}
{\bf X}^{\mu}({\bf z},\overline{\bf z})
=\sqrt{\frac{2}{\alpha^{\prime}}}X^{\mu}(z,\overline{z})
 +i \theta \psi^{\mu}(z,\overline{z})
 +i \overline{\theta} \widetilde{\psi}^{\mu}(z,\overline{z})
 +i\theta\overline{\theta}F^{\mu}(z,\overline{z})~.
\end{equation} 
In terms of the component fields the action (\ref{eq:action-1})
is given by
\begin{equation}
 S=\frac{1}{2\pi} \int d^{2}{\xi}
  \left(g_{\mu\nu}+2\pi\alpha^{\prime}B_{\mu\nu}\right)
  \left(\frac{2}{\alpha^{\prime}}
        \overline{\partial}X^{\mu}\partial X^{\nu}
      {}-\overline{\partial}\psi^{\mu} \, \psi^{\nu}
        +\widetilde{\psi}^{\mu}\partial\widetilde{\psi}^{\nu}
   \right)~.
\end{equation}
Here
we have eliminated the auxiliary field $F^{\mu}(z,\overline{z})$
by using the equation of motion $F^{\mu}=0$, and
the operators $\partial$ and $\overline{\partial}$ denote
$\frac{\partial}{\partial z}$ and 
$\frac{\partial}{\partial\overline{z}}$ respectively.
Since the $B$ dependent terms do not couple to the worldsheet metric,
the energy-momentum tensor of this system has the same form
as that of the string without $B$ field background:
\begin{eqnarray}
&& {\rm T}({\bf z})
\equiv T_{F}(z)+\theta T_{B}(z)
\equiv -\frac{1}{2}g_{\mu\nu}D{\bf X}^{\mu}D^{2}{\bf X}^{\nu}~,\\
&&\hspace{2em}\mbox{with}\left\{ 
\begin{array}{l}
  T_{F}(z)=-\displaystyle\frac{i}{2}\sqrt{\frac{2}{\alpha^{\prime}}}
    g_{\mu\nu}\psi^{\mu}\partial X^{\nu}\\
  T_{B}(z)=-\displaystyle\frac{1}{\alpha^{\prime}}
     g_{\mu\nu}\partial X^{\mu}\partial X^{\nu}
     {}-\frac{1}{2}g_{\mu\nu}\psi^{\mu}\partial\psi^{\nu}
\end{array}
\right.~.\label{eq:emtensor}
\end{eqnarray}

Let us consider a $p$-$p^{\prime}$ open string
in the type IIA theory with $p<p^{\prime}$
and $p$ and $p^{\prime}$ being even integers.
We concentrate on the situation in which a D$p$-brane extends in
the $(x^{0},x^{1},\ldots,x^{p})$-directions and a D$p^{\prime}$-brane
extends in the $(x^{0},x^{1},\ldots,x^{p^{\prime}})$-directions
with the D$p$-brane inside.
The worldsheet of the open string corresponds to the upper half-plane:
${\rm Im}z \geq 0$ $(\Leftrightarrow 0\leq \sigma \leq \pi)$.
The D$p$-brane worldvolume contains the boundary $\sigma=0$ 
while the D$p^{\prime}$-brane
worldvolume contains the boundary $\sigma=\pi$.
As the space-time is flat with the metric
\begin{equation}
 g_{\mu\nu} = \left(
\begin{array}{ccccccc}
 \multicolumn{1}{c|}{-1}& & & & & & \\ \cline{1-4}
 \multicolumn{1}{c|}{ }&\multicolumn{3}{c|}{ }& & &   \\
 \multicolumn{1}{c|}{ }& &{g_{ij}}&\multicolumn{1}{c|}{}& & &  \\
 \multicolumn{1}{c|}{ }& &{ }&\multicolumn{1}{c|}{}& & & \\ \cline{2-7}
     & & &\multicolumn{1}{c|}{ } &1 & & \\
     & & &\multicolumn{1}{c|}{ } & & \ddots & \\
     & & &\multicolumn{1}{c|}{ } & & & 1 
\end{array}
\right)~,
\qquad\quad
g_{ij}=\varepsilon \delta_{ij}\quad(i,j=1,\ldots,p^{\prime})~,
\label{eq:g-munu}
\end{equation}
we can bring
$B_{\mu\nu}$ into a canonical form 
\begin{equation}
 B_{ij}=\frac{\varepsilon}{2\pi\alpha^{\prime}}\left(
 \begin{array}{ccccc}
  0&b_{1} & & & \\
  {}-b_{1}&0& & & \\
   & & 0 & b_{2}& \\
   & & -b_{2} & 0 & \\
   & & & & \ddots
 \end{array}\right)
\quad (i,j=1,\ldots,p^{\prime})~,
\quad\mbox{otherwise $B_{\mu\nu}=0$}~.  \label{eq:B-ij}
\end{equation}
In what follows we will investigate the system on this background.

The boundary conditions for the string coordinates in the NS sector
are \cite{CIMM}\cite{SW}
\begin{eqnarray}
&& \left. D{\bf X}^{0}-\overline{D}{\bf X}^{0}
\right|_{\sigma=0,\pi~\theta=\overline{\theta}}=0~,
\quad
\left. D{\bf X}^{p^{\prime}+1,\ldots,9}
       +\overline{D}{\bf X}^{p^{\prime}+1,\ldots,9}
\right|_{\sigma=0,\pi~\theta=\overline{\theta}}=0~,\nonumber\\
&& \left. g_{kl}(D{\bf X}^{l}-\overline{D}{\bf X}^{l})
          +2\pi\alpha^{\prime}B_{kl}
            (D{\bf X}^{l}+\overline{D}{\bf X}^{l})
   \right|_{\sigma=0,\pi~\theta=\overline{\theta}}=0
\quad (k,l=1,\ldots,p)\nonumber\\
&&\left. D{\bf X}^{i}+\overline{D}{\bf X}^{i}
  \right|_{\sigma=0~\theta=\overline{\theta}}
 =\left.g_{ij}(D{\bf X}^{j}-\overline{D}{\bf X}^{j})
  +2\pi \alpha^{\prime}B_{ij}(D{\bf X}^{j}+\overline{D}{\bf X}^{j})
  \right|_{\sigma=\pi~\theta=\overline{\theta}}=0\nonumber\\
&& \hspace{24em}(i,j=p+1,\ldots,p^{\prime})~.
\end{eqnarray}
For the bosonic components these conditions read
\begin{eqnarray}
&&\left. (\partial-\overline{\partial})X^{0}\right|_{\sigma=0,\pi}=0~,
 \qquad
\left.(\partial+\overline{\partial})
      X^{p^{\prime}+1,\ldots,9}\right|_{\sigma=0,\pi}
=0~,\nonumber\\
&& \left. g_{kl}(\partial-\overline{\partial})X^{l}
       +2\pi \alpha^{\prime}
        B_{kl}(\partial+\overline{\partial})
                 X^{l}\right|_{\sigma=0,\pi}=0
\quad (k,l=1,\ldots,p)~,\nonumber\\
&&\left.(\partial+\overline{\partial})X^{i}\right|_{\sigma=0}=
\left.g_{ij}(\partial-\overline{\partial})X^{j}
       +2\pi \alpha^{\prime} B_{ij}
       (\partial+\overline{\partial})X^{j}\right|_{\sigma=\pi}=0
\quad (i,j=p+1,\ldots,p^{\prime})~,
\label{eq:bc-boson}
\end{eqnarray}
and for the fermionic components 
\begin{eqnarray}
&& \left.\psi^{0}-\widetilde{\psi}^{0}\right|_{\sigma=0,\pi}=0~,
\qquad
 \left.\psi^{p^{\prime}+1,\ldots,9}+
  \widetilde{\psi}^{p^{\prime}+1,\ldots,9}\right|_{\sigma=0,\pi}=0~,
 \nonumber\\
&& \left.g_{kl}(\psi^{l}-\widetilde{\psi}^{l})
         +2\pi\alpha^{\prime}B_{kl}
               (\psi^{l}+\widetilde{\psi}^{l})
   \right|_{\sigma=0,\pi}=0
\quad (k,l=1,\ldots,p)\nonumber \\
&& \left.\psi^{i}+\widetilde{\psi}^{i}\right|_{\sigma=0}
   =\left.g_{ij}(\psi^{j}-\widetilde{\psi}^{j})
          +2\pi\alpha^{\prime}B_{ij}(\psi^{j}+\widetilde{\psi}^{j})
    \right|_{\sigma=\pi}=0
\quad (i,j=p+1,\ldots,p^{\prime})~.
\label{eq:bc-nsfermion}
\end{eqnarray}
The boundary conditions for the R fermions are
\begin{eqnarray}
&& \left.\psi^{0}-\widetilde{\psi}^{0}\right|_{\sigma=0}=
 \left.\psi^{0}+\widetilde{\psi}^{0}\right|_{\sigma=\pi}=0~,
\quad \left.\psi^{p^{\prime}+1,\ldots,9}
      +\widetilde{\psi}^{p^{\prime}+1,\ldots,9}\right|_{\sigma=0}=
      \left.\psi^{p^{\prime}+1,\ldots,9}
      {}-\widetilde{\psi}^{p^{\prime}+1,\ldots,9}\right|_{\sigma=\pi}=0~,
\nonumber\\
&&\left. g_{kl}(\psi^{l}-\widetilde{\psi}^{l})
          +2\pi\alpha^{\prime}B_{kl}
               (\psi^{l}+\widetilde{\psi}^{l})
      \right|_{\sigma=0}
=\left. g_{kl}(\psi^{l}+\widetilde{\psi}^{l})
          +2\pi\alpha^{\prime}B_{kl}(\psi^{l}-\widetilde{\psi}^{l})
      \right|_{\sigma=\pi}=0\nonumber\\
&& \hspace{30em}(k,l=1,\ldots,p)~,\nonumber\\
&&\left.\psi^{i}+\widetilde{\psi}^{i}\right|_{\sigma=0}=
\left.g_{ij}(\psi^{j}+\widetilde{\psi}^{j})
      +2\pi\alpha^{\prime}B_{ij}(\psi^{j}-\widetilde{\psi}^{j})
      \right|_{\sigma=\pi}=0
\quad (i,j=p+1,\ldots,p^{\prime})~.
\label{eq:bc-rfermion}
\end{eqnarray}
It should be noted that these boundary conditions are
written on the complex plane,
while the boundary conditions in \cite{CIMM}
are written on the strip.
%%%%%%%%%%%%%%%%%%%%%%%%%%%%%%%%%%%%%%%%%%%%%%%%%%%%%%%%%%
\subsection{Quantization of a $p$-$p^{\prime}$ string}
In this subsection we will give the mode expansions of the
string coordinates of a $p$-$p^{\prime}$ open string
and the commutation relations among their modes.

Let us first consider the $x^{0}$-direction.
In this direction the string coordinate obeys the Neumann boundary
condition on both ends.
The coordinate $X^{0}(z,\overline{z})$ is expanded as
\begin{equation}
 X^{0}(z,\overline{z})=x^{0}-i\alpha^{\prime}p^{0}\ln(z\overline{z})
     +i\sqrt{\frac{\alpha^{\prime}}{2}}\sum_{m\neq 0}
        \frac{\alpha^{0}_{m}}{m}\left(z^{-m}+\overline{z}^{-m}\right)~.
\end{equation}
The modes satisfy the following commutation relations,
\begin{equation}
 [x^{0},p^{0}]=ig^{00}~,
\quad
 [x^{0},x^{0}]=[p^{0},p^{0}]=0~;
\quad
 [\alpha^{0}_{m},\alpha^{0}_{n}]=g^{00}m\delta_{m+n}~.
\end{equation}
The mode expansions of the NS fermions $\psi^{0}$ and
$\widetilde{\psi}^{0}$ become
\begin{equation}
 \psi^{0}(z)=\sum_{r\in {\bf Z}+1/2} b^{0}_{r}z^{-r-\frac{1}{2}}~,
\quad
 \widetilde{\psi}^{0}(\overline{z})=\sum_{r\in {\bf Z}+1/2}
     b^{0}_{r}\overline{z}^{-r-\frac{1}{2}}~.
\end{equation}
The oscillators satisfy
\begin{equation}
 \{b^{0}_{r},b^{0}_{s}\}=g^{00}\delta_{r+s}~.
\end{equation}
In the R sector, we have
\begin{equation}
 \psi^{0}(z)=\sum_{m\in{\bf Z}}d^{0}_{m}z^{-m-\frac{1}{2}}~,
\quad \widetilde{\psi}^{0}(\overline{z})
    =\sum_{m\in{\bf Z}}d^{0}_{m}\overline{z}^{-m-\frac{1}{2}}~,
\end{equation}
and
\begin{equation}
 \{d^{0}_{m},d^{0}_{n}\}=g^{00}\delta_{m+n}~.
\end{equation}

Next we consider the $x^{i}$-directions, $i=1,\ldots,p$.
In these directions, the boundary conditions on the string
coordinates are the same as those on the string coordinates
along the D$p$-branes in the D$p$-D$p$ system with $B$ field.
This implies that the mode expansions and the commutation
relations among the oscillators take the same form
as those in the D$p$-D$p$ system in the $B$ field background.
The mode expansions of the bosonic coordinates
$X^{i}(z,\overline{z})$ are
\begin{eqnarray}
\lefteqn{ X^{i}(z,\overline{z})
=x^{i}
 {}-i\alpha^{\prime}{\left[g^{-1}(g-2\pi\alpha^{\prime}B)\right]^{i}}_{j}
   \,p^{j}\ln z
 {}-i\alpha^{\prime}{\left[g^{-1}(g+2\pi\alpha^{\prime}B)\right]^{i}}_{j}
   \,p^{j}\ln \overline{z} }\nonumber\\
&&\hspace{3em}+i\sqrt{\frac{\alpha^{\prime}}{2}}
   \sum_{m\neq 0}
\left[ {\left(g^{-1}(g-2\pi\alpha^{\prime}B) \right)^{i}}_{j}z^{-m}
        +{\left(g^{-1}(g+2\pi\alpha^{\prime}B) \right)^{i}}_{j}
                                                  \overline{z}^{-m}
\right]\frac{\alpha^{j}_{m}}{m}~.\label{eq:mode-x-SW}
\end{eqnarray}
Under this expansion,
the commutation relations among the oscillators are
given \cite{CH}\cite{Jabba} by
\begin{equation}
 [x^{i},x^{j}]=i\theta^{ij}~,  \quad [p^{i},p^{j}]=0~,
\quad [x^{i},p^{j}]=i G^{ij}~;
\quad [\alpha^{i}_{m},\alpha^{j}_{n}]=G^{ij}m\delta_{m+n}~,
\end{equation}
where $\theta^{ij}$ is the noncommutativity parameter \cite{SW}
and $G^{ij}$ is the inverse of the open string metric $G_{ij}$
\cite{SW} defined respectively as
\begin{equation}
 \theta^{ij}=-(2\pi\alpha^{\prime})^{2}
    \left(\frac{1}{g+2\pi\alpha^{\prime}B}B
          \frac{1}{g-2\pi\alpha^{\prime}B}\right)^{ij}~,
\quad
 G^{ij}=\left(\frac{1}{g+2\pi\alpha^{\prime}B}g
              \frac{1}{g-2\pi\alpha^{\prime}B}\right)^{ij}~.
\end{equation}
We will further generalize $G^{ij}$ to include the time direction.
This is denoted by $G^{\mu\nu}$
and will simplify our formulas in the subsequent sections.
Taking eqs.~(\ref{eq:g-munu}) and (\ref{eq:B-ij}) into account,
we obtain
\begin{equation}
G^{\sigma\rho}=\left(
  \begin{array}{c|ccc}
   g^{00}& & & \\ \hline
       {}& & & \\
       {}& &G^{ij}& \\
       {}& & &
  \end{array}
\right)
=\left( \begin{array}{c|ccccc}
-1 & & & & &\\ \hline
 &\frac{1}{\varepsilon(1+b_1^2)}&0& & & \\
 &0&\frac{1}{\varepsilon(1+b_1^2)}& & & \\
 & & &\frac{1}{\varepsilon(1+b_2^2)}&0&  \\
 & & &0&\frac{1}{\varepsilon(1+b_2^2)}& \\
 & & & & &\ddots \\
 \end{array} \right).\label{eq:G0-p}
\end{equation}

We obtain for the NS fermions
\begin{equation}
\left\{
 \begin{array}{l}
   \displaystyle
     \psi^{i}(z)=\sum_{r\in {\bf Z}+1/2}
      {\left[g^{-1}(g-2\pi\alpha^{\prime} B)\right]^{i}}_{j}b^{j}_{r}
        z^{-r-\frac{1}{2}}\\
  \displaystyle
     \widetilde{\psi}^{i}(\overline{z})
      =\sum_{r\in{\bf Z}+1/2}
       {\left[g^{-1}(g+2\pi\alpha^{\prime}B)\right]^{i}}_{j}
       b^{j}_{r}\overline{z}^{-r-\frac{1}{2}}
 \end{array}
\right.~, 
  \quad\mbox{with \  $\{b^{i}_{r},b^{j}_{s}\}=G^{ij}\delta_{r+s}$}~,
\label{eq:mode-ns-SW}
\end{equation}
and for the R fermions
\begin{equation}
 \left\{
  \begin{array}{l}
  \displaystyle
    \psi^{i}(z)=\sum_{m\in {\bf Z}}
      {\left[g^{-1}(g-2\pi\alpha^{\prime} B)\right]^{i}}_{j}d^{j}_{m}
        z^{-m-\frac{1}{2}}\\
  \displaystyle
     \widetilde{\psi}^{i}(\overline{z})
      =\sum_{m\in{\bf Z}}
       {\left[g^{-1}(g+2\pi\alpha^{\prime}B)\right]^{i}}_{j}
       d^{j}_{m}\overline{z}^{-m-\frac{1}{2}}\\
 \end{array}
 \right.~,\quad
 \mbox{with \  $\{d^{i}_{m},d^{j}_{n}\}=G^{ij}\delta_{m+n}$}~.
\end{equation}

Finally we investigate the $x^{i}$-directions
$(i=p+1,\ldots,p^{\prime})$.
We complexify the string coordinates
${\bf X}^{i}({\bf z},\overline{\bf z})$ in these directions as
\begin{eqnarray}
 {\bf Z}^{I}({\bf z,\overline{z}})&=&
    {\bf X}^{2I-1}({\bf z,\overline{z}})
    +i{\bf X}^{2I}({\bf z,\overline{z}})
  = \sqrt{\frac{2}{\alpha^{\prime}}}Z^{I}(z,\bar{z})
      +i\theta\Psi^{I}(z)
      +i\overline{\theta}\widetilde{\Psi}^{I}(\overline{z})~,\nonumber\\
\overline{\bf Z}^{\overline{I}}({\bf z,\overline{z}})&=&
     {\bf X}^{2I-1}({\bf z,\overline{z}})
    {}-i{\bf X}^{2I}({\bf z,\overline{z}})
 = \sqrt{\frac{2}{\alpha^{\prime}}}
         \overline{Z}^{\overline{I}}(z,\overline{z})
   +i \theta\overline{\Psi}^{\overline{I}}(z)
   +i \overline{\theta}
         \widetilde{\overline{\Psi}}^{\overline{I}}(\overline{z})~,
\end{eqnarray}
where $I,\overline{I}=\frac{p+2}{2},\ldots,\frac{p^{\prime}}{2}$
and we have eliminated the auxiliary field $F^{i}$.
%In terms of $Z^{I}(z,\overline{z})$ and $\overline{Z}^{I}(z,\overline{z})$,
%the boundary conditions (\ref{eq:bc-boson}) on the bosonic coordinates in
%the $x^{i}$-directions with $i=p+1,\ldots,p^{\prime}$ are rewritten into
%\begin{eqnarray}
%&&\left.\partial_{\tau}Z^{I}\right|_{\sigma=0}=
%\left.\partial_{\sigma}Z^{I}
%       +b_{I}\partial_{\tau}Z^{I}\right|_{\sigma=\pi}=0~,\nonumber\\
%&&\left.\partial_{\tau}\overline{Z}^{I}\right|_{\sigma=0}=
%\left.\partial_{\sigma}\overline{Z}^{I}
%       {}-b_{I}\partial_{\tau}\overline{Z}^{I}\right|_{\sigma=\pi}=0~.
%\end{eqnarray}
{}From the boundary conditions eq.~(\ref{eq:bc-boson}) and equations
of motion, we find that
the mode expansions of $Z^{I}$ and $\overline{Z}^{\overline{I}}$
are given by
\begin{eqnarray}
 Z^{I}(z,\overline{z})
  =i\sqrt{\frac{\alpha^{\prime}}{2}}\sum_{n\in {\bf Z}}
     \frac{\alpha^{I}_{n-\nu_{I}}}{n-\nu_{I}}
     \left(z^{-(n-\nu_{I})}-\overline{z}^{-(n-\nu_{I})}\right)~,\nonumber\\
 \overline{Z}^{\overline{I}}(z,\overline{z})
   =i\sqrt{\frac{\alpha^{\prime}}{2}}\sum_{m\in {\bf Z}}
     \frac{\overline{\alpha}^{\overline{I}}_{m+\nu_{I}}}{m+\nu_{I}}
     \left( z^{-(m+\nu_{I})}-\overline{z}^{-(m+\nu_{I})} \right)~,
\end{eqnarray}
where $\nu_{I}$ are defined by
\begin{equation}
 e^{2\pi i \nu_{I}}=-\frac{1+ib_{I}}{1-ib_{I}}~,
\quad 0< \nu_{I} < 1~.
\end{equation}
Now we can introduce the open string metric
$G^{IJ}$, $G^{\overline{I}\overline{J}}$, $G^{I\overline{J}}$
and $G^{\overline{I}J}$
concerning the $x^{p+1},\ldots,x^{p^\prime}$ directions,
\begin{equation}
G^{IJ}=G^{\overline{I}\overline{J}}=0~,\quad
G^{I\overline{J}}=G^{\overline{J}I}
  =\frac{2}{\varepsilon(1+b^2_I)}\delta^{I\overline{J}}~.
\end{equation}

Similarly, the boundary conditions eq.~(\ref{eq:bc-nsfermion}),
eq.~(\ref{eq:bc-rfermion}) and equations of motion 
lead us to the mode expansions of the NS-fermions,
\begin{eqnarray}
 \Psi^{I}(z)=\sum_{r\in {\bf Z}+1/2}b^{I}_{r-\nu_{I}}
             z^{-(r-\nu_{I})-\frac{1}{2}}~,
 &&
  \widetilde{\Psi}^{I}(\overline{z})=-\sum_{r\in{\bf Z}+1/2}
              b^{I}_{r-\nu_{I}}\overline{z}^{-(r-\nu_{I})-\frac{1}{2}}~,
    \nonumber\\
\overline{\Psi}^{\overline{I}}(z)
    =\sum_{s\in{\bf Z}+1/2}\overline{b}^{\overline{I}}_{s+\nu_{I}}
            z^{-(s+\nu_{I})-\frac{1}{2}}~,
  &&
   \widetilde{\overline{\Psi}}^{\overline{I}}(\overline{z})
       =-\sum_{s\in{\bf Z}+1/2}
            \overline{b}^{\overline{I}}_{s+\nu_{I}}
            \overline{z}^{-(s+\nu_{I})-\frac{1}{2}}~,
\end{eqnarray}
and that of R-fermions,
\begin{eqnarray}
 \Psi^{I}(z)=\sum_{n\in{\bf Z}}d^{I}_{n-\nu_{I}}
             z^{-(n-\nu_{I})-\frac{1}{2}}~,
    && \widetilde{\Psi}^{I}(\overline{z})=-\sum_{n\in{\bf Z}}
            d^{I}_{n-\nu_{I}}
            \overline{z}^{-(n-\nu_{I})-\frac{1}{2}}~,\nonumber\\
 \overline{\Psi}^{\overline{I}}(z)
         =\sum_{m\in{\bf Z}}\overline{d}^{\overline{I}}_{m+\nu_{I}}
                z^{-(m+\nu_{I})-\frac{1}{2}}~,
    && \widetilde{\overline{\Psi}}^{\overline{I}}(\overline{z})
          =-\sum_{m\in{\bf Z}}\overline{d}^{\overline{I}}_{m+\nu_{I}}
              \overline{z}^{-(m+\nu_{I})-\frac{1}{2}}~.
\end{eqnarray}
The commutation relations
are
\begin{eqnarray}
&&[\alpha^{I}_{n-\nu_{I}} \, , \,
        \overline{\alpha}^{\overline{J}}_{m+\nu_{J}}]
   =\frac{2}{\varepsilon}\delta^{I\overline{J}}(n-\nu_{I})
                   \delta_{n+m}~,\nonumber\\
&&\{b^{I}_{r-\nu_{I}}\, , \,\overline{b}^{\overline{J}}_{s+\nu_{J}}\}
     =\frac{2}{\varepsilon} \, \delta^{I\overline{J}}\delta_{r+s}~,
 \qquad
  \{d^{I}_{n-\nu_{I}}\, , \,\overline{d}^{\overline{J}}_{m+\nu_{J}}\}
      =\frac{2}{\varepsilon} \, \delta^{I\overline{J}}\delta_{n+m}~.
\end{eqnarray}

As for the $x^{p^{\prime}+1},\ldots,x^{9}$-directions, 
the string coordinates obey the Dirichlet boundary condition. 
Our analysis
in the remaining part of this paper
does not involve these directions.

%%%%%%%%%%%%%%%%%%%%%%%%%%%%%%%%%%%%%%%%%%%%%%%%%%%%%%%%%%%%%%%%%%%%%%%%%%
\subsection{Two-point functions on superspace}
In this subsection we construct two-point functions of the 
$p$-$p^{\prime}$ open string coordinates on superspace.
For this purpose, we begin by defining the oscillator vacuum of the
system.

As shown in the last subsection, the mode expansions
in the $x^{0}$ and the $x^{i}$-directions $(i=1,\ldots,p)$
are  similar to those of the usual open strings obeying Neumann boundary
conditions in the sense that the bosons and the R fermions
have integral moding oscillators
and the NS fermions have half-integral moding ones.
Therefore we can define the vacuum in these directions
in the same way as the usual open string.
In the NS sector the vacuum $| 0 \rangle$ is
defined by
\begin{equation}
 \left\{
  \begin{array}{ccl}
      \alpha^{0}_{m}|0\rangle=0~, & \alpha^{i}_{m}|0\rangle=0~,
          & \mbox{for $m\geq 0$}\\
      b^{0}_{r}|0\rangle=0~,      & b^{i}_{r}|0\rangle=0~,
          & \mbox{for $r\geq\frac{1}{2}$}
  \end{array}
 \right.~,
\end{equation}
where $\alpha^{\mu}_{0}=\sqrt{2\alpha^{\prime}}p^{\mu}$
($\mu=0,1,\ldots,p$).
In the R sector the vacuum $| S^{\alpha} \rangle$
belongs to the spinor representation of the $SO(p,1)$ group.
Now that the commutation relations and the vacuum $|0\rangle$
are determined,
we can evaluate the two-point functions of the string coordinates
in these directions \cite{SW}\cite{Callan1}\cite{GM},
\begin{eqnarray}
 \lefteqn{
   {\bf G}^{00}\left({\bf z}_{1},\overline{\bf z}_{1}|
                   {\bf z}_{2},\overline{\bf z}_{2}\right)
  \equiv \langle 0 | 
             {\cal R}{\bf X}^{0}({\bf z}_{1},\overline{\bf z}_{1})
             {\bf X}^{0}({\bf z}_{2},\overline{\bf z}_{2})
           |0\rangle
   }\nonumber\\
  && = -g^{00}\left[ \ln (z_{1}-z_{2}-\theta_{1}\theta_{2})
                     (\overline{z}_{1}-\overline{z}_{2}
                        {}-\overline{\theta}_{1}\overline{\theta}_{2})
                    + \ln (z_{1}-\overline{z}_{2}
                        {}-\theta_{1}\overline{\theta}_{2})
                        (\overline{z}_{1}-z_{2}
                        {}-\overline{\theta}_{1}\theta_{2})
          \right]~,\nonumber\\
 \lefteqn{{\bf G}^{ij}\left({\bf z}_{1},\overline{\bf z}_{1}|
                   {\bf z}_{2},\overline{\bf z}_{2}\right)
  \equiv \langle 0 | 
             {\cal R}{\bf X}^{i}({\bf z}_{1},\overline{\bf z}_{1})
             {\bf X}^{j}({\bf z}_{2},\overline{\bf z}_{2})
           |0\rangle
   }\nonumber\\
  &&=-g^{ij} \ln (z_{1}-z_{2}-\theta_{1}\theta_{2})
                     (\overline{z}_{1}-\overline{z}_{2}
                        {}-\overline{\theta}_{1}\overline{\theta}_{2})
     +(g^{ij}-2G^{ij}) \ln (z_{1}-\overline{z}_{2}
                        {}-\theta_{1}\overline{\theta}_{2})
                        (\overline{z}_{1}-z_{2}
                        {}-\overline{\theta}_{1}\theta_{2})
    \nonumber\\
   && \hspace{1em}-2\frac{\theta^{ij}}{2\pi\alpha^{\prime}}
      \ln \frac{z_{1}-\overline{z}_{2}-\theta_{1}\overline{\theta}_{2}}
               {\overline{z}_{1}-z_{2}-\overline{\theta}_{1}\theta_{2}}
       {}-2D^{ij}~,\label{eq:propSW-1}
\end{eqnarray}
where ${\cal R}$ stands for the radial ordering.
$D^{ij}$ are the contributions from the zero modes $x^{i}$
and these will be fixed conveniently as is done in \cite{SW}.
When we restrict these two-point functions onto the D$p^{\prime}$-brane
worldvolume, i.e.\ the worldsheet boundary characterized by
$z=e^{\tau+i\pi}=-e^{\tau}$ and $\theta=\overline{\theta}$,
they become
\begin{eqnarray}
 {\bf G}^{00}\left(-e^{\tau_{1}},\theta_{1}
                     | -e^{\tau_{2}},\theta_{2}\right)
  &\equiv& \left.{\bf G}^{00}\left({\bf z}_{1},\overline{\bf z}_{1}
                          | {\bf z}_{2},\overline{\bf z}_{2}\right)
         \right|_{\sigma=\pi,\theta=\overline{\theta}}
      =-2g^{00}\ln(e^{\tau_{1}}-e^{\tau_{2}}+\theta_{1}\theta_{2})^{2}~,
    \nonumber\\
 {\bf G}^{ij}\left(-e^{\tau_{1}},\theta_{1}|-e^{\tau_{2}},\theta_{2}\right)
  &\equiv& \left.{\bf G}^{ij}\left({\bf z}_{1},\overline{\bf z}_{1}
                          | {\bf z}_{2},\overline{\bf z}_{2}\right)
       \right|_{\sigma=\pi,\theta=\overline{\theta}}
    \nonumber\\
  &=&-2G^{ij}\ln(e^{\tau_{1}}-e^{\tau_{2}}+\theta_{1}\theta_{2})^{2}
    -\frac{i}{\alpha^{\prime}}\theta^{ij}\epsilon(\tau_{1}-\tau_{2})~,
\label{eq:propSW-2}
\end{eqnarray}
where $\epsilon (x)$ is the sign function.

In the $x^{i}$-directions $(i=p+1,\ldots,p^{\prime})$, the situation
is more complex as the string coordinates are expanded in
non-integer power of $z$ and $\overline{z}$.
We define the oscillator vacuum $\left|\sigma\right\rangle$
for the bosonic sector so that this should be annihilated by
the negative energy modes:
\begin{equation}
\left| \sigma \right\rangle 
  = \bigotimes_{I}\left|\sigma_{I}\right\rangle
\qquad
\mbox{with}
\qquad
\left\{
\begin{array}{ll}
 \alpha^{I}_{n-\nu_{I}}\left|\sigma_{I}\right\rangle=0
&n>\nu_{I} \\
 \overline{\alpha}^{\overline{I}}_{m+\nu_{I}}
        \left|\sigma_{I}\right\rangle=0
&m>-\nu_{I}
\end{array}
\right.~. \label{eq:osci-vac-z}
\end{equation}
For the fermions in the NS sector
we define the oscillator vacuum $\left|s\right\rangle$
by\footnote{The vacuum $\left|s\right\rangle$ is defined
in order that the negative energy
and the positive energy modes for $0<\nu_{I}<1/2$
should be the annihilation and the creation modes respectively.
The energy carried by the lowest creation mode
$\overline{b}^{\overline{I}}_{-\frac{1}{2}+\nu_{I}}$ becomes negative
when $\nu_{I}$ becomes greater than $1/2$.
We could define another oscillator vacuum
$\left|\widetilde{s}\right\rangle$ by
\[
 \left|\widetilde{s}\right\rangle
      = \bigotimes_{I}\left|\widetilde{s}_{I}\right\rangle
\qquad
\mbox{with}
\qquad
\left\{
\begin{array}{ll}
 b^{I}_{r-\nu_{I}}\left| \widetilde{s}_{I} \right\rangle =0
& r \geq \frac{3}{2} \\
 \overline{b}^{\overline{I}}_{s+\nu_{I}}
       \left|\widetilde{s}_{I}\right\rangle =0
& s \geq -\frac{1}{2}
\end{array}
\right.~,
\]
which makes the negative energy
and the positive energy modes for $1/2< \nu_{I} < 1$
to be the annihilation and the creation modes respectively.\label{s-tilde}}
\begin{equation}
 \left|s\right\rangle = \bigotimes_{I}\left|s_{I}\right\rangle
\qquad
\mbox{with}
\qquad
\left\{
\begin{array}{ll}
 b^{I}_{r-\nu_{I}}\left| s_{I} \right\rangle =0
& r \geq \frac{1}{2} \\
 \overline{b}^{\overline{I}}_{s+\nu_{I}}\left|s_{I}\right\rangle =0
& s \geq \frac{1}{2}
\end{array}
\right.~,\label{eq:osci-vac-ns}
\end{equation}
and for the R sector we define the oscillator vacuum
$\left| S \right\rangle $ by
\begin{equation}
 \left|S\right\rangle
  =\bigotimes_{I}\left|S_{I}\right\rangle
\qquad  \mbox{with} \qquad
\left\{
 \begin{array}{ll}
  d^{I}_{n-\nu_{I}}\left|S_{I}\right\rangle=0
& n>\nu_{I} \\
  \overline{d}^{\overline{I}}_{m+\nu_{I}}
        \left|S_{I}\right\rangle=0
& m>-\nu_{I}
 \end{array}
\right.~.\label{eq:osci-vac-r}
\end{equation}
By using the commutation relations and the defining relations of
the vacua, we can calculate the two-point functions%%%%%%%
\footnote{If we adopted
$\left|\widetilde{s}\right\rangle$ as the vacuum of the NS fermions
instead of $\left|s\right\rangle$,
the two-point function of the supercoordinates ${\bf Z}^{I}$ and
$\overline{\bf Z}^{\overline{I}}$ would be
\begin{eqnarray*}
\lefteqn{
 \widetilde{\mbox{\boldmath${\cal G}$}}^{I\overline{J}}
 \left({\bf z}_{1},\overline{\bf z}_{1}
       | {\bf z}_{2},\overline{\bf z}_{2}  \right) \equiv
  \left\langle \sigma,\widetilde{s}\right|
        {\cal R}{\bf Z}^{I}({\bf z}_{1},\overline{\bf z}_{1})
        \overline{\bf Z}^{\overline{J}}
                   ({\bf z}_{2},\overline{\bf z}_{2})
 \left| \sigma,\widetilde{s}\right\rangle}\\
&&=\Theta(|z_{1}|-|z_{2}|)\frac{2\delta^{I\overline{J}}}{\varepsilon}
   \left[  {\cal F}\left(1-\nu_{I}\,;\,
                       \frac{z_{2}}{z_{1}-\theta_{1}\theta_{2}}
                   \right)
    + {\cal F}\left(1-\nu_{I}\,;\,
      \frac{\overline{z}_{2}}
          {\overline{z}_{1}-\overline{\theta}_{1}\overline{\theta}_{2}}
              \right)\right.\\
  &&\hspace{3em}\left.- {\cal F}\left(1-\nu_{I}\,;\,
      \frac{\overline{z}_{2}}
          {z_{1}-\theta_{1}\overline{\theta}_{2}}
              \right)
    {}-{\cal F}\left(1-\nu_{I}\,;\,
      \frac{z_{2}}
          {\overline{z}_{1}-\overline{\theta}_{1}\theta_{2}}
              \right)\right]\\
&&+\Theta(|z_{2}|-|z_{1}|)\frac{2\delta^{I\overline{J}}}{\varepsilon}
    \left[ {\cal F}\left(\nu_{I}\,;\,
             \frac{z_{1}-\theta_{1}\theta_{2}}{z_{2}}
                   \right)
      +{\cal F}\left(\nu_{I}\,;\,
             \frac{\overline{z}_{1}
                 {}-\overline{\theta}_{1}\overline{\theta}_{2}}
                  {\overline{z}_{2}}\right)
    {}-{\cal F}\left(\nu_{I}\,;\,
             \frac{z_{1}
                 {}-\theta_{1}\overline{\theta}_{2}}
                  {\overline{z}_{2}}\right)
     {}-{\cal F}\left(\nu_{I}\,;\,
             \frac{\overline{z}_{1}
                 {}-\overline{\theta}_{1}\theta_{2}}{z_{2}}\right)
     \right]~.
\end{eqnarray*}
}: %%%
\begin{eqnarray}
&&\mbox{\boldmath$\cal G$}^{I\overline{J}}
   \left({\bf z}_{1},\overline{\bf z}_{1}
         | {\bf z}_{2},\overline{\bf z}_{2}\right)
 \equiv
   \left\langle \sigma,s \right|{\cal R}
     {\bf Z}^{I}({\bf z}_{1},\overline{\bf z}_{1})
      \overline{\bf Z}^{\overline{J}}({\bf z}_{2},\overline{\bf z}_{2})
  \left| \sigma,s\right\rangle\nonumber\\
&&=\Theta(|z_{1}|-|z_{2}|)\frac{2\delta^{I\overline{J}}}{\varepsilon}
  \left[ 
  {\cal F}\left(1-\nu_{I} \, ;
                 \, \frac{z_{2}+\theta_{1}\theta_{2}}{z_{1}}\right)
 +{\cal F}\left(1-\nu_{I}\, ; \,
     \frac{\overline{z}_{2}+\overline{\theta}_{1}\overline{\theta_{2}}}
          {\overline{z_{1}}}\right)   
    \right.\nonumber \\
 && \hspace{9em}\left.
 {}-{\cal F}\left(1-\nu_{I}\,;\,
     \frac{\overline{z}_{2}+\theta_{1}\overline{\theta_{2}}}{z_{1}}
            \right)
 {}-{\cal F}\left(1-\nu_{I};
     \frac{z_{2}+\overline{\theta}_{1}\theta_{2}}
          {\overline{z_{1}}}\right)\right]\nonumber\\
&& +\Theta(|z_{2}|-|z_{1}|)\frac{2\delta^{I\overline{J}}}{\varepsilon}
   \left[ 
  {\cal F}\left(\nu_{I} \, ;
                 \, \frac{z_{1}}{z_{2}+\theta_{1}\theta_{2}}\right)
 +{\cal F}\left(\nu_{I}\, ; \,
     \frac{\overline{z_{1}}}
          {\overline{z}_{2}+\overline{\theta}_{1}\overline{\theta_{2}}}
    \right)   \right.\nonumber \\
 && \hspace{9em}\left.
 {}-{\cal F}\left(\nu_{I}\,;\,
     \frac{z_{1}}{\overline{z}_{2}+\theta_{1}\overline{\theta_{2}}}
            \right)
 {}-{\cal F}\left(\nu_{I};
     \frac{\overline{z_{1}}}
          {z_{2}+\overline{\theta}_{1}\theta_{2}}  \right)\right]~,
\label{eq:propCIMM-1}
\end{eqnarray}
where $\Theta(x)$ is the step function,
${\cal F}(\nu\,;\,z)$ is defined as
\begin{equation}
 {\cal F}(\nu\,;\, z) =\frac{z^{\nu}}{\nu}\, F(1,\nu;1+\nu;z)
 =\sum_{n=0}^{\infty}\frac{1}{n+\nu}z^{n+\nu}~,
\end{equation}
and $F(a,b;c;z)$ is the hypergeometric function.
When we restrict this two-point function onto the worldsheet boundary on
the D$p^{\prime}$-brane worldvolume,
this becomes
\begin{eqnarray}
\lefteqn{
\mbox{\boldmath$\cal G$}^{I \overline{J}}
    \left(-e^{\tau_{1}},\theta_{1} | -e^{\tau_{2}},\theta_{2}\right)
  \equiv
  \left.\mbox{\boldmath$\cal G$}^{I\overline{J}}
    \left({\bf z}_{1},\overline{\bf z}_{1}|{\bf z}_{2},\overline{\bf z}_{2}
    \right)
  \right|_{\sigma=\pi,\theta=\overline{\theta}}
}\nonumber\\
&&
   =4 G^{I\overline{J}}
  \left[\Theta(\tau_{1}-\tau_{2})
     {\cal F}\left(1-\nu_{I}\,;\,
               \frac{e^{\tau_{2}}-\theta_{1}\theta_{2}}{e^{\tau_{1}}}
             \right)
       +\Theta(\tau_{2}-\tau_{1})
     {\cal F}\left(\nu_{I}\,;\,
                \frac{e^{\tau_{1}}}{e^{\tau_{2}}-\theta_{1}\theta_{2}}
             \right)\right]~.\label{eq:propCIMM-2}
\end{eqnarray}

Now we would like to study the two-point function
eq.~(\ref{eq:propCIMM-2}) more closely.
Let us recast the right hand side of  eq.~(\ref{eq:propCIMM-2}) into
\begin{eqnarray}
&& 4 G^{I\overline{J}}
\left[ \frac{1}{2} \left\{
{\cal F}\left(1-\nu_{I};\frac{e^{\tau_{2}}}{e^{\tau_{1}}} \right)
 +{\cal F}\left(\nu_{I};\frac{e^{\tau_{1}}}{e^{\tau_{2}}}
 \right)\right\}-\theta_{1}\theta_{2}
    \frac{\left(\frac{e^{\tau_{1}}}{e^{\tau_{2}}}\right)^{\nu_{I}}}
         {e^{\tau_{1}} - e^{\tau_{2}}} \right]\nonumber\\
&&
+\epsilon(\tau_{1}-\tau_{2})\,
   \frac{4}{\varepsilon}
   \frac{\delta^{I\overline{J}}}{1+b_{I}^{2}} \left\{
    {\cal F}\left(1-\nu_{I};\frac{e^{\tau_{2}}}{e^{\tau_{1}}}\right)
    {}-{\cal F}\left(\nu_{I};\frac{e^{\tau_{1}}}{e^{\tau_{2}}}\right)
\right\}~.\label{eq:epsilon}
\end{eqnarray}
As noncommutativity of the D$p^{\prime}$-brane worldvolume
originates from the term proportional to the sign function
$\epsilon(\tau_{1}-\tau_{2})$ in the above equation
\cite{CIMM}\cite{SW},
we will also refer to this term as noncommutativity term
in what follows.
Here it should be noted that
by using the relations,
\begin{eqnarray}
 \frac{d}{dz}\left[ z^{c-1} F(a,b;c;z)\right]
&=&(c-1)z^{c-2}F(a,b;c-1;z)~,\nonumber\\
F(a,b;b;z)&=&(1-z)^{-a}~.\label{eq:power}
\end{eqnarray}
we can obtain
\begin{equation}
 \frac{d}{dz}\left[ {\cal F}\left(1-\nu_{I};\frac{1}{z}\right)
                    {}-{\cal F}\left(\nu_{I};z\right)\right]=0.
\label{eq:bibun}
\end{equation}
This implies that the noncommutativity term
in eq.~(\ref{eq:epsilon}) is constant.
The value of this constant can be fixed by evaluating
the noncommutativity term at a certain point on the real axis,
such as $\frac{e^{\tau_{1}}}{e^{\tau_{2}}}=1$.
By using the hypergeometric series, we find that
\begin{equation}
 {\cal F}(1-\nu_{I};1)-{\cal F}(\nu_{I};1)
=- \sum_{n=-\infty}^{\infty}\frac{1}{n+\nu_{I}}
=-\pi \cot \left( \pi\nu_{I} \right)=\pi b_{I}~.
\label{eq:hyperseries}
\end{equation}
Thus we find that the noncommutativity term becomes
\begin{equation}
 \frac{4}{\varepsilon}\frac{\delta^{I\overline{J}}}{1+b_{I}^{2}}
\left\{{\cal F}\left(1-\nu_{I};\frac{e^{\tau_{2}}}{e^{\tau_{1}}}\right)
 {}-{\cal F}\left(\nu_{I};\frac{e^{\tau_{1}}}{e^{\tau_{2}}}\right)
 \right\}
=  \frac{4}{\varepsilon}
  \frac{\delta^{I\overline{J}}}{1+b_{I}^{2}}  \pi b_{I}~.
\label{eq:samenc}
\end{equation}
When we rewrite the complex string coordinates ${\bf Z}^{I}$
and $\overline{\bf Z}^{\overline{I}}$ into the real one
${\bf X}^{i}$ $(i=p+1,\ldots,p^{\prime})$,
this noncommutativity term takes the same form
as that in eq.~(\ref{eq:propSW-2}).
This means that, as is pointed out in \cite{CIMM},
the noncommutativity on the D-brane worldvolume in the $p$-$p^{\prime}$
system is the same as that in the $p$-$p$ system
\cite{SW}\cite{CH}\cite{Jabba}.
{}From eq.~(\ref{eq:samenc}), we conclude that
\begin{equation}
 \mbox{\boldmath${\cal G}$}^{I\overline{J}}
 \left(-e^{\tau_{1}},\theta_{1} | -e^{\tau_{2}},\theta_{2}\right)
=4 G^{I\overline{J}}
  \,{\cal H}\left(\nu_{I};
         \frac{e^{\tau_{1}}}{e^{\tau_{2}}-\theta_{1}\theta_{2}}\right)
+\epsilon(\tau_{1}-\tau_{2}) \frac{4}{\varepsilon}
 \frac{\delta^{I\overline{J}}}{1+b_{I}^{2}}\pi b_{I}~,
\label{eq:propCIMM-3}
\end{equation}
where ${\cal H}(\nu;z)$ is defined by using the hypergeometric
series as
\begin{equation}
 {\cal H}(\nu;z)=\left\{
\begin{array}{ll}
 \displaystyle{\cal F} \left(1-\nu_{I};\frac{1}{z}\right)
     {}-\frac{\pi}{2} b_{I}
   =\sum_{n=0}^{\infty}\frac{z^{-n-1+\nu_{I}}}{n+1-\nu_{I}}
     {}-\frac{\pi}{2} b_{I}&
     \mbox{ for $|z| > 1$}\\[3ex]
 \displaystyle{\cal F} \left( \nu_{I};z \right)+\frac{\pi}{2}b_{I}
  =\sum_{n=0}^{\infty}\frac{z^{n+\nu_{I}}}{n+\nu_{I}}
   +\frac{\pi}{2}b_{I} &
     \mbox{ for $|z| <1$}
\end{array}
\right. ~. \label{eq:h-prop}
\end{equation}
The two infinite series in the above defining relation
should be analytically continued to each other.

In Appendix
%~\ref{sect:hyper}
we give another derivation of the
two-point function (\ref{eq:propCIMM-3}) from eq.~(\ref{eq:propCIMM-2}).

%%%%%%%%%%%%%%%%%%%%%%%%%%%%%%%%%%%%%%%%%%%%%%%%%%%%%%%%%%%%%%%%%%%%%%
\subsection{Twist field and spin field}\label{sect:twist}
Here we would like to make further consideration on the
oscillator vacuum consisting of the bosonic sector $|\sigma\rangle$ and 
the fermionic sector $|s\rangle$~.
As is explained in the last subsection,
the primary fields $D{\bf Z}^{I}$,
$\overline{D}{\bf Z}^{I}$,
$D\overline{\bf Z}^{\overline{I}}$
and $\overline{D}\overline{\bf Z}^{\overline{I}}$
 defined on the upper half plane
are expanded in non-integer powers of $z$ and $\overline{z}$.
It follows that when we extend the defining region
of these fields to the whole complex plane through
the doubling trick these fields become multi-valued functions
on the whole plane.
For example, when 
the primary fields $\partial Z^{I}(z)$,
$\overline{\partial}Z^{I}(\overline{z})$,
$\partial \overline{Z}^{\overline{I}}(z)$
and $\overline{\partial}\overline{Z}^{\overline{I}}(\overline{z})$
on the whole plane
are transported once around the origin,
they gain phase factors:
\begin{eqnarray}
 \partial Z^{I}(e^{2\pi i}z)=e^{2\pi i\nu_{I}}\partial Z^{I}(z)~,
&&
\overline{\partial}Z^{I}(e^{2\pi i}\overline{z})
   =e^{2\pi i \nu_{I}}\overline{\partial}Z^{I}(\overline{z})
\nonumber\\
\partial\overline{Z}^{\overline{I}}(e^{2\pi i}z)
    =e^{-2\pi i\nu_{I}}\partial\overline{Z}^{\overline{I}}(z)~,
&&\overline{\partial}\overline{Z}^{\overline{I}}(e^{2\pi i}\overline{z})
   =e^{-2\pi i\nu_{I}}
    \overline{\partial}\overline{Z}^{\overline{I}}(\overline{z})~.
\end{eqnarray}
This implies that a twist field $\sigma^{+}_{I}(\xi^{1})$
and an anti-twist field $\sigma^{-}_{I}(\xi^{1})$, both of
which are mutually non-local with respect to
$Z^{I}$ and $\overline{Z}^{I}$,
are located at the origin and at  infinity on the plane respectively.
They create a branch cut between themselves.
The twist field $\sigma^{+}$
serves as a boundary changing operator from the $p^{\prime}$-brane
to the $p$-brane and the anti-twist field $\sigma^{-}$ acts
in the opposite way  \cite{AH}\cite{GNS}.
The incoming vacuum $\left|\sigma_{I}\right\rangle$
defined in eq.~(\ref{eq:osci-vac-z}) should be interpreted
as being excited from the $SL(2,{\bf R})$-invariant vacuum
$|0\rangle$ by the twist field $\sigma^{+}_{I}$:
\begin{equation}
 \left|\sigma_{I}\right\rangle
 = \lim_{\xi^{1}\rightarrow 0}\sigma^{+}_{I}(\xi^{1})
   \,|0\rangle~.
\end{equation}
In the same way, the outgoing vacuum $\langle \sigma_{I}|$
should be regarded as
\begin{equation}
 \langle \sigma_{I} |
  = \lim_{\widetilde{\xi}^{1}\rightarrow 0} 
     \left(\frac{1}{\widetilde{\xi}^{1}}\right)^{2h_{\sigma_{I}}}
     \,  \langle 0 |
    ~\sigma^{-}_{I}\left(-\frac{1}{\widetilde{\xi}^{1}}\right)~,
\end{equation}
where $h_{\sigma_{I}}$ denotes the weight
of the (anti-) twist field. We will later explain that
$h_{\sigma_{I}} = \frac{1}{2}\nu_{I}(1-\nu_{I})$ \cite{DFMS}.
We can read off the OPE's of $Z^{I}$ and $\overline{Z}^{I}$ with
$\sigma^{\pm}_{I}$ from eq.~(\ref{eq:osci-vac-z}):
\begin{eqnarray}
&& \left\{
     \begin{array}{ll}
        \partial Z^{I}(z) \sigma^{+}_{J}(0)
           \sim 
         \delta^{I}_{J}z^{-(1-\nu_{I})}\tau_{I}^{+}(0)~,
   & \overline{\partial}Z^{I}(\overline{z})
     \sigma^{+}_{J}(0)
           \sim   \delta^{I}_{J}\overline{z}^{-(1-\nu_{I})}
                  \widetilde{\tau}^{+}_{I}(0)~,
      \\
        \partial\overline{Z}^{\overline{I}}(z)
              \sigma^{+}_{J}(0)
            \sim
         \delta^{\overline{I}}_{J}z^{-\nu_{I}}
           \tau^{\prime +}_{I}(0)~,
    & \overline{\partial}\overline{Z}^{\overline{I}}(\overline{z})
      \sigma^{+}_{J}(0)
          \sim   \delta^{\overline{I}}_{J}\overline{z}^{-\nu_{I}}
                 \widetilde{\tau}^{\prime +}_{I}(0)~,
     \end{array}
    \right.\\
&& \left\{
     \begin{array}{ll}
        \partial Z^{I}(z) \sigma^{-}_{J}(0)
           \sim 
         \delta^{I}_{J}z^{-\nu_{I}}\tau_{I}^{-}(0)~,
     & \overline{\partial}Z^{I}(\overline{z})\sigma^{-}_{J}(0)
          \sim \delta^{I}_{J}\overline{z}^{-\nu_{I}}
               \widetilde{\tau}^{-}_{I}(0)~,\\
        \partial\overline{Z}^{\overline{I}}(z)\sigma^{-}_{J}(0)
            \sim
         \delta^{\overline{I}}_{J}z^{-(1-\nu_{I})}
           \tau^{\prime -}_{I}(0)~,
   & \overline{\partial}\overline{Z}^{\overline{I}}(\overline{z})
       \sigma^{-}_{J}(0)
      \sim  \delta^{I}_{\overline{J}}\overline{z}^{-(1-\nu_{I})}
            \widetilde{\tau}^{\prime -}_{I}(0)~,
     \end{array}
    \right.
\end{eqnarray}
where $\tau$~'s are excited twist fields.

Similar argument holds for the fermionic coordinates.
In the NS sector, the spin fields $s^{+}_{I}$ and $s^{-}_{I}$
are mutually non-local with respect to the fermions.
They are located at the origin and at the infinity
on the worldsheet respectively.
They exchange the boundary conditions corresponding to
the $p$-brane and those to the $p^{\prime}$-brane
by generating a branch cut between themselves.
The incoming vacuum $\left|s_{I}\right\rangle$
and the outgoing vacuum $\langle s_{I} |$ should be regarded as
being excited from the $SL(2,{\bf R})$-invariant vacuum by
spin fields:
\begin{equation}
 \left|s_{I}\right\rangle
   = \lim_{\xi^{1}\rightarrow 0}s^{+}_{I}(\xi^{1})|0\rangle~,
\quad
 \langle s_{I} |
  = \lim_{\widetilde{\xi}^{1} \rightarrow 0}
      \left(\frac{1}{\widetilde{\xi}^{1}}\right)^{2h_{s_{I}}}
      \,\langle 0 |~s^{-}_{I}\left(-\frac{1}{\widetilde{\xi}^{1}}
      \right)~,
\end{equation}
where $h_{s_{I}}$ is the weight of the spin fields
which will be found to be $h_{s_{I}}=\frac{1}{2}{\nu_{I}}^{2}$.
The defining relation eq.~(\ref{eq:osci-vac-ns}) yields the OPE's,
\begin{eqnarray}
&& \left\{
    \begin{array}{ll}
     \Psi^{I}(z)s^{+}_{J}(0)
       \sim \delta^{I}_{J} z^{+\nu_{I}} t^{\prime +}_{I}(0)~,
     & \widetilde{\Psi}^{I}(\overline{z}) s^{+}_{J}(0)
       \sim \delta^{I}_{J} \overline{z}^{+\nu_{I}}
              \widetilde{t}^{\prime +}_{I}(0)~, \\
     \overline{\Psi}^{\overline{I}}(z)s^{+}_{J}(0)
        \sim \delta^{\overline{I}}_{J} z^{-\nu_{I}} t^{+}_{I}(0)~,
     & \widetilde{\overline{\Psi}}^{\overline{I}}(\overline{z})
        s^{+}_{J}(0)
       \sim \delta^{\overline{I}}_{J}\overline{z}^{-\nu_{I}}
        \widetilde{t}^{+}(0)~,
    \end{array}
 \right.\\
&&
 \left\{
    \begin{array}{ll}
     \Psi^{I}(z)s^{-}_{J}(0)
       \sim \delta^{I}_{J} z^{-\nu_{I}} t^{\prime -}_{I}(0)~,
     & \widetilde{\Psi}^{I}(\overline{z})s^{-}_{J}(0)
       \sim \delta^{I}_{J} \overline{z}^{-\nu_{I}}
       \widetilde{t}^{\prime -}_{I}(0) \\
     \overline{\Psi}^{\overline{I}}(z)s^{-}_{J}(0)
        \sim \delta^{\overline{I}}_{J} z^{+\nu_{I}} t^{-}_{I}(0)~,
    & \widetilde{\overline{\Psi}}^{\overline{I}}(\overline{z})
        s^{-}_{J}(0)
       \sim \delta^{\overline{I}}_{J}\overline{z}^{+\nu_{I}}
       \widetilde{t}^{-}_{I}(0)~,
    \end{array}~,
 \right.
\label{eq:ope-spin}
\end{eqnarray}
In the fermionic sector, the bosonization simplifies
the treatment of the spin fields.
While we will not invoke this treatment here,
we include it here for the sake of completeness.
We write
\begin{equation}
 \Psi^{I}(z) \cong \sqrt{\frac{2}{\varepsilon}}e^{iH^{I}(z)}~,
\quad
 \overline{\Psi}^{\overline{I}}(z)
             \cong \sqrt{\frac{2}{\varepsilon}}e^{-iH^{I}(z)}~,
\end{equation}
where $H^{I}(z)$ are free bosons normalized as
$H^{I}(z)H^{J}(w)\sim -\delta^{IJ}\ln (z-w)$.
Then the OPE (\ref{eq:ope-spin}) tells us that
the spin fields $s^{\pm}_{I}(z)$ should be bosonized as
\footnote{We are also able to perform the same analysis
on the other incoming and the outgoing vacua,
$|\widetilde{s}_{I}\rangle$ and $\langle \widetilde{s}|$,
in the NS sector defined in the footnote \ref{s-tilde}.
These states are excited from the $SL(2,{\bf R})$-invariant
vacuum by $\widetilde{s}^{+}_{I}(z)$ and $\widetilde{s}^{-}_{I}(z)$
respectively which are bosonized as
\[
\widetilde{s}^{+}_{I}(z) \cong e^{+i(-1+\nu_{I})H^{I}(z)}~,\quad
\widetilde{s}^{-}_{I}(z) \cong e^{-i(-1+\nu_{I})H^{I}(z)}~.
\] }
\begin{equation}
 s^{+}_{I}(z) \cong e^{+i \nu_{I} H^{I}(z)}~,
\quad
 s^{-}_{I}(z) \cong e^{-i\nu_{I} H^{I}(z)}~.
\end{equation}
We can repeat the same analysis in the R sector.
We find that the incoming vacuum eq.~(\ref{eq:osci-vac-r})
and the outgoing vacuum
are excited from the $SL(2,{\bf R})$-invariant vacuum
by the spin fields $S^{+}_{I}(z)$ and $S^{-}(z)$ respectively.
They are bosonized as
\begin{equation}
 S^{+}_{I}(z) \cong e^{i\left(-\frac{1}{2}+\nu_{I}\right)H^{I}(z)}~,
\quad
 S^{-}_{I}(z) \cong e^{-i\left(-\frac{1}{2}+\nu_{I}\right)H^{I}(z)}~.
\end{equation}

%%%%%%%%%%%%%%%%%%%%%%%%%%%%%%%%%%%%%%%%%%%%%%%%%%%%%%%%%%%%%%%%%%
\subsection{Subtracted two-point functions and weights of twist
            and spin fields}

In the $x^{i}$-directions $(i=p+1,\ldots,p^{\prime})$, 
we have two types of vacuum:
the one is the $SL(2,{\bf R})$-invariant vacuum
and the other is the oscillator vacuum.
We can define the normal ordering
corresponding to each one.
We will use the symbols $:\ :$ and $\maru~\maru$
to denote the normal orderings
with respect to the $SL(2,{\bf R})$-invariant vacuum
and the oscillator vacuum respectively.

In the other directions we have a single type of vacuum,
namely $SL(2,{\bf R})$-invariant vacuum.
We have $:\ :$-normal ordered product only.
For free bosons and free fermions in these directions,
it is defined by a subtraction:
\begin{equation}
 :{\bf X}^{\mu}({\bf z}_{1},\overline{\bf z}_{1})
  {\bf X}^{\nu}({\bf z}_{2},\overline{\bf z}_{2}):\,
  ={\cal R}{\bf X}^{\mu}({\bf z}_{1},\overline{\bf z}_{1})
     {\bf X}^{\nu}({\bf z}_{2},\overline{\bf z}_{2})
  {}-{\bf G}^{\mu\nu}\left({\bf z}_{1},\overline{\bf z}_{1}
                 |{\bf z}_{2},\overline{\bf z}_{2}\right)~,
\label{eq:subSW}
\end{equation}
for $\mu,\nu=0,1,\ldots,p$. Here
 ${\bf G}^{\mu\nu}({\bf z}_{1},\overline{\bf z}_{1}|
                         {\bf z}_{2},\overline{\bf z}_{2})$
is the two-point function defined in eq.~(\ref{eq:propSW-1}).

In the same way we can define $\maru~\maru$-normal ordered product
for the free fields in the $x^{i}$-directions
$(i=p+1,\ldots,p^{\prime})$  as
\begin{equation}
 \maru~{\bf Z}^{I}({\bf z}_{1},\overline{\bf z}_{1})
    \overline{\bf Z}^{\overline{J}}({\bf z}_{2},\overline{\bf z}_{2})~
 \maru
   ={\cal R}{\bf Z}^{I}({\bf z}_{1},\overline{\bf z}_{1})
    \overline{\bf Z}^{\overline{J}}({\bf z}_{2},\overline{\bf z}_{2})
   {}-\mbox{\boldmath${\cal G}$}^{I\overline{J}}
           ({\bf z}_{1},\overline{\bf z}_{1}|
            {\bf z}_{2},\overline{\bf z}_{2})~,
\end{equation}
for $I,\overline{J}=\frac{p+2}{2},\ldots,\frac{p^{\prime}}{2}$.
Here $\mbox{\boldmath${\cal G}$}^{I\overline{J}}
           ({\bf z}_{1},\overline{\bf z}_{1}|
            {\bf z}_{2},\overline{\bf z}_{2})$
is the two-point function defined in eq.~(\ref{eq:propCIMM-1}).
In addition to the $\maru~\maru$-normal ordered product,
we would like to define the $:\ :$-normal ordered product
for these free fields ${\bf Z}^{I}$ and
$\overline{\bf Z}^{\overline{I}}$.
In order to apply the definition (\ref{eq:subSW}) directly to these
fields, we have to evaluate their two-point functions
on the $SL(2,{\bf R})$-invariant vacuum.
As is pointed out, the twist and the spin fields play
the role of boundary changing operators.
This implies that if we delete the twist and the spin fields
and thus remove the cut generated by them, the boundary conditions
imposed on the positive real axis of the $z$-plane
is expected to be identical to those imposed on the negative
real axis.
This leads us to conclude that the two point function
of ${\bf Z}^{I}$ and $\overline{\bf Z}^{\overline{I}}$
evaluated on the $SL(2,{\bf R})$-invariant vacuum
takes the same form as that of the $p^{\prime}$-$p^{\prime}$
system with $B$ field:
\begin{eqnarray}
&& {\bf G}^{I\overline{J}}\left(
     {\bf z}_{1},\overline{\bf z}_{1}|
     {\bf z}_{2},\overline{\bf z}_{2}\right)
\equiv \langle 0| {\cal R}{\bf Z}^{I}({\bf z}_{1},\overline{\bf z}_{1})
         \overline{\bf Z}^{\overline{J}}({\bf z}_{2},\overline{\bf z}_{2})
       |0\rangle  \\
&&=\frac{2\delta^{I\overline{J}}}{\varepsilon}
  \left[-\ln(z_1-z_2-\theta_1\theta_2)
            (\bar{z}_1-\bar{z}_2-\bar{\theta}_1\bar{\theta}_2)
        +\ln(z_1-\bar{z}_2-\theta_1\bar{\theta}_2)
            (\bar{z}_1-z_2-\bar{\theta}_1\theta_2) \right.\nonumber\\
& &\left.-\frac{2}{1+b_I^2}\ln(z_1-\bar{z}_2-\theta_1\bar{\theta}_2)
                        (\bar{z}_1-z_2-\bar{\theta}_1\theta_2)
   {}-2i\frac{b_I}{1+b^2_I}
       \ln \frac{z_1-\bar{z}_2-\theta_1\bar{\theta}_2}
                {\bar{z}_1-z_2-\bar{\theta}_1\theta_2}\right]
+\mbox{D-term}\nonumber 
\end{eqnarray}
Computing this two point function at the
boundary $\sigma=\pi$ and $\theta=\overline{\theta}$,
we obtain
\begin{eqnarray}
\lefteqn{ {\bf G}^{I\overline{J}}\left(-e^{\tau_{1}},\theta_{1}
         |-e^{\tau_{2}},\theta_{2}\right)
 \equiv 
 \left.
 {\bf G}^{I\overline{J}}\left(
     {\bf z}_{1},\overline{\bf z}_{1}|
     {\bf z}_{2},\overline{\bf z}_{2}\right)
 \right|_{\sigma=\pi,\theta=\overline{\theta}}}\nonumber\\
&&=-\frac{4\delta^{I\overline{J}}}{\varepsilon(1+b^{2}_{I})}
   \ln \left(-e^{\tau_{1}}+e^{\tau_{2}}-\theta_{1}\theta_{2}\right)^{2}
   +\frac{4\pi\delta^{I\overline{J}} b_{I}}{\varepsilon(1+b^{2}_{I})}
     \epsilon(\tau_{1}-\tau_{2})~.
\label{eq:propSW-3}
\end{eqnarray}

{}From the fact that the normal ordered product
is defined by a subtraction,
we can readily find that for an arbitrary functional
$\mbox{\boldmath $\cal O$}$
of the free fields ${\bf Z}^{I}$ and $\overline{\bf Z}^{\overline{I}}$
the normal ordering is formally expressed as
(see e.g.\ \cite{Pol})
\begin{eqnarray}
&&:\mbox{\boldmath $\cal O$}:
=\exp\left(-\int d^{2}{\bf z}_{1}
           d^{2}{\bf z}_{2} \,
           {\bf G}^{I\overline{J}}({\bf z}_{1},\overline{\bf z}_{1}|
                                   {\bf z}_{2},\overline{\bf z}_{2})
      \frac{\delta}{\delta{\bf Z}^{I}({\bf z}_{1},\overline{\bf z}_{1})}
      \frac{\delta}{\delta \overline{\bf Z}^{\overline{J}}
                   ({\bf z}_{2},\overline{\bf z}_{2})}
   \right)
  \mbox{\boldmath $\cal O$}~,
\nonumber\\
&&\maru~
  \mbox{\boldmath $\cal O$}~\maru
=\exp\left(-\int d^{2}{\bf z_{1}}
           d^{2}{\bf z}_{2} \,
           \mbox{\boldmath $\cal G$}^{I\overline{J}}
                  ({\bf z}_{1},\overline{\bf z}_{1}|
                   {\bf z}_{2},\overline{\bf z}_{2})
    \frac{\delta}{\delta{\bf Z}^{I}({\bf z}_{1},\overline{\bf z}_{1})}
    \frac{\delta}{\delta \overline{\bf Z}^{\overline{J}}
                   ({\bf z}_{2},\overline{\bf z}_{2})}
   \right)
  \mbox{\boldmath $\cal O$}~,
\end{eqnarray}
where $d^{2}{\bf z}$ is defined as
$d^{2}{\bf z}=d^2\xi d\theta d\overline{\theta}$.
{}From these general definitions of the normal orderings,
we can read off the formula of the reordering between them:
\begin{equation}
 :\mbox{\boldmath$\cal O$}:
  =\exp\left(\int d^{2}{\bf z_{1}}
           d^{2}{\bf z}_{2} \,
           {\mbox{\boldmath ${\cal G}$}_{\rm sub}}^{I\overline{J}}
                  ({\bf z}_{1},\overline{\bf z}_{1}|
                   {\bf z}_{2},\overline{\bf z}_{2})
   \frac{\delta}{\delta{\bf Z}^{I}({\bf z}_{1},\overline{\bf z}_{1})}
   \frac{\delta}{\delta \overline{\bf Z}^{\overline{J}}
                   ({\bf z}_{2},\overline{\bf z}_{2})}
   \right)
  \maru~\mbox{\boldmath $\cal O$}~\maru~.
\label{eq:reorder}
\end{equation}
Here 
${\mbox{\boldmath$\cal G$}_{\rm sub}}^{I\overline{J}}
    ({\bf z}_{1},\overline{\bf z}_{1}|{\bf z}_{2},\overline{\bf z}_{2})$
is a subtracted two-point function defined as
\begin{eqnarray}
&& {\mbox{\boldmath$\cal G$}_{\rm sub}}^{I\overline{J}}
  ({\bf z}_{1},\overline{\bf z}_{1}|{\bf z}_{2},\overline{\bf z}_{2})
\equiv
  \langle \sigma,s|:{\bf Z}^{I}({\bf z}_{1},\overline{\bf z}_{1})
    \overline{\bf Z}^{\overline{J}}({\bf z}_{2},\overline{\bf z}_{2}):
  |\sigma,s\rangle     \nonumber\\
&& \hspace{1em}=  \mbox{\boldmath$\cal G$}^{I\overline{J}}
     ({\bf z}_{1},\overline{\bf z}_{1}|{\bf z}_{2},\overline{\bf z}_{2})
  {}-{\bf G}^{I\overline{J}}
     ({\bf z}_{1},\overline{\bf z}_{1}|{\bf z}_{2},\overline{\bf z}_{2})~.
\label{eq:G-sub}
\end{eqnarray}

Let us compute the subtracted two-point function
$ {\mbox{\boldmath$\cal G$}_{\rm sub}}^{I\overline{J}}$
at the worldsheet boundary $\sigma=\pi$ and
$\theta=\overline{\theta}$.
{}From eqs.~(\ref{eq:propCIMM-2}) and (\ref{eq:propSW-3}),
we find that
\begin{eqnarray}
&& {\mbox{\boldmath${\cal G}$}_{\rm sub}}^{I\overline{J}}
   \left(-e^{\tau_{1}},\theta_{1}|-e^{\tau_{2}},\theta_{2}\right)
\equiv
  \langle \sigma,s | : {\bf Z}^{I}(-e^{\tau_{1}},\theta_{1})
  \overline{\bf Z}^{\overline{J}}(-e^{\tau_{2}},\theta_{2}) :
  |\sigma,s\rangle \nonumber\\
&&=\frac{8\delta^{I\overline{J}}}{\varepsilon(1+b^{2}_{I})}
 \left[-\gamma
   {}-\frac{\mbox{\boldmath$\psi$}(\nu_{I})
            +\mbox{\boldmath$\psi$}(1-\nu_{I})}{2}
    +\frac{1}{2}\ln e^{\tau_{1}+\tau_{2}}\right.\nonumber\\
&& \hspace{1em} +\Theta(\tau_{1}-\tau_{2})\left\{
    \frac{1}{2}\ln\frac{e^{\tau_{1}}}{e^{\tau_{2}}}
     {}-\left(\frac{e^{\tau_{2}}}{e^{\tau_{1}}}\right)^{1-\nu_{I}}
        \sum_{n=1}^{\infty}\frac{\left(1-\nu_{I}\right)_{n}}{n!}
            \sum_{m=0}^{n-1}
             \frac{\nu_{I}}{(m+1)(m+1-\nu_{I})}
        \left(1-\frac{e^{\tau_{2}}}{e^{\tau_{1}}}\right)^{n}
     \right\}\nonumber\\
&&\hspace{1em}\left. +\Theta(\tau_{2}-\tau_{1})
     \left\{\frac{1}{2}\ln\frac{e^{\tau_{2}}}{e^{\tau_{1}}}
    {}-\left(\frac{e^{\tau_{1}}}{e^{\tau_{2}}}\right)^{\nu_{I}}
       \sum_{n=1}^{\infty}
       \frac{\left(\nu_{I}\right)_{n}}{n!}
         \sum_{m=0}^{n-1}\frac{1-\nu_{I}}{(m+1)(m+\nu_{I})}
          \left(1-\frac{e^{\tau_{1}}}{e^{\tau_{2}}}\right)^{n}
     \right\}
\right]\nonumber\\
&&+\theta_{1}\theta_{2}
    \frac{8\delta^{I\overline{J}}}{\varepsilon(1+b^{2}_{I})}
    \left\{ \left(\frac{e^{\tau_{1}}}{e^{\tau_{2}}}\right)^{\nu_{I}}
    {}-1\right\}
   \frac{1}{-e^{\tau_{1}}+e^{\tau_{2}}}~,
\end{eqnarray}
where $\gamma$ is Euler's constant,
$\mbox{\boldmath$\psi$}(w)$ denotes the digamma function
defined as $\mbox{\boldmath$\psi$}(w)=\frac{d}{dw}\ln \Gamma(w)$
and $(a)_{n}\equiv \frac{\Gamma(a+n)}{\Gamma(a)}$.
Here we have used the formulas (\ref{eq:loghyper}) and (\ref{eq:psib}).

Let us compute
the conformal weights of the twist and spin fields. This helps us appreciate
 these subtracted two-point functions better.
The relevant part of the energy-momentum tensor in this computation
consists of the terms depending on the string coordinates
in the $x^{i}$-directions, $i=p+1,\ldots,p^{\prime}$:
$T^{(Z,\Psi)}_{B}(z)
=T^{(Z,\overline{Z})}_{B}(z)+T^{(\Psi,\overline{\Psi})}_{B}(z)$
with
\begin{equation}
   T^{(Z,\overline{Z})}_{B}(z)
 =-\displaystyle\frac{\varepsilon\,\delta_{I\overline{J}}}{\alpha^{\prime}}
   : \partial Z^{I}\partial\overline{Z}^{\overline{J}}(z):~,
\quad
 T^{(\Psi,\overline{\Psi})}_{B}(z)
 =-\frac{\varepsilon\,\delta_{I\overline{J}}}{4}
      \left[ :\Psi^{I}\partial\overline{\Psi}^{\overline{J}}(z):
       {}-:\partial\Psi^{I}\, \overline{\Psi}^{\overline{J}}(z):\right]~.
\end{equation}
Using the subtracted two point function, we obtain
\begin{eqnarray}
\lefteqn{
\langle \sigma_{I} | T_{B}(z) | \sigma_{I} \rangle
= -\frac{\varepsilon}{\alpha^{\prime}}  \lim_{w\rightarrow z}
   \langle \sigma_{I}| : \partial Z^{I}(w)
          \partial\overline{Z}^{\overline{I}}(z) :
   | \sigma_{I}\rangle
}  \nonumber\\
%&&= \lim_{w\rightarrow z}\left[ -\frac{\varepsilon}{\alpha^{\prime}}
%\langle \sigma_{I}|{\cal R}
%     \partial Z^{I}(w)\partial\overline{Z}^{\overline{I}}(z)
%|\sigma_{I}\rangle     - \frac{1}{(w-z)^{2}}\right]\nonumber\\
&&
=\lim_{w\rightarrow z}\left[
 \left(\frac{w}{z}\right)^{\nu_{I}-1}\frac{1}{w-z}
 \left(\frac{1-\nu_{I}}{z}+\frac{1}{w-z}\right)
 {}-\frac{1}{(w-z)^{2}}\right]
 =\frac{1}{z^{2}}\frac{\nu_{I}(1-\nu_{I})}{2}~.
\end{eqnarray}
This implies that the twist fields $\sigma^{\pm}_{I}$ have the
weights $h_{\sigma_{I}}=\frac{1}{2}\nu_{I}(1-\nu_{I})$.
In the same way, we obtain
\begin{eqnarray}
&&\langle s_{I}| T_{B}(z)|s_{I}\rangle
   =-\frac{\varepsilon}{4}   \lim_{w\rightarrow z}
    \langle s_{I}|:
      \left(\Psi^{I}(w)\partial\overline{\Psi}^{\overline{I}}(z)
      {}-\partial\Psi^{I}(w)\, \overline{\Psi}^{\overline{I}}(z)\right)
    :  |s_{I}\rangle
\nonumber\\
%&&\hspace{2em}
% =\lim_{w\rightarrow z}
%  \left[-\frac{\varepsilon}{4}\langle s_{I}|{\cal R}
%      \left(\Psi^{I}(w)\partial\overline{\Psi}^{\overline{I}}(z)
%       {}-\partial\Psi^{I}(w)\, \overline{\Psi}^{\overline{I}}(z)\right)
%      |s_{I}\rangle
%       +\frac{1}{(w-z)^{2}}\right]\nonumber\\
&&\hspace{2em}=\lim_{w\rightarrow z}\left[
    \left(\frac{w}{z}\right)^{\nu_{I}}\frac{1}{w-z}\left\{
     \frac{\nu_{I}}{2}\left(\frac{1}{w}+\frac{1}{z}\right)
     {}-\frac{1}{w-z}\right\}+\frac{1}{(w-z)^{2}}\right]
  =\frac{1}{z^{2}}\frac{{\nu_{I}}^{2}}{2}~.
\end{eqnarray}
{}From this equation we can read
\footnote{This result can also be obtained from the fact
that $T^{(\Psi,\overline{\Psi})}_{B}(z)$ is bosonized as
$T^{(\Psi,\overline{\Psi})}_{B}(z)
  \cong -\frac{1}{2}\delta_{IJ}:~\!\! \partial H^{I}\partial H^{J}(z)
    \!\!~:$~.
} 
that
the spin fields $s^{\pm}_{I}$ have the weights
$h_{s_{I}}=\frac{1}{2} {\nu_{I}}^{2}$.

%%%%%%%%%%%%%%%%%%%%%%%%%%%%%%%%%%%%%%%%%%%%%%%%%%%%%%%%%%%%%%%%
\section{Vertex operators}

Let us pay some attention to vertex operators of our system
before we start calculating scattering amplitudes.
We focus on two types of vertex operators.
The one is the tachyon vertex operators,
and the other is the massless vector vertex operators.
These are the relevant ones in order for us to find out the 
spacetime processes occurring on the D$p^\prime$-brane worldvolume.

Here we make a comment on the GSO projection.
In this paper we take the GSO projection in the NS sector
so that the oscillator vacuum corresponding to
the tachyon vertex operator survives.
It follows that the GSO projection adopted in this paper
is not always the same as that in our previous work \cite{CIMM}:
in the cases of $p^{\prime}=p+2$, $p+6$ they are opposite
to each other, while
in the cases of $p^{\prime}=p+4$, $p+8$ they are the same.
This is attributed to
the sign convention of the D$p$-brane charge:
in the cases of $p^{\prime}=p+2$, $p+6$
the D$p$-branes in this paper should be referred to
as anti-D$p$-branes in  the convention of \cite{CIMM}.

\subsection{Tachyon vertex operator of $p$-$p^\prime$ string}

First let us investigate vertex operators of the $p$-$p^{\prime}$
open string which contain the twist and the spin fields.
We will focus on the vertex operator
that corresponds to the ground state in the NS sector
of this open string.
This vertex operator is seen for instance in \cite{GNS}\cite{DFMS}:
\begin{equation}
 {\bf V}^{\pm}_{T}(\xi^{1},\theta)
  =V^{\pm\,(-1)}_{T}(\xi^{1})+\theta V^{\pm\,(0)}_{T}(\xi^{1})
= \mbox{\boldmath${\cal T}$}^{\pm}(\xi^{1},\theta)
    \,:\exp\left( i\sqrt{\frac{\alpha^{\prime}}{2}}
       \sum_{\mu=0}^{p}k_{\mu}{\bf X}^{\mu}(\xi^{1},\theta)\right)
:~,
\label{eq:t-vo-1}
\end{equation}
where 
${\bf X}^{\mu}(\xi^{1},\theta) $ is the boundary
value of the superfield ${\bf X}^{\mu}({\bf z},\overline{\bf z})$:
$ {\bf X}^{\mu}(\xi^{1},\theta) \equiv\left.
  {\bf X}^{\mu}({\bf z},\overline{\bf z})
  \right|_{z=\overline{z}=\xi^{1},\theta=\overline{\theta}}$,
and 
$\mbox{\boldmath$\cal T$}^{\pm}(\xi^{1},\theta)$
is the superfield whose lowest component
${\cal T}^{\pm}_{0}(\xi^{1})$ consists of the twist and
the spin fields,
\begin{equation}
 {\cal T}^{\pm}_{0}(\xi^{1})
  =\prod_{I} \sigma^{\pm}_{I}(\xi^{1})s^{\pm}_{I}(\xi^{1})~.
\end{equation}
The upper component is obtained by applying the supercurrent
$T_{F}(z)$ to ${\cal T}^{\pm}_{0}(\xi^{1})$ \cite{DFMS}.
In eq.~(\ref{eq:t-vo-1}), $V^{\pm\,(0)}_{T}(\xi^{1})$
denotes the $0$-picture vertex and $V^{\pm\,(-1)}_{T}(\xi^{1})$
is the matter field contribution to the $(-1)$-picture
vertex
\begin{equation}
 {\cal V}^{\pm\,(-1)}_{T}(\xi^{1})=e^{-\phi}(\xi^{1})
  \, V^{\pm\,(-1)}_{T}(\xi^{1})
=e^{-\phi}\,\prod_{I}\sigma^{\pm}_{I}s^{\pm}_{I}
 \,:\exp\left( i\sum_{\mu=0}^{p}k_{\mu}X^{\mu}\right):~,
\end{equation}
where $e^{-\phi}$ comes from the $\beta\gamma$-ghost sector.

It is worth noting that space-time momentum $k_{\mu}$
of the tachyon vertex operator eq.~(\ref{eq:t-vo-1}) is not 
$(p^{\prime}+1)$ dimensional but $(p+1)$ dimensional.
This is because the $p$-$p^{\prime}$ string coordinates
in $x^{p+1},\ldots,x^{p^{\prime}}$-directions
do not possess zero-modes
and thus the momenta in these directions are not defined.
This implies that an initial/final tachyon field corresponding to
this vertex operator is frozen
in these space-time directions.

Let us study the physical state conditions for this vertex operator.
If the $(-1)$-picture vertex satisfies the physical state condition,
the $0$-picture vertex is automatically physical
because of the worldsheet supersymmetry.
We will therefore concentrate on the $(-1)$-picture vertex.
Through the operator-state mapping
this vertex operator corresponds to the state%%%%%
\footnote{Here we ignore the ghost sector.},
\begin{equation}
 \left| V^{(-1)}_{T} \right\rangle
  \equiv\lim_{\xi^{1}\rightarrow 0}V^{(-1)}_{T}(\xi^{1})|0\rangle
  =|0;k_{\mu}\rangle \otimes |\sigma,s\rangle~,
\end{equation}
where the state $|0;k_{\mu}\rangle$ is defined as
$|0;k_{\mu}\rangle = \exp\left(i\displaystyle\sum_{\mu=0}^{p}
                                k_{\mu}x^{\mu}\right) |0\rangle$.
Here $x^{\mu}$ are the zero modes of the bosonic coordinates.
It is evident that $\left|V_{T\,(-1)}\right\rangle$ is a primary state.
We just require that this state should have weight
$\frac{1}{2}$.
{}From the calculation in the last subsection, we find that
the state $|\sigma,s\rangle$ has a weight
\begin{equation}
 h\Big[ \, |\sigma,s\rangle \, \Big]
 =\sum_{I}\left(\frac{\nu_{I}(1-\nu_{I})}{2}
          +\frac{ {\nu_{I}}^{2} }{2}\right)
 =\sum_{I}\frac{\nu_{I}}{2}~.
\end{equation}
Substituting the mode expansions eqs.~(\ref{eq:mode-x-SW}) and
(\ref{eq:mode-ns-SW}) into the defining relation eq.~(\ref{eq:emtensor}),
we see that the terms in $T_{B}(z)$
which depend on the string coordinates in $x^{\mu}$-directions
$(\mu=0,1,\ldots,p)$ are 
\begin{eqnarray}
&& T^{(X,\psi)}_{B}(z)\equiv
 {}-\sum_{\mu=0}^{p}\left(\frac{1}{\alpha^{\prime}}
         g_{\mu\nu}\partial X^{\mu}\partial X^{\nu}(z)
   +\frac{1}{2}g_{\mu\nu}\psi^{\mu}\partial \psi^{\nu}(z)\right)
 \equiv \sum_{m\in {\bf Z}}L_{m}z^{-m-2}~,\nonumber\\
&&\hspace{-1em}\mbox{with}\quad
  L_{m}=\frac{1}{2}\sum_{n\in{\bf Z}} \sum_{\sigma,\rho=0}^{p}
        G_{\sigma\rho}:\alpha^{\sigma}_{m-n}\alpha^{\rho}_{n}:
+\frac{1}{4}\sum_{r\in{\bf Z}+1/2}(2r-m)
       \sum_{\sigma,\rho=0}^pG_{\sigma\rho}:b^{\sigma}_{m-r}b^{\rho}_{r}:~,
       \label{eq:emtensor-SW}
\end{eqnarray}
where $\alpha^{\mu}_{0}=\sqrt{2\alpha^{\prime}}p^{\mu}$.
Here $G_{\sigma\rho}$ is the open string metric
including time direction, its inverse is given in eq.~(\ref{eq:G0-p}).
This yields
\begin{equation}
 L_{0}|0;k_{\mu}\rangle
=\alpha^{\prime} \sum_{\sigma,\rho=0}^pG^{\sigma\rho}k_{\sigma}k_{\rho}
                 |0;k_{\mu}\rangle~.
\end{equation}
Gathering all results obtained above, we conclude that
the weight of the state $|V^{(-1)}_{T}\rangle$ is
$L_{0}= {\displaystyle\sum_{I}\frac{1}{2}\nu_{I}
  +\alpha^{\prime}\sum_{\sigma,\rho=0}^pG^{\sigma\rho}k_{\sigma}k_{\rho}}$.
Thus the on-shell condition $L_{0}=\frac{1}{2}$
requires that the mass squared of this state should be
\begin{equation}
 \alpha^{\prime}m_{T}^{2}
 \equiv -\alpha^{\prime}
 \sum_{\sigma,\rho=0}^pG^{\sigma\rho}k_{\sigma}k_{\rho}
 =-\frac{1}{2} \left(1-\sum_{I}\nu_{I} \right)~.
\label{eq:tachyonmass}
\end{equation}

%%%%%%%%%%%%%%%%%%%%%%%%%%%%%%%%%%%%%%%%%%%%%%%%%%%%%%%%%%%%%%%%%%%%%%
%%%%%%%%%%%%%%%%%%%%%%%%%%%%%%%%%%%%%%%%%%%%%%%%%%%%%%%%%%%%%%%%%%%%%%
\subsection{Massless vector vertex operator of $p^{\prime}$-$p^{\prime}$
         string}

The vector emission vertex operator takes the form of
\begin{equation}
 {\bf V}_{\rm vec}(\xi^{1},\theta)
 \equiv V^{(-1)}_{\rm vec}(\xi^{1})+\theta V^{(0)}_{\rm vec}(\xi^{1})
\equiv
     \frac{i}{2}\sum_{\mu=0}^{p^{\prime}}:
      \zeta_{\mu}(k)
       {\bf \dot{X}}^{\mu}(\xi^{1},\theta)
        \,\exp\left( i\sqrt{\frac{\alpha^{\prime}}{2}}
           \sum_{\rho=0}^{p^{\prime}}
           k_{\rho}{\bf X}^{\rho}(\xi^{1},\theta)\right):~,
\end{equation}
where $\zeta_{\mu}(k)$ denotes the polarization vector
and ${\bf \dot{X}}^{\mu}(\xi^{1},\theta)\equiv
\left. (D+\overline{D}){\bf X}^{\mu}({\bf z},\overline{\bf z})
\right|_{z=\overline{z}=\xi^{1},\theta=\overline{\theta}}$.
The operator $V^{(0)}_{\rm vec}(\xi^{1})$ is the $0$-picture vertex
and $V^{(-1)}_{\rm vec}(\xi^{1})$ is the matter field contribution
to the $(-1)$-picture vertex
 ${\cal V}^{(-1)}_{\rm vec}(\xi^{1})= e^{-\phi}
   V^{(-1)}_{\rm vec}(\xi^{1}) $.
Their explicit forms are
\begin{eqnarray}
&&
   V^{(-1)}_{\rm vec}(\xi^{1})
  =-\frac{1}{2}\sum_{\mu=0}^{p^{\prime}}:
    \zeta_{\mu}(k)\left(\psi^{\mu}+\widetilde{\psi}^{\mu}\right)
    \exp\left( i\sum^{p^{\prime}}_{\rho=0}
         k_{\rho}X^{\rho}\right):~,\\
&&
  V^{(0)}_{\rm vec}(\xi^{1})=\frac{1}{\sqrt{2\alpha^{\prime}}}
  \sum_{\mu=0}^{p^{\prime}} :  \zeta_{\mu}(k)
  \left[i \dot{X}^{\mu}+\frac{\alpha^{\prime}}{2}
     \left(\sum_{\rho=0}^{p^{\prime}}k_{\rho}
      \left(\psi^\rho+\widetilde{\psi}^{\rho}\right)\right)
      \left(\psi^{\mu}+\widetilde{\psi}^{\mu}\right)
  \right]
\exp\left( i\sum^{p^{\prime}}_{\lambda=0}
              k_{\lambda}X^{\lambda}\right):~,
\nonumber
\end{eqnarray}
where $\dot{X}^{\mu}(\xi^{1})$ is defined as
$\dot{X}^{\mu}(\xi^{1})
=\left.(\partial +\overline{\partial})X^{\mu}(z,\overline{z})
 \right|_{z=\overline{z}=\xi^{1}}$.

The string coordinates of a $p^{\prime}$-$p^{\prime}$ open string
in the $x^{i}$-directions $(i=1,\ldots,p^{\prime})$
obey the same boundary conditions
as the $x^{i}$-directions with $i=1,\ldots,p$ of the $p$-$p^{\prime}$
string.
They have the same 
mode expansions and the commutators among the oscillating modes.
Therefore the vector emission vertex operator
in each picture corresponds to the respective state
\begin{eqnarray}
&& \left|V^{(-1)}_{\rm vec}\right\rangle
 \equiv \lim_{\xi^{1}\rightarrow 0}V^{(-1)}_{\rm vec}(\xi^{1})
   |0\rangle
= -\sum_{\mu=0}^{p^{\prime}}
   \zeta_{\mu}(k)\,b^{\mu}_{-\frac{1}{2}}|0;k_{\rho}\rangle^{\prime}~,
\nonumber\\
&&
 \left| V^{(0)}_{\rm vec}\right\rangle
\equiv \lim_{\xi^{1}\rightarrow 0}V^{(0)}_{\rm vec}(\xi^{1})|0\rangle
=\sum_{\mu =0}^{p^{\prime}}\zeta_{\mu}(k)\left(
    \alpha^{\mu}_{-1}+\sqrt{2\alpha^{\prime}}
    \left(\sum_{\lambda =0}^{p^{\prime}}
          k_{\lambda}b^{\lambda}_{-\frac{1}{2}}\right)
          b^{\mu}_{-\frac{1}{2}}
 \right)|0;k_{\rho}\rangle^{\prime}~,
\end{eqnarray}
where $|0;k_{\rho}\rangle^{\prime}$ is defined as
$|0;k_{\rho}\rangle^{\prime}
=\exp \left({\displaystyle \sum_{\rho=0}^{p^{\prime}}}
           k_{\rho}x^{\rho}\right)|0\rangle$.
Let us consider the physical state conditions on these states.
The relevant part of the energy-momentum tensor for this analysis
is eq.~(\ref{eq:emtensor-SW}) with $p$ being replaced by
$p^{\prime}$.
This yields
\begin{eqnarray}
&& L_{0}\left|V^{(0)}_{\rm vec}\right\rangle
=\left[\alpha^{\prime}\sum_{\sigma,\rho=0}^{p^{\prime}}
       G^{\sigma\rho}k_{\sigma}k_{\rho} +1 \right]
  \left|V^{(0)}_{\rm vec}\right\rangle~,\nonumber\\
&&  L_{1}\left|V^{(0)}_{\rm vec}\right\rangle
   =\sqrt{2\alpha^{\prime}}  \sum_{\sigma,\rho=0}^{p^{\prime}}
     G^{\sigma\rho}k_{\sigma}\zeta_{\rho}(k) |0;k_{\rho}\rangle^{\prime}~.
\end{eqnarray}
{}From these relations we find that the physical state conditions,
$L_{0}=1$ and $L_{1}=0$, require that
\begin{equation}
\alpha^{\prime} m^{2}_{\rm vec}
  \equiv -\alpha^{\prime}\sum_{\sigma,\rho=0}^{p^{\prime}}
    G^{\sigma\rho}k_{\sigma}k_{\rho}=0~,
\qquad \sum_{\sigma,\rho=0}^{p^{\prime}}
   G^{\sigma\rho}k_{\sigma}\zeta_{\rho}(k)=0~.
\label{eq:vectormass}
\end{equation}

We will write the vector emission vertex operator
${\bf V}_{\rm vec}(\xi^{1},\theta)$ in an
exponential form \cite{IM},
\begin{equation}
 {\bf V}_{\rm vec}(\xi^{1},\theta)
 = \int d\eta : \left. 
   \exp \left[i\sum_{\mu=0}^{p^{\prime}}
     \left\{\sqrt{\frac{\alpha^{\prime}}{2}} k_{\mu} 
          +\eta \zeta_{\mu}(k)
           \frac{1}{2}\left( D+\overline{D} \right)
     \right\}{\bf X}^{\mu}({\bf z},\overline{\bf z})\right]
   :~\right|_{\scriptstyle z=\overline{z}=\xi^{1}
              \atop\scriptstyle\theta=\overline{\theta}}~,
\label{eq:expvec}
\end{equation}
by introducing a Grassmann parameter $\eta$.

We will write the part of vertex operator eq.~(\ref{eq:expvec})
which depends on the string coordinates
in the $x^{i}$-directions $(i=p+1,\ldots,p^{\prime})$ as
\begin{equation}
 : \left.\exp\left[
        \sum_{I=\frac{p+2}{2}}^{p^{\prime}/2}
            {\bf E}_{I}(\zeta,k,\eta){\bf Z}^{I}({\bf z},\overline{\bf z})
       +\sum_{\overline{J}=\frac{p+2}{2}}^{p^{\prime}/2}
          \overline{\bf E}_{\overline{J}}(\zeta,k,\eta)
        \overline{\bf Z}^{\overline{J}}({\bf z},\overline{\bf z})
      \right]:~\right|_{z=\overline{z}=\xi^{1},
                               \theta=\overline{\theta}}~,
\label{eq:veczz}
\end{equation}
using the complex variables.
Here ${\bf E}_{I}(\zeta,k,\eta)$ and
$\overline{\bf E}_{\overline{I}}(\zeta,k,\eta)$
are differential operators on the superspace defined as
\begin{eqnarray}
 &&{\bf E}_{I}(\zeta,k,\eta)
     =i\sqrt{\frac{\alpha^{\prime}}{2}}\kappa_{I}
      +i \eta e_{I}(k)\frac{1}{2}\left(D+\overline{D}\right)~,\nonumber\\
 &&
 \overline{\bf E}_{\overline{I}}(\zeta,k,\eta)
     =i \sqrt{\frac{\alpha^{\prime}}{2}}
         \overline{\kappa}_{\overline{I}}
      +i \eta \overline{e}_{\overline{I}}(k)
         \frac{1}{2}\left(D+\overline{D}\right)~,
\end{eqnarray}
with
\begin{eqnarray}
\kappa_{I} = \frac{1}{2}\left(k_{2I-1}-ik_{2I}\right)~,
&&
\overline{\kappa}_{\overline{I}}
 =\frac{1}{2}\left(k_{2I-1}+ik_{2I}\right)~;\nonumber\\
e_{I}(k)=\frac{1}{2}\Big(\zeta_{2I-1}(k)-i\zeta_{2I}(k)\Big)~,
&&
\overline{e}_{\overline{I}}(k)
  =\frac{1}{2}\Big(\zeta_{2I-1}(k)+i\zeta_{2I}(k)\Big)~.
\end{eqnarray}
\section{Scattering Amplitudes}

\begin{figure}[htb]
 \epsfxsize=13em
 \centerline{\epsfbox{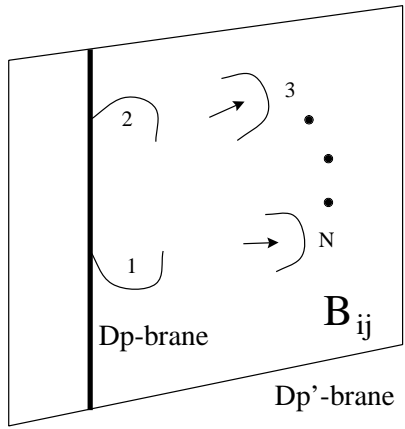}}
   \caption{The space-time picture of the process.}
   \label{fig:spacetime}
\end{figure}

\begin{figure}[htb]
 \epsfxsize=25em
   \centerline{\epsfbox{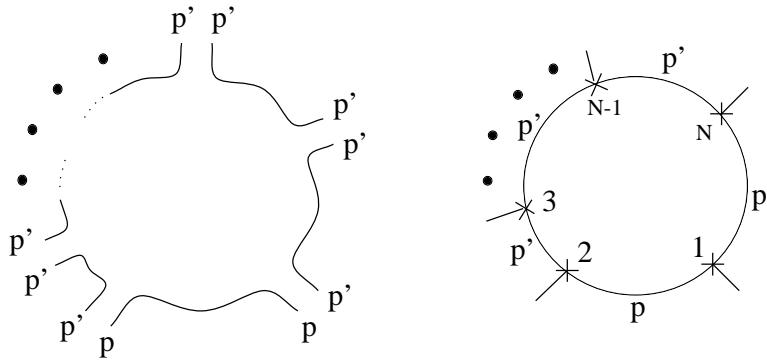}}
   \caption{$N$ point string diagram.}
   \label{fig:n-ptamp}
\end{figure}

We would like to  find out a proper description of the physical
processes taking place on the worldvolume of the D$p^{\prime}$-brane
with the D$p$-brane inside ($p<p^{\prime}$)  in the case where
the $B$ field  is nonvanishing.
In this section,  we will consider multiparticle tree scattering
amplitudes consisting of the external states of $N-2$ vectors
obtained from the mode of the $p^{\prime}$-$p^{\prime}$ open string
and two tachyons from  the mode of the $p$-$p^{\prime}$ open string.
The spacetime picture of this string scattering process is
depicted in Fig.\ref{fig:spacetime}.
That this process is possible is easy to  see once we draw a spacetime
diagram and map the end points of the open strings onto a circle.
See Fig.\ref{fig:n-ptamp}.

Open string tree amplitudes in general are obtained by
placing vertex operators  on the boundary of the upper half plane,
namely, the real axis, integrating over the positions of the vertex 
operators and dividing by the volume of the (super)conformal killing
vectors.
To obtain the amplitudes of our concern, we first locate each of
the two kinds of the tachyon vertex operators 
${\bf V}^{+}_{T}(\xi,\theta)$, ${\bf V}^{-}_{T}(\xi,\theta)$ discussed
in the last section at $\xi=\xi_{1}$ and at $\xi=\xi_{2}$ respectively.
A cut is generated on the interval between these two locations
as ${\bf V}^{+}_{T}$ and ${\bf V}^{-}_{T}$
contain the twist field and the anti-twist field respectively.
The worldvolume of the D$p^{\prime}$-brane  contains this interval
on which we place the $N-2$ vector emission vertex operators
${\bf V}_{\rm vec}(\xi,\theta)$ of the $p^{\prime}$-$p^{\prime}$
open string.
In what follows we will obtain the integral (Koba-Nielsen)
representation of the amplitudes.
The explicit expressions for the $N=3, 4$ cases that we obtain
will be exploited to determine the form of the low energy effective
field theory in the subsequent section.

In manifestly supersymmetric formulation on superspace,
the $N$ point tree amplitude in question reads
\begin{eqnarray}
\frac{c}{V_{\rm SCKV}} \int \prod^{N}_{a=1}d\xi_{a}d\theta_{a}
  \langle 0 |~{\bf V}^{+}_{T}(\xi_{1},\theta_{1};k_{1\mu})
             {\bf V}^{-}_{T}(\xi_{2},\theta_{2};k_{2 \mu})
    \prod^{N}_{c=3}
       {\bf V}_{\rm vec}(\xi_{c},\theta_{c}; k_{c \mu},\zeta_{c \mu})
    ~|0\rangle ~, \label{eq:amp-1}
\end{eqnarray}
where
$V_{\rm SCKV}$ denotes the volume of the isometry group
generated by the superconformal Killing vectors,
namely, the graded extension of the $SL(2,{\bf R})$ group.
We have denoted by $c$ the overall constant which does not
concern us in this paper.
In eq.~(\ref{eq:amp-1}), the domain of  $\xi_{a}$ integrations is
not restricted except that  $\xi_3, \xi_{4}, \cdots, \xi_{N}$
are located  on the cut created on  the interval
between  $\xi_{1}$ and $\xi_{2}$.
This domain falls into  a sum of the $(N-2)!$ regions.
In each region,
an ordering among $\xi_3, \xi_{4}, \cdots, \xi_{N}$ is specified
and  integrals over each region give a contribution
corresponding to a respective
open string (dual) diagram\footnote{ Recall that we have $(N-1)!$
open string (dual) diagrams in the case of the $N$ point
amplitude of a $p$-$p$ open string.}.
We will evaluate  the contribution from the region
$\xi_{2}< \xi_3 < \xi_{4}< \cdots <\xi_{N} <\xi_1$  and
this is denoted by $A_{N}$.
In most cases below, we will not write the region of
integrations explicitly.

Eq.~(\ref{eq:amp-1}) is invariant under the graded
$SL(2,{\bf R})$ transformations
after the physical state conditions are invoked at each vertex operator.
For actual evaluation of the amplitude, 
we  first set $\theta_{1}=\theta_{2}=0$ to fix the odd elements of
the transformations.  We then fix the even elements 
by giving fixed values to three of the locations of the vertex operators. 
These locations are chosen  as $\xi_1, \xi_2,$ and $\xi_3$.
This amounts to factoring out the following volume element from
the integration,
\begin{equation}
 d^{3}F\left(\xi_{1},\xi_{2},\xi_{3}\right)
        d\theta_{1}d\theta_{2}\,(\xi_{1}-\xi_{2})~,
\label{eq:volosp}
\end{equation}
 where
\begin{equation}
     d^{3}F(\xi_{1},\xi_{2},\xi_{3}) \equiv
     \frac{d\xi_{1}d\xi_{2}d\xi_{3}}
          {(\xi_{1}-\xi_{2})(\xi_{2}-\xi_{3})(\xi_{3}-\xi_{1})}~.
\end{equation}
Having done this, we obtain
\begin{equation}
  A_{N}=c\int \frac{\displaystyle \prod_{a=1}^{N}d\xi_{a}}
                     {d^{3}F(\xi_{1},\xi_{2},\xi_{3})}
         \frac{\displaystyle \prod_{c=3}^{N}d\theta_{c}}{\xi_{1}-\xi_{2}} 
    \langle 0 |~V^{+(-1)}_{T}(\xi_{1};k_{1\mu})
                V^{-(-1)}_{T}(\xi_{2};k_{2\mu})
      \prod_{c^{\prime}=3}^{N}{\bf V}_{\rm vec}(\xi_{c^{\prime}},
      \theta_{c^{\prime}};k_{c^{\prime}\mu},\zeta_{c^{\prime}\mu})
    ~|0\rangle~.
\end{equation}
The  component corresponding to the $(-1)$-picture
has been selected at each of the tachyon vertex operators. 
Let us choose $\xi_1=0$, $\xi_2=-\infty$ and $\xi_3=-1$,
so that the negative real axis becomes the worldsheet boundary
ending on the D$p^{\prime}$-brane.
Introducing positive real variables
$x_{a}\equiv -\xi_{a}\left(=e^{\tau_{a}}\right) >0$ and
adopting eq.~(\ref{eq:expvec}) for
the vector emission vertex operators,  we find
\begin{eqnarray}
 \lefteqn{ A_{N}=c \int \frac{\displaystyle \prod_{a=1}^{N}dx_{a}}
                     {d^{3}F(x_{1},x_{2},x_{3})}
     \frac{\displaystyle \prod_{a^{\prime}=3}^{N}d\theta_{a^{\prime}}
      d\eta_{a^{\prime}}}{x_{1}-x_{2}}
      \left(\frac{1}{x_{2}}\right)^{\displaystyle 
                              \sum_{I=\frac{p+2}{2}}^{p^{\prime}/2}
                                     \nu_{I}}
} \label{eq:amp-3} \\
&& \times 
   \left\langle 0 \right|\,\prod_{f=1}^{N}
   :\left.\exp\left[i\sum_{\mu=0}^{p}\left\{
      \sqrt{\frac{\alpha^{\prime}}{2}}
                   k_{f\mu}{\bf X}^{\mu}(-x_{f},\theta_{f})
    +\frac{1}{2}\eta_{f}\zeta_{f \mu}
            {\bf \dot{X}}^{\mu}(-x_{f},\theta_{f})\right\}\right]
      :\ \right|_{\scriptstyle \zeta_{1\mu}=\zeta_{2\mu}=0
                   \atop \scriptstyle\theta_{1}=\theta_{2}=0}
  \left| 0 \right\rangle 
\nonumber\\
&&\times  \left\langle \sigma,s \right| \,\prod_{c=3}^{N} 
   :\exp\left[
        \sum_{I=\frac{p+2}{2}}^{p^{\prime}/2}
        {\bf E}_{I}\left(\zeta_{c},k_{c},\eta_{c}\right)
                {\bf Z}^{I}(-x_{c},\theta_{c})
      +\sum_{\bar{J}=\frac{p+2}{2}}^{p^{\prime}/2}
        \overline{\bf E}_{\overline{J}}
       \left(\zeta_{c},k_{c},\eta_{c}\right)
       \overline{\bf Z}^{\overline{J}}(-x_{c},\theta_{c})
      \right]  : \left|\sigma,s\right\rangle~, \nonumber 
\end{eqnarray}
where $x_{1}=0$, $x_{2}=\infty$ and $x_{3}=1$.
Here the right hand side  consists of
  two of the expectation values of the exponential
  operators: the one is obtained from the $x^{\mu}$-directions
 $(\mu=0,\ldots,p)$
and the other is from the $x^{i}$-directions
$(i=p+1,\ldots,p^{\prime})$, and  we have used 
\begin{equation}
 \left\langle \sigma_{I},s_{I} \right|
= \lim_{x\rightarrow\infty}x^{\nu_{I}}\,
   \langle 0|\,\sigma_{I}s_{I}(-x)~.
\end{equation}

Let us examine the contribution from the
$x^{i}$-directions $(i=p+1,\ldots,p^{\prime})$.
As is explained in section \ref{sect:pprime}, the operators inside
$\langle \sigma,s | \cdots | \sigma,s\rangle$ in eq.~(\ref{eq:amp-3})
are normal ordered
with respect to the $SL(2,{\bf R})$ invariant vacuum
and are not with respect to the oscillator vacuum.
Applying the reordering formula  eq.~(\ref{eq:reorder}), we obtain
\begin{eqnarray}
\lefteqn{
  \left\langle \sigma,s \right| \,\prod_{a=3}^{N}
   :\exp\left( \sum_{I}{\bf E}_{aI}{\bf Z}^{I}(-x_{a},\theta_{a})
      +\sum_{\bar{J}}\overline{\bf E}_{a\overline{J}}
       \overline{\bf Z}^{\overline{J}}(-x_{a},\theta_{a})
        \right): \left|\sigma,s\right\rangle
}\nonumber\\
&&=\prod^{N}_{a=3} \exp\left[\sum_{I,\overline{J}}
    {\bf E}_{aI}\overline{\bf E}_{a\overline{J}}
    {\mbox{\boldmath$\cal G$}_{\rm sub}}^{I\overline{J}}
        (\!(-x_{a},\theta_{a})\!)\right]
 \nonumber\\
&&\hspace{3em}\times
  \left\langle \sigma,s \right| \,\prod_{c=3}^{N}
   \maru  \exp\left[\,\sum_{I}{\bf E}_{cI}{\bf Z}^{I}(-x_{c},\theta_{c})
      +\sum_{\overline{J}}\overline{\bf E}_{c\overline{J}}
       \overline{\bf Z}^{\overline{J}}(-x_{c},\theta_{c})
       \right]\maru \left|\sigma,s\right\rangle~,\nonumber\\
&&=\prod_{a=3}^{N}
    x_{a}^{ -2\alpha^{\prime} \sum_{I,\overline{J}}  
           \kappa_{aI}\overline{\kappa}_{a\overline{J}} G^{I\overline{J}}}
\exp\left[{\cal C}_{a}\left(\nu_{I}\right) 
           +\sqrt{2\alpha^{\prime} }\eta_{a} \sum_{I\overline{J}}
             e_{aI}\overline{\kappa}_{a\overline{J}}G^{I\overline{J}}
            \frac{\theta_{a}}{x_{a}}\right]\nonumber\\
&& \hspace{3em}\times
   \left\langle \sigma,s \right| \,\prod_{c=3}^{N}
   \maru  \exp\left[\,\sum_{I}{\bf E}_{cI}{\bf Z}^{I}(-x_{c},\theta_{c})
      +\sum_{\overline{J}}\overline{\bf E}_{c\overline{J}}
       \overline{\bf Z}^{\overline{J}}(-x_{c},\theta_{c})
       \right]\maru \left|\sigma,s\right\rangle~,
\label{eq:amp-z-factor}
\end{eqnarray}
where
\begin{equation}
{\cal C}_{a}(\nu_I)=\alpha^{\prime}\sum_{I,\overline{J}}
    2 \kappa_{aI}\overline{\kappa}_{a\overline{J}} G^{I\overline{J}}   
   \left\{ \gamma + \frac{1}{2}\Big(\mbox{\boldmath${\psi}$}(\nu_{I})
            +\mbox{\boldmath${\psi}$}(1-\nu_{I})\Big)\right\}~,
\label{eq:C}
\end{equation}
and ${\bf E}_{cI}$ and $\overline{\bf E}_{c\overline{J}}$
stand for ${\bf E}_{I}(\zeta_{c},k_{c},\eta_{c})$
and $\overline{\bf E}_{\overline{J}}(\zeta_{c},k_{c},\eta_{c})$
respectively. Note that the part in eq.~(\ref{eq:amp-z-factor})
that corresponds to self-contractions  has been given by the subtracted
Green function at the coincident point:
\begin{eqnarray}
&&\sum_{I\overline{J}} {\bf E}_{aI} \overline{\bf E}_{a\overline{J}}
  {\mbox{\boldmath$\cal G$}_{\rm sub}}^{I\overline{J}}
   (\!(-x_{a},\theta_{a})\!)
\equiv \lim_{\theta_{c}\rightarrow \theta_{a}
               \atop \scriptstyle x_{c}\rightarrow x_{a} }
    \sum_{I\overline{J}}
    {\bf E}_{cI}\overline{\bf E}_{a\overline{J}}
          {\mbox{\boldmath$\cal G$}_{\rm sub}}^{I\overline{J}}
          \left(-x_{c},\theta_{c}|-x_{a},\theta_{a}\right)
   \nonumber\\
&& \hspace{2em}
 ={\cal C}_{a}(\nu_{I}) + \sum_{I,\overline{J}}
  \left[-2\alpha^{\prime}\kappa_{aI}\overline{\kappa}_{a\overline{J}}
        G^{I\overline{J}} \ln x_{a}
        +\sqrt{2\alpha^{\prime}}\eta_{a}e_{aI}
         \overline{\kappa}_{a\overline{J}} G^{I\overline{J}}
         \frac{\theta_{a}}{x_{a}}
\right]~.
\end{eqnarray}
The last factor $\langle \sigma,s|\cdots |\sigma,s \rangle$
in eq.~(\ref{eq:amp-z-factor}) is now calculated using
the two-point function $\mbox{\boldmath${\cal G}$}^{I\overline{J}}
                   (-x_{c},\theta_{c}|-x_{c'},\theta_{c'})$:
\begin{eqnarray}
 \lefteqn{\langle \sigma,s | \prod_{c=3}^{N}
  \maru \exp\left[\sum_{I}
       {\bf E}_{cI}{\bf Z}^{I}(-x_{c},\theta_{c})
                +\sum_{\overline{J}}\overline{\bf E}_{c\overline{J}}
                 \overline{\bf Z}^{\overline{J}}(-x_{c},\theta_{c})
        \right]\maru |\sigma,s\rangle
}\nonumber\\
&&=\prod_{3\leq c<c'\leq N}
 \exp\left[ \sum_{I,\overline{J}}
   \left\{  {\bf E}_{cI}\overline{\bf E}_{c'\overline{J}}
        \mbox{\boldmath${\cal G}$}^{I\overline{J}}
              (-x_{c},\theta_{c}|-x_{c'},\theta_{c'})
    +\overline{\bf E}_{c\overline{J}}{\bf E}_{c'I}
         \mbox{\boldmath${\cal G}$}^{I\overline{J}}
              (-x_{c'},\theta_{c'}|-x_{c},\theta_{c})  \right\}
      \right]\nonumber\\
&&=\prod_{3\leq c<c'\leq N}
  \exp\left[ \sum_{I,\overline{J}} G^{I\overline{J}}
      \left[-2\alpha^{\prime}
         \kappa_{cI}\overline{\kappa}_{c'\overline{J}}
            \,{\cal H}\left(\nu_{I}; 
             \frac{x_{c}}{x_{c^{\prime}}-\theta_{c}\theta_{c^{\prime}}}
                    \right)
             {}-2\alpha^{\prime}
          \overline{\kappa}_{c\overline{J}}\kappa_{c^{\prime}I}
            \, {\cal H}\left(\nu_{I};
              \frac{x_{c^{\prime}}}{x_{c}+\theta_{c}\theta_{c^{\prime}}}
                     \right)
 \right. \right.\nonumber\\
&&  \hspace{4.5em} 
   +\eta_{c} \sqrt{ 2 \alpha^{\prime}}
    \left\{
     e_{cI}\overline{\kappa}_{c^{\prime}\overline{J}}
     \frac{ \left(\frac{x_{c}}{x_{c^{\prime}}}\right)^{\nu_{I}}
            \theta_{c^{\prime}} 
            {}-\left(\frac{x_{c}}{x_{c^{\prime}}}\right)^{\nu_{I}-1}
               \theta_{c}}
          { x_{c}-x_{c^{\prime}} }
    +\overline{e}_{c\overline{J}} \kappa_{c^{\prime}I}
      \left(\frac{x_{c^{\prime}}}{x_{c}}\right)^{\nu_{I}}
            \frac{\theta_{c^{\prime}}-\theta_{c}}{x_{c}-x_{c^{\prime}}}
     \right\}  \nonumber\\
&& \hspace{4.5em}
  +\eta_{c^{\prime}} \sqrt{ 2\alpha^{\prime}}
    \left\{
      \kappa_{cI} \overline{e}_{c^{\prime}\overline{J}}
       \left(\frac{x_{c}}{x_{c^{\prime}}}\right)^{\nu_{I}}
       \frac{\theta_{c^{\prime}}-\theta_{c}}{x_{c}-x_{c^{\prime}}}
  +\overline{\kappa}_{c\overline{J}} e_{c^{\prime}I}
     \frac{\left(\frac{x_{c^{\prime}}}{x_{c}}\right)^{\nu_{I}-1}
            \theta_{c^{\prime}}
           {}-\left(\frac{x_{c^{\prime}}}{x_{c}}\right)^{\nu_{I}}
            \theta_{c}}
          {x_{c}-x_{c^{\prime}}}
    \right\}  \nonumber\\
&&   \hspace{4.5em} 
  +\eta_{c}\eta_{c^{\prime}}
     \left\{
      e_{cI}\overline{e}_{c^{\prime}\overline{J}}
      \,\frac{ \left(\frac{x_{c}}{x_{c^{\prime}}}\right)^{\nu_{I}} }
           { x_{c} - x_{c^{\prime}} }
     \left(
        1-\theta_{c}\theta_{c^{\prime}}
           \frac{ (1-\nu_{I}) +\nu_{I}\frac{x_{c^{\prime}}}{x_{c}} }
                {x_{c}-x_{c^{\prime}}}
     \right)
\right.\nonumber\\
&& \left.\left.
\hspace{10em}
    +\overline{e}_{c\overline{J}}e_{c^{\prime}I}
     \frac{ \left(\frac{x_{c^{\prime}}}{x_{c}}\right)^{\nu_{I}} }
          {x_{c}-x_{c^{\prime}}}
     \left(1-\theta_{c}\theta_{c^{\prime}}
              \frac{(1-\nu_{I})+\nu_{I}\frac{x_{c}}{x_{c^{\prime}}} }
                   { x_{c}-x_{c^{\prime}}}\right)~\right\}~\right]
\nonumber\\
&& \left. \hspace{2.5em}
 {}-\sum_{I,\overline{J}} \epsilon(x_{c}-x_{c^{\prime}})
 \frac{2\delta^{I\overline{J}} \pi b_{I}}
   {\varepsilon(1+b_{I}^{2})}\alpha^{\prime}
   \left( \kappa_{cI}\overline{\kappa}_{c^{\prime \overline{J}}}
          {}-\overline{\kappa}_{c\overline{J}}\kappa_{c^{\prime}I}
  \right)
\right]~.
\end{eqnarray}

The factor coming from the $x^{\mu}$-directions
$(\mu=0,\ldots,p)$
is handled by  the two-point function ${\bf G}^{\mu\mu^{\prime}}$:
\begin{eqnarray}
\lefteqn{ \langle 0 | \prod_{f=1}^{N}:\exp\left[\,\sum_{\mu=0}^{p}
      {\cal E}_{\mu}\left(\zeta_{f},k_{f},\eta_{f}\right)
      {\bf X}^{\mu}(-x_{f},\theta_{f})\right]:
 | 0 \rangle
}\nonumber\\
&&=\left.\exp\left[\,\sum_{1\leq f < f^{\prime}\leq N}
             \sum_{\mu,\mu^{\prime}=0}^{p}
                  {\cal E}_{\mu}\left(\zeta_{f},k_{f},\eta_{f}\right)
       {\cal E}_{\mu^{\prime}}\left(\zeta_{f^{\prime}},k_{f^{\prime}},
                 \eta_{\prime}\right)
       {\bf G}^{\mu\mu^{\prime}}({\bf z}_{f},\overline{\bf z}_{f}|
                   {\bf z}_{f^{\prime}},\overline{\bf z}_{f^{\prime}})
        \right]\right|_{\scriptstyle z_{f}=\overline{z}_{f}=-x_{f},
                         \atop
                        \scriptstyle \theta_{f}=\overline{\theta}_{f}}
\nonumber\\
&&\hspace{1em}\times \langle 0 |
  :\exp \left[\, \sum_{c=1}^{N}\sum_{\rho=0}^{p}
   {\cal E}_{\rho}(\zeta_{c},k_{c},\eta_{c}){\bf X}^{\rho}
   (-x_{c},\theta_{c})\right]:|0\rangle\nonumber\\
&&=\exp\left[\,\sum_{1\leq f < f^{\prime}\leq N}
             \left\{ \sum_{\sigma,\rho=0}^{p}\Bigg\{
 \alpha^{\prime}G^{\sigma\rho}k_{f\sigma}k_{f^{\prime}\rho}
 \ln\left(x_{f}-x_{f^{\prime}}+\theta_{f}\theta_{f^{\prime}}\right)^{2}
\right. \right. \nonumber\\
&&
\hspace{5em} -\sqrt{2\alpha^{\prime}}\left(
   \eta_{f}G^{\sigma\rho}\zeta_{f\sigma}k_{f^{\prime}\rho}
   +\eta_{f^{\prime}}G^{\sigma\rho}k_{f\sigma}\zeta_{f^{\prime}\rho}\right)
      \frac{\theta_{f}-\theta_{f^{\prime}}}{x_{f}-x_{f^{\prime}}}
\nonumber\\
&&
\hspace{5em}
\left.\left.
+\eta_{f}\eta_{f^{\prime}}G^{\sigma\rho}
           \zeta_{f\sigma}\zeta_{f^{\prime}\rho}
  \frac{1}{(x_{f}-x_{f^{\prime}}+ \theta_{f}\theta_{f^{\prime}})}\Bigg\}
+\sum_{i,j=1}^p\frac{i}{2}\,\theta^{ij}k_{fi}k_{f^{\prime}j}
         \epsilon\left(x_{f}-x_{f^{\prime}}\right)
\right\}
\right]\nonumber\\
&&\hspace{1em}\times(2\pi)^{p+1}
\prod_{\mu=0}^{p}\delta\left(k_{1\mu}+\cdots+k_{N\mu}\right)~,
\end{eqnarray}
where ${\cal E}_{\mu}(\zeta_{f},k_{f},\eta_{f})$ is a differential
operator defined as
\begin{eqnarray}
&& {\cal E}_{\mu}\left(\zeta_{f},k_{f},\eta_{f}\right)
  =i\sqrt{\frac{\alpha^{\prime}}{2}}k_{f\mu}
   +i\frac{1}{2}\eta_{f}\zeta_{f\mu}(k)
\left(D_{f}+\overline{D}_{f} \right)~,\nonumber\\
&&
   {\cal E}_{\mu}\left(\zeta_{f},k_{f},\eta_{f}\right)
      {\bf X}^{\mu}(-x_{f},\theta_{f})
 \equiv {\cal E}_{\mu}\left(\zeta_{f},k_{f},\eta_{f}\right)
      {\bf X}^{\mu}({\bf z}_{f},\overline{\bf z}_{f})
 \bigg|_{\scriptstyle z_{f}=\overline{z}_{f}=-x_{f}
           \atop\scriptstyle \theta_{f}=\overline{\theta}_{f}}~.
\end{eqnarray}

We have now evaluated the two of the expectation values in
eq.~(\ref{eq:amp-3}) and are ready to present
the integral representation of $A_N$.
Let us first introduce several shorthand notations. 
We denote by $\pdot$ the inner product of two $(p+1)$-dimensional
vectors $A_i$ and $B_j$ lying on the $Dp$-brane worldvolume
with respect to the open string metric. Namely
\begin{equation}
A\pdot B
=\sum^p_{\sigma,\rho=0}G^{\sigma\rho}A_{\sigma} B_{\rho} \;.
\end{equation}
Similarly, we denote by $\prdot$ the inner product of
two $(p^\prime+1)$-dimensional vectors $A_i$ and $B_j$
lying on the $Dp^\prime$-brane worldvolume
with respect to the open string metric. 
We will also write $A\ppdot B$ to denote the inner product
of the last $(p^\prime-p)$ components of the two vectors.
For example, we have
\begin{eqnarray}
&& k \ppdot \zeta 
=\sum_{I,\overline{J}=\frac{p+2}{2}}^{\frac{p^{\prime}}{2}}
  \left(G^{I\overline{J}}\kappa_{I}\overline{e}_{\overline{J}}
        +G^{\overline{J}I}\overline{\kappa}_{\overline{J}}e_{I}\right)
=\sum_{I,\overline{J}=\frac{p+2}{2}}^{\frac{p^{\prime}}{2}}
   G^{I\overline{J}}\left(\kappa_{I}\overline{e}_{\overline{J}}
                           +\overline{\kappa}_{\overline{J}}e_{I}
     \right)~,\nonumber\\
&& k \ppdot k 
  =\sum_{I,\overline{J}=\frac{p+2}{2}}^{\frac{p^{\prime}}{2}}
    2G^{I\overline{J}}\kappa_{I}\overline{\kappa}_{\overline{J}}~,
 \qquad
  \zeta \ppdot \zeta 
   =\sum_{I,\overline{J}=\frac{p+2}{2}}^{\frac{p^{\prime}}{2}}
     2G^{I\overline{J}}e_{I}\overline{e}_{\overline{J}}~.
\end{eqnarray}
We will use the notations
$\kuroten$ and $\peke$ which denote
\begin{equation}
 \left(k \kuroten \zeta\right)_{I}=\sum_{\overline{J}}
G^{I\overline{J}}(\kappa_{I}\overline{e}_{\overline{J}}
                  +\overline{\kappa}_{\overline{J}}e_{I})~,
\quad
\left(k \peke \zeta \right)_{I}=\sum_{\overline{J}}
  \frac{2\delta^{I\overline{J}}(\kappa_{I}\overline{e}_{\overline{J}}
        {}-\overline{\kappa}_{\overline{J}}e_{I})}
       {\varepsilon (1+b_{I}^{2})}~,
\end{equation}
etc. {}From these defining relations one can find that
\begin{equation}
 \sum_{I}\left(k\kuroten\zeta\right)_{I}
  =k\ppdot\zeta~,
\quad
 \sum_{I}\left(k\peke \zeta\right)_{I}
 =ik\ppdot J\zeta~,
\end{equation}
where $J$ is a $(p^{\prime}+1)\times (p^{\prime}+1)$
antisymmetric matrix defined as
\begin{equation}
  J =\left( {J_{\mu}}^{\rho}\right)\equiv 
   \begin{array}{r@{}l}
    \left(\begin{array}{ccc|ccccc}
	 0&      & &  & &      & & \\
          &\ddots& &  & &      & & \\
          &      &0&  & &      & & \\ \hline
          &      & &0 &1&      & & \\
          &      & &-1&0&      & & \\
          &      & &  & &\ddots& & \\
          &      & &  & &      &0&1\\
          &      & &  & &      &-1&0
        \end{array}\right)
             &   \begin{array}{l}
                 \mbox{\scriptsize $0$} \\  \vdots\\ 
                 \mbox{\scriptsize $p$}\\ \mbox{\scriptsize $p+1$}\\
                 \mbox{\scriptsize $p+2$}\\  \vdots\\
                 \mbox{\scriptsize $p^{\prime}-1$}\\
                 \mbox{\scriptsize $p^{\prime}$}
                 \end{array}        
   \end{array}~.
\end{equation}
We will group the terms in the exponent
by the number of $\eta_a$'s and by the number of $\theta_a$'s,
using the notation $[0,2]$, $[2,0]$, $[1,1]$, $[2,2]$.
The first number in the bracket indicates the number of 
$\eta_a$'s and the second number the number of $\theta_a$'s.

Having prepared these, we write eq.~(\ref{eq:amp-3}) as 
\begin{eqnarray}
\lefteqn{ A_{N} = c(2\pi)^{p+1}\prod^p_{\mu=0}
  \delta\left(\sum^N_{e=1}k_{e\mu}\right) \int \prod^N_{a=4}dx_a
  \prod^N_{a^\prime=3}d\theta_{a^\prime}d\eta_{a^\prime}
     \,\exp {{\cal C}_{a^\prime}(\nu_I)}}\nonumber\\
&&  \ \times  x_{2}^{-\sum_{I}\nu_I} (x_2-x_3)(x_3-x_1) 
 \prod^N_{c^{\prime\prime}=3}
    x_{c^{\prime\prime}}^{-\alpha^\prime k_{c^{\prime\prime}}\ppdot
                            k_{c^{\prime\prime}}}
 \prod_{1\leq c<c^\prime\leq N}
     (x_{c^\prime}-x_c)^{2\alpha^\prime k_{c^\prime}\pdot k_c} 
\nonumber\\
 && \ \times
 \prod_{3\leq c<c^\prime\leq N}
 \exp\left[ -2\alpha^{\prime}\sum_{I,\bar{J}}G^{I\overline{J}}
 \left\{ \kappa_{cI} \overline{\kappa}_{c^{\prime}\overline{J}} 
         {\cal H}\left(\nu_{I};\frac{x_{c}}{x_{c^{\prime}}}\right)
        +\overline{\kappa}_{c \overline{J}} \kappa_{c^{\prime}I}
          {\cal H}\left(\nu_{I};\frac{x_{c^{\prime}}}{x_{c}}\right)
  \right\}\right] \nonumber\\
&& \ \times \exp\left({\rm NC}\right)
   \, \exp\bigg([0,2]+[2,0]+[1,1]+[2,2]\bigg)~.
\label{anbcr}
\end{eqnarray}
Here
\begin{eqnarray}
%%%%%%%
%%%%%%%     [0,2]
%%%%%%%
\lefteqn{[0,2] = 2\alpha^{\prime}\sum_{3\leq c<c^{\prime}\leq N}
     \frac{\theta_c\theta_{c^\prime}}{x_{c}-x_{c^{\prime}}}
     \Bigg[ k_{c}\pdot k_{c^\prime}
}\nonumber\\
&&\hspace{5em}\left.
  + \frac{1}{2}\sum_{I}
     \left\{\left( k_{c}\kuroten k_{c^{\prime}}\right)_{I}
      \left[ \left(\frac{x_{c}}{x_{c^{\prime}}}\right)^{\nu_{I}}
               +\left(\frac{x_{c^{\prime}}}{x_{c}}\right)^{\nu_{I}}
      \right]
      + \left( k_{c}\peke k_{c^{\prime}} \right)_{I}
      \left[ \left(\frac{x_{c}}{x_{c^{\prime}}}\right)^{\nu_{I}}
          {}-\left(\frac{x_{c^{\prime}}}{x_{c}}\right)^{\nu_{I}}
      \right]\,\right\}\,\right]~,
\nonumber\\
%%%%%%%%
%%%%%%%%  [1,1]
%%%%%%%%
\lefteqn{ [1,1] = \sqrt{2\alpha^{\prime}}\sum_{c=3}^{N} \eta_{c}\theta_{c}
    \left[\frac{1}{2}\zeta_{c} \ppdot (1+iJ) k_{c}\,\frac{1}{x_{c}}
          +k_{1}\pdot\zeta_{c}\,\frac{1}{x_{1}-x_{c}}
          +k_{2}\pdot\zeta_{c}\,\frac{1}{x_{2}-x_{c}} \right]
}\nonumber\\
 && -\sqrt{2\alpha^{\prime}}\sum_{3\leq c<c^\prime\leq N}
    \frac{1}{x_{c}-x_{c^{\prime}}}\Bigg[
    \left( \eta_{c} \zeta_{c}\pdot k_{c^\prime}
   +\eta_{c^\prime}k_{c}\pdot\zeta_{c^\prime}\right)
     \left(  \theta_{c}-\theta_{c^\prime} \right)
\nonumber\\
&&\hspace{3em} +\frac{1}{2}
  \sum_{I}\Bigg\{
   \eta_{c}\left( \zeta_{c}\kuroten k_{c^{\prime}}\right)_{I}
     \left(
       \left\{ \left(\frac{x_{c}}{x_{c^{\prime}}}\right)^{\nu_{I}-1}
           +\left(\frac{x_{c^{\prime}}}{x_{c}}\right)^{\nu_{I}}
       \right\}\theta_{c}
    {}-\left\{\left(\frac{x_{c}}{x_{c^{\prime}}}\right)^{\nu_{I}}
           +\left(\frac{x_{c^{\prime}}}{x_{c}}\right)^{\nu_{I}}
       \right\}\theta_{c^{\prime}}
      \right) \nonumber\\
&&\hspace{7em}
  + \eta_{c^{\prime}}\left( k_{c}\kuroten \zeta_{c^{\prime}}\right)_{I}
   \left(
       \left\{\left(\frac{x_{c}}{x_{c^{\prime}}}\right)^{\nu_{I}}
             +\left(\frac{x_{c^{\prime}}}{x_{c}}\right)^{\nu_{I}}
       \right\}\theta_{c}
    {}-\left\{\left(\frac{x_{c}}{x_{c^{\prime}}}\right)^{\nu_{I}}
            +\left(\frac{x_{c^{\prime}}}{x_{c}}\right)^{\nu_{I}-1}
       \right\}\theta_{c^{\prime}}
   \right)\nonumber\\
&&\hspace{7em}
  +\eta_{c}\left(\zeta_{c}\peke k_{c^{\prime}}\right)_{I}
     \left(
        \left\{\left(\frac{x_{c}}{x_{c^{\prime}}}\right)^{\nu_{I}-1}
             {}-\left(\frac{x_{c^{\prime}}}{x_{c}}\right)^{\nu_{I}}
        \right\}\theta_{c}
     {}-\left\{\left(\frac{x_{c}}{x_{c^{\prime}}}\right)^{\nu_{I}}
           {}-\left(\frac{x_{c^{\prime}}}{x_{c}}\right)^{\nu_{I}}
        \right\}\theta_{c^{\prime}}
     \right)  \nonumber\\
&&\hspace{7em} 
  +\eta_{c^{\prime}}\left( k_{c}\peke\zeta_{c^{\prime}}\right)_{I}
     \left(
        \left\{\left(\frac{x_{c}}{x_{c^{\prime}}}\right)^{\nu_{I}}
           {}-\left(\frac{x_{c^{\prime}}}{x_{c}}\right)^{\nu_{I}}
        \right\}\theta_{c}
     {}-\left\{\left(\frac{x_{c}}{x_{c^{\prime}}}\right)^{\nu_{I}}
          {}-\left(\frac{x_{c^{\prime}}}{x_{c}}\right)^{\nu_{I}-1}
        \right\}\theta_{c^{\prime}}
     \right)
\,\Bigg\} \, \Bigg]~,  \nonumber\\
%%%%%%%%
%%%%%%%%  [2,0]
%%%%%%%%
\lefteqn{
[2,0]=\sum_{3\leq c<c^\prime\leq N}
   \frac{\eta_{c}\eta_{c^{\prime}}}{x_{c}-x_{c^{\prime}} }
   \Bigg[
   \zeta_{c} \pdot \zeta_{c^\prime} 
}\nonumber\\
&& \hspace{6em}
 +\frac{1}{2}\sum_{I}
   \left\{ \left(\zeta_{c}\kuroten\zeta_{c^{\prime}}\right)_{I}
     \left[
        \left(\frac{x_{c}}{x_{c^{\prime}}}\right)^{\nu_{I}}
           +\left(\frac{x_{c^{\prime}}}{x_{c}}\right)^{\nu_{I}}
        \right]
  +\left(\zeta_{c}\peke\zeta_{c^{\prime}}\right)_{I}
    \left[
        \left(\frac{x_{c}}{x_{c^{\prime}}}\right)^{\nu_{I}}
          {}-\left(\frac{x_{c^{\prime}}}{x_{c}}\right)^{\nu_{I}}
    \right]  \right\}\Bigg]~, \nonumber\\
%%%%%%%%%
%%%%%%%%%   [2,2]
%%%%%%%%%
\lefteqn{[2,2]=\sum_{3\leq c<c^\prime\leq N}
   \frac{\eta_c\theta_c\eta_{c^\prime}\theta_{c^\prime}}
        {(x_{c}-x_{c^{\prime}})^{2}}
   \Bigg[\zeta_c\pdot \zeta_{c^\prime}
}\nonumber \\
&&\hspace{2em}
   +\frac{1}{2} \sum_{I}  \Bigg\{
       \left(\zeta_{c}\kuroten\zeta_{c^{\prime}}\right)_{I}
   \left(
       (1-\nu_{I})\left\{
       \left(\frac{x_{c}}{x_{c^{\prime}}}\right)^{\nu_{I}}
      +\left(\frac{x_{c^{\prime}}}{x_{c}}\right)^{\nu_{I}}\right\}
     +\nu_{I}\left\{
     \left(\frac{x_{c}}{x_{c^{\prime}}}\right)^{\nu_{I}-1}
      +\left(\frac{x_{c^{\prime}}}{x_{c}}\right)^{\nu_{I}-1}\right\}
   \right)
        \nonumber\\
&&\hspace{5em} +\left(\zeta_{c}\peke\zeta_{c^{\prime}}\right)_{I}
   \left(
     (1-\nu_{I})\left\{
     \left(\frac{x_{c}}{x_{c^{\prime}}}\right)^{\nu_{I}}
      {}-\left(\frac{x_{c^{\prime}}}{x_{c}}\right)^{\nu_{I}}\right\}
     +\nu_{I}\left\{
     \left(\frac{x_{c}}{x_{c^{\prime}}}\right)^{\nu_{I}-1}
      {}-\left(\frac{x_{c^{\prime}}}{x_{c}}\right)^{\nu_{I}-1}\right\}
   \right)
\Bigg\}\Bigg]~,\nonumber\\
\end{eqnarray}
and $({\rm NC})$ denotes the terms containing the sign function:
\begin{eqnarray}
({\rm NC}) &=&\sum_{1\leq a<a^{\prime}\leq N}\frac{i}{2}
              \epsilon(x_{a}-x_{a^{\prime}})
          \sum_{i,j=1}^{p}\theta^{ij}k_{ai}k_{a^{\prime}j}
\nonumber\\
 &&-\sum_{3\leq c<c^{\prime} \leq N}
   \epsilon(x_{c}-x_{c^{\prime}})
   \sum_{I,\overline{J}}\alpha^{\prime}
    \frac{2\delta^{I\overline{J}}\pi b_{I}}{\varepsilon(1+b_{I}^{2})}
    \left(\kappa_{cI}\overline{\kappa}_{c^{\prime}\overline{J}}
          {}-\overline{\kappa}_{c\overline{J}}\kappa_{c^{\prime}I}
    \right)\nonumber\\
 &=&\sum_{1\leq a<a^{\prime}\leq N}\frac{i}{2}
      \epsilon (x_{a}-x_{a^{\prime}})\sum_{\mu,\lambda=0}^{p^{\prime}}
      \theta^{\mu\lambda}k_{a\mu}k_{a^{\prime}\lambda}~,
\label{eq:realnc}
\end{eqnarray}
with $k_{1j}=k_{2j}=0$ for $(j=p+1,\ldots,p^{\prime})$.
Here we have written the noncommutativity term
in terms of the real variables and generalized the notation
$\theta^{\mu\lambda}$ to include the time components
$\theta^{0i}=0$.
We also remind the readers that
${\cal H}\left(\nu_{I};\frac{x_{c}}{x_{c^{\prime}}}\right)$
is given in eq.~(\ref{eq:h-prop}).

So far, we have not exploited that
$x_1=0, x_2=\infty$ and $x_3=1$ except  that
the oscillator vacuum  $|\sigma,s\rangle$, $\langle \sigma,s|$ is
obtained from the tachyon vertex operators.
Firstly, by sending $x_2=\infty$, all factors in eq.~(\ref{anbcr})
containing $x_2$ are removed. In fact, this is ensured by an equality 
\begin{equation}
1-\sum_{I=\frac{p+2}{2}}^{\frac{p^\prime}{2}}{\nu_I}
  +2\alpha^{\prime}\sum_{c\neq 2}k_2 \pdot k_c=0~,
\end{equation}
which is obtained from the momentum conservation
$\displaystyle
  \prod_{\mu =0}^{p}\delta\left( \sum_{c=1}^{N}k_{c\mu}\right)$
and the on-shell condition (eq.~(\ref{eq:tachyonmass})) for the tachyon.
  Secondly, setting $x_1=0$, we find
\begin{eqnarray}
\lefteqn{\prod_{\scriptstyle 1\leq c<c^\prime\leq N \atop
                \scriptstyle c,c^{\prime} \neq 2}
    (x_{c^\prime}-x_c)^{2\alpha^{\prime} k_{c^\prime}\pdot k_c}
  \prod^N_{c^{\prime\prime}=3}
     x_{c^{\prime\prime}}^{-\alpha^{\prime} k_{c^{\prime\prime}}
                             \ppdot k_{c^{\prime\prime}}}}
\nonumber\\
&&
=\prod^N_{c=3}x_c^{-\alpha^{\prime} s_c+\alpha^{\prime} m^2_T}
  \prod_{3\leq c<c^\prime\leq N}
     (x_c-x_{c^\prime})^{2\alpha^{\prime} k_{c^\prime}\pdot k_c}~.
\end{eqnarray}
Here $s_c\equiv -(k_c+k_1)\pdot (k_c+k_1)$ and we have used
the on-shell condition for the tachyon  (eq.~(\ref{eq:tachyonmass}))
and that for the vector (eq.~(\ref{eq:vectormass})).
Finally we would like to  convert  the integrations at eq.~(\ref{anbcr})
  into those over a set of  $N-3$  $SL(2,{\bf R})$ invariant cross ratios. 
We choose these cross ratios as
\begin{equation}
x^{(a+3)}\equiv \frac{(x_1-x_{a+3})(x_2-x_3)}{(x_1-x_3)(x_2-x_{a+3})}
  = \frac{x_{a+3}}{x_3}, \;\;\;  a=1,  \ldots N-3 \;\;.
\end{equation}
We can therefore accomplish this conversion
by rescaling $x_{a+3}$ by $x_3$ and setting $x_3=1$ in eq.~(\ref{anbcr})
without changing the form of the integrand.

Putting all these considerations together,
we obtain  from eq.~(\ref{anbcr})
\begin{eqnarray}
\lefteqn{A_N=c (2\pi)^{p+1}\prod^p_{\mu=0}
\delta\left(\sum^N_{e=1}k_{e\mu}\right) \int \prod^N_{a=4}dx_a
\prod^N_{a^\prime=3}d\theta_{a^\prime}d\eta_{a^\prime}
\exp{\cal C}_{a^\prime}(\nu_I)} \nonumber\\
& &\times \prod^N_{c=4}\left[
          x_c^{-\alpha^\prime s_c+\alpha^{\prime} m^2_T}
          (1-x_c)^{2\alpha^{\prime} k_3\pdot k_c}\right]
    \prod_{4\leq c<c^\prime\leq N}
       (x_c-x_{c^\prime})^{2\alpha^{\prime} k_{c^\prime}\pdot k_c}
\nonumber \\
& &\times \prod_{3\leq c<c^\prime\leq N}
   \exp\left[-2\alpha^{\prime}\sum_{I,\overline{J}}G^{I\overline{J}}
     \left\{ \kappa_{cI}\overline{\kappa}_{c^\prime\overline{J}}
            {\cal H}\left(\nu_I; \frac{x_c}{x_{c^\prime}}\right)
           +\overline{\kappa}_{c\overline{J}} \kappa_{c^{\prime}I}
          {\cal H}\left(\nu_{I};\frac{x_{c^{\prime}}}{x_{c}}\right)
     \right\} \right]
\nonumber\\
& &\times \exp\left({\rm NC}\right)
 \exp\Big([0,2]+[2,0]+[1,1]+[2,2]\Big)
\Bigg|_{x_{1}=0,x_{2}=\infty,x_3=1}~.
\end{eqnarray}
This expression is  regarded as an $SL(2,{\bf R})$ invariant integral
(Koba-Nielsen) representation for the amplitude of our concern. 
Let us list several features which are distinct from the corresponding
formula in the case of a $p$-$p$ open string. (See \cite{IM}).
\begin{enumerate}
\item The term denoted by  $\exp\left({\rm NC}\right)$ 
      which originated from the noncommutativity of the worldvolume
      extends into both the $x^{1},\ldots, x^{p}$ directions
      and the remaining $x^{p+1},\ldots,x^{p^\prime}$ directions.
\item To each external vector leg, we have a momentum dependent
      multiplicative factor $\exp {{\cal C}(\nu_I)}$.
\item A new tensor $J$ has appeared.
\item There are parts in the expression
      which are expressible in terms of the momenta of the tachyons, 
      the momenta and the polarization tensors of the vectors
      and $J$ alone, using the inner product with respect to
      the open string metric.
      These parts come, however, with a host of other parts
      which do not permit such generic description in terms of
      the inner product.
\end{enumerate}

Let us finally compute $N=3, 4$ cases explicitly.
For $N=3$ case, we need to pick up $\theta_3$ and
$\eta_3$ from $[1,1]$.
Using the transversality of the polarization
vector (eq.~(\ref{eq:vectormass})) 
and the $(p+1)$ dimensional momentum conservation, we find that
\begin{equation}
A_{3}= c (2\pi)^{p+1}\prod_{\mu=0}^{p}
  \delta\left(\sum^3_{a=1}k_{a\mu}\right)
\sqrt{\frac{\alpha^{\prime}}{2}}
\left\{(k_2-k_1)\pdot \zeta_3-ik_3\ppdot J \zeta_3
\right\} \nonumber\\
\,e^{{\cal C}_{3}(\nu_I)}e^{\frac{i}{2}\theta^{ij}k_{1i}k_{2j}}~.
\label{eq:A3}
\end{equation}

For $N=4$ case, we have
\begin{eqnarray}
%%%%%%%
%%%%%%%     [0,2]
%%%%%%%
\lefteqn{
[0,2] = 2\alpha^{\prime}
     \frac{\theta_3\theta_{4}}{1-x}
     \left[ k_{3}\pdot k_{4}
  + \frac{1}{2}\sum_{I}
     \left\{\left( k_{3}\kuroten k_{4}\right)_{I}
      \left[ x^{-\nu_{I}}+x^{\nu_{I}}
      \right]
      + \left( k_{3}\peke k_{4} \right)_{I}
      \left[x^{-\nu_{I}}-x^{\nu_{I}} \right]\,\right\}\,\right]~,}
\nonumber\\
%%%%%%%%
%%%%%%%%  [1,1]
%%%%%%%%
\lefteqn{ [1,1] = 
 \sqrt{2\alpha^{\prime}}\left[
  \frac{\eta_{3}\theta_{3}}{2}
  \left\{\left( (k_{2}-k_{1})\pdot\zeta_{3}
          {}-ik_{3}\ppdot J\zeta_{3}\right)+k_{4}\pdot\zeta_{3}\right\}
\right.}\nonumber\\
&&\hspace{5em}
  +\frac{\eta_{4}\theta_{4}}{2x}
  \left\{\left( (k_{2}-k_{1})\pdot\zeta_{4}
          {}-ik_{4}\ppdot J\zeta_{4}\right)+k_{3}\pdot\zeta_{4}\right\}
\nonumber\\
&&\hspace{5em} +\frac{\theta_{3}-\theta_{4}}{1-x}
\left(\eta_{3}k_{4}\pdot\zeta_{3}
   +\eta_{4}k_{3}\pdot\zeta_{4}\right)  \nonumber\\
&&\hspace{3em} -\frac{1}{2}\frac{1}{1-x}
  \sum_{I}\Bigg\{
   \eta_{3}\left( \zeta_{3}\kuroten k_{4}\right)_{I}
     \left\{
       \left( x^{-\nu_{I}+1} +x^{\nu_{I}}\right)\theta_{3}
    {}-\left( x^{-\nu_{I}}+x^{\nu_{I}}\right)\theta_{4}
      \right\} \nonumber\\
&&\hspace{8em}
  + \eta_{4}\left( k_{3}\kuroten \zeta_{4}\right)_{I}
   \left\{
       \left(  x^{-\nu_{I}}+x^{\nu_{I}}\right) \theta_{3}
    {}-\left( x^{-\nu_{I}}+x^{\nu_{I}-1}\right)\theta_{4}
   \right\}\nonumber\\
&&\hspace{8em}
  +\eta_{3}\left(\zeta_{3}\peke k_{4}\right)_{I}
     \left\{
        \left(  x^{-\nu_{I}+1}-x^{\nu_{I}} \right) \theta_{3}
     {}-\left( x^{-\nu_{I}}-x^{\nu_{I}}\right)\theta_{4}
     \right\}  \nonumber\\
&&\hspace{8em} 
  +\eta_{4}\left( k_{3}\peke\zeta_{4}\right)_{I}
     \left\{
        \left( x^{-\nu_{I}}-x^{\nu_{I}}\right)\theta_{3}
     {}-\left( x^{-\nu_{I}}-x^{\nu_{I}-1}\right)\theta_{4}
     \right\}
\,\Bigg\} \, \Bigg]~,  \nonumber\\
%%%%%%%%
%%%%%%%%  [2,0]
%%%%%%%%
\lefteqn{
[2,0]= \frac{\eta_{3}\eta_{4}}{ 1-x }
   \left[
   \zeta_{3} \pdot \zeta_{4} 
 +\frac{1}{2}\sum_{I}
   \left\{ \left(\zeta_{3}\kuroten\zeta_{4}\right)_{I}
       \left(   x^{-\nu_{I}}+x^{\nu_{I}}   \right)
     +\left(\zeta_{3}\peke\zeta_{4}\right)_{I}
       \left( x^{-\nu_{I}}-x^{\nu_{I}}\right) 
   \right\}\right]~,} \nonumber\\
%%%%%%%%%
%%%%%%%%%   [2,2]
%%%%%%%%%
\lefteqn{[2,2]=
   \frac{\eta_{3}\theta_{3}\eta_{4}\theta_{4}}
        {(1-x)^{2}}
   \Bigg[\zeta_{3}\pdot \zeta_{4}
   +\frac{1}{2} \sum_{I}  \Bigg\{
       \left(\zeta_{3}\kuroten\zeta_{4}\right)_{I}
   \left\{
     (1-\nu_{I})\left(x^{-\nu_{I}}+x^{\nu_{I}} \right)
     +\nu_{I}  \left( x^{-\nu_{I}+1}+x^{\nu_{I}-1}\right)
   \right\}
}\nonumber\\
&&\hspace{5.5em} +\left(\zeta_{3}\peke\zeta_{4}\right)_{I}
   \left\{
     (1-\nu_{I})\left( x^{-\nu_{I}}-x^{\nu_{I}}\right)
     +\nu_{I}\left( x^{-\nu_{I}+1}-x^{\nu_{I}-1} \right)
   \right\}
\Bigg\}\Bigg]~,
\end{eqnarray} 
where we have set $x_{3}=1$ and written $x_{4}=x$.
By picking up terms from
$[2,2]+[0,2][2,0]+\frac{1}{2}[1,1]^2$, we obtain
\begin{eqnarray}
 \lefteqn{  A_{4}=c(2\pi)^{p+1}\prod_{\mu=0}^{p}
  \delta\left(\sum_{a=1}^{4} k_{a\mu} \right)
  \int^1_0 dx x^{-\alpha^{\prime}t+\alpha^{\prime}m_{T}^{2}}
  (1-x)^{2\alpha^{\prime} k_{3}\pdot k_{4}}
  \exp \Big({\cal C}_{3}(\nu_{I})+{\cal C}_{4}(\nu_{I})
              + ({\rm NC}) \Big) }
\nonumber\\
&&\times \exp \left[ -\alpha^{\prime}\sum_{I}
     \left\{ \left(k_{3}\kuroten k_{4}+k_{3}\peke k_{4}\right)_{I}
          {\cal H}\left(\nu_{I};\frac{1}{x}\right)
     +\left(k_{3}\kuroten k_{4}-k_{3}\peke k_{4}\right)_{I}
          {\cal H}\left(\nu_{I};x \right)\right\}\right]
\nonumber\\
&&\times \Bigg[ \frac{1}{(1-x)^{2}}
   \zeta_{3}\pdot \zeta_{4}
   \left(1-2\alpha^{\prime}k_{3}\pdot k_{4}\right)
\nonumber\\
&&\hspace{0.5em}
   +\frac{\alpha^{\prime}}{2}  \frac{1}{x} \left\{\left[
      (k_{2}-k_{1})\pdot\zeta_{3} -ik_{3}\ppdot J\zeta_{3} \right]
    {}- k_{4}\pdot\zeta_{3}   \right\}
% \nonumber\\
% &&\hspace{4em}\times
\left\{\left[(k_{2}-k_{1})\pdot\zeta_{4}
    {}-ik_{4}\ppdot J\zeta_{4}\right]+k_{3}\pdot\zeta_{4}\right\}
\nonumber\\
&&\hspace{0.5em}+\alpha^{\prime} \frac{1}{1-x}\left\{
   \left[
       (k_{2}-k_{1})\pdot\zeta_{3}-ik_{3}\ppdot J\zeta_{3}\right]
   k_{3}\pdot \zeta_{4}
%  \right.\nonumber\\
%  &&\hspace{7em}\left.
 {} -k_{4}\pdot\zeta_{3}
      \left[(k_{2}-k_{1})\pdot\zeta_{4}-ik_{4}\ppdot J\zeta_{4}\right]
\right\}\nonumber\\
&&\hspace{0.5em}+\sum_{I}\frac{x^{-\nu_{I}} }{(1-x)^{2}}
   \left\{ -\alpha^{\prime}
  \left(k_{3}\kuroten k_{4} + k_{3}\peke k_{4}\right)_{I}
              \zeta_{3}\pdot\zeta_{4}
  \right.\nonumber\\
  &&\hspace{8em}\left.
     +\left(\frac{1-\nu_{I}}{2}-\alpha^{\prime}k_{3}\pdot k_{4}\right)
        \left(\zeta_{3}\kuroten\zeta_{4} 
               + \zeta_{3}\peke\zeta_{4}\right)_{I}
      \right\} \nonumber\\
&&\hspace{0.5em}+\sum_{I}\frac{x^{\nu_{I}}}{(1-x)^{2}}
   \left\{ -\alpha^{\prime}
    \left(k_{3}\kuroten k_{4} - k_{3}\peke k_{4}\right)_{I}
              \zeta_{3}\pdot\zeta_{4}
   \right.\nonumber\\
   &&\hspace{8em}\left.
     +\left(\frac{1-\nu_{I}}{2}-\alpha^{\prime}k_{3}\pdot k_{4}\right)
       \left(\zeta_{3}\kuroten\zeta_{4}
              {}-\zeta_{3}\peke\zeta_{4}\right)_{I}
   \right\}  \nonumber\\
&&\hspace{0.5em} +\sum_{I}\frac{\nu_{I}}{2}
    \frac{x^{-\nu_{I}+1}}{(1-x)^{2}}
    \left(\zeta_{3}\kuroten\zeta_{4}
          + \zeta_{3}\peke\zeta_{4}\right)_{I}
  %%%%%%%  
    +\sum_{I}\frac{\nu_{I}}{2}
     \frac{x^{\nu_{I}-1}}{(1-x)^{2}}
     \left(\zeta_{3}\kuroten\zeta_{4}
           {}-\zeta_{3}\peke\zeta_{4}\right)_{I} \nonumber\\
&&\hspace{0.5em}-\frac{\alpha^{\prime}}{2}\sum_{I,L}
   \frac{x^{-\nu_{I}-\nu_{L}}}{(1-x)^{2}}
   \left( k_{3}\kuroten k_{4} + k_{3}\peke k_{4}\right)_{I}
   \left( \zeta_{3}\kuroten\zeta_{4} 
         +\zeta_{3}\peke\zeta_{4}\right)_{L}
\nonumber\\
&&\hspace{0.5em}-\frac{\alpha^{\prime}}{2}\sum_{I,L}
   \frac{x^{\nu_{I}+\nu_{L}}}{(1-x)^{2}}
   \left( k_{3}\kuroten k_{4} - k_{3}\peke k_{4}\right)_{I}
   \left( \zeta_{3}\kuroten\zeta_{4} 
         {}-\zeta_{3}\peke\zeta_{4}\right)_{L}
\nonumber\\
&&\hspace{0.5em}-\frac{\alpha^{\prime}}{2}\sum_{I,L}
   \frac{x^{-\nu_{I}+\nu_{L}}}{(1-x)^{2}}
   \left( k_{3}\kuroten k_{4} + k_{3}\peke k_{4}\right)_{I}
   \left( \zeta_{3}\kuroten\zeta_{4} 
         {}-\zeta_{3}\peke\zeta_{4}\right)_{L}
\nonumber\\
&&\hspace{0.5em}-\frac{\alpha^{\prime}}{2}\sum_{I,L}
   \frac{x^{\nu_{I}-\nu_{L}}}{(1-x)^{2}}
   \left( k_{3}\kuroten k_{4} - k_{3}\peke k_{4}\right)_{I}
   \left( \zeta_{3}\kuroten\zeta_{4} 
          +\zeta_{3}\peke\zeta_{4}\right)_{L}
\nonumber\\
&&\hspace{0.5em}+\frac{\alpha^{\prime}}{2}\sum_{I,L}
   \frac{ x^{-\nu_{I}-\nu_{L}} }{ 1-x }
   \left(k_{4}\kuroten\zeta_{3}-k_{4}\peke\zeta_{3}\right)_{I}
   \left(k_{3}\kuroten\zeta_{4}+k_{3}\peke\zeta_{4}\right)_{L}
\nonumber\\
&&\hspace{0.5em}-\frac{\alpha^{\prime}}{2}\sum_{I,L}
   \frac{ x^{\nu_{I}+\nu_{L}-1} }{ 1-x }
   \left(k_{4}\kuroten\zeta_{3} + k_{4}\peke\zeta_{3}\right)_{I}
   \left(k_{3}\kuroten\zeta_{4} - k_{3}\peke\zeta_{4}\right)_{L}
\nonumber\\
&&\hspace{0.5em}+\frac{\alpha^{\prime}}{2}\sum_{I}
    \frac{ x^{\nu_{I}-1} }{ 1-x }\left\{\left(
    \left[(k_{2}-k_{1})\pdot\zeta_{3}-ik_{3}\ppdot J\zeta_{3}\right]
    {}-k_{4}\pdot\zeta_{3}    \right)
    \left( k_{3}\kuroten \zeta_{4}-k_{3}\peke\zeta_{4}\right)_{I}
 \right.\nonumber\\
   && \hspace{5em}\left.-\left(
     \left[(k_{2}-k_{1})\pdot\zeta_{4}-ik_{4}\ppdot J\zeta_{4}\right]
      +k_{3}\pdot\zeta_{4}\right)
          \left(k_{4}\kuroten\zeta_{3}+k_{4}\peke\zeta_{3}\right)_{I}
     \right\}\nonumber\\
&&\hspace{0.5em}+\frac{\alpha^{\prime}}{2}\sum_{I}
   \frac{ x^{-\nu_{I}} }{1-x}\left\{\left(
   \left[(k_{2}-k_{1})\pdot\zeta_{3}-ik_{3}\ppdot J\zeta_{3}\right]
      +k_{4}\pdot\zeta_{3}\right)
  \left(k_{3}\kuroten\zeta_{4}+k_{3}\peke\zeta_{4}\right)_{I}
 \right.\nonumber\\
   &&\hspace{5em}-\left.\left(
  \left[(k_{2}-k_{1})\pdot\zeta_{4}-ik_{4}\ppdot J\zeta_{4}\right]
      {}-k_{3}\pdot\zeta_{4}\right)
  \left(k_{4}\kuroten\zeta_{3}-k_{4}\peke\zeta_{3}\right)_{I}
  \right\}
\Bigg]~,
\label{eq:A4}
\end{eqnarray}
where $t$ is defined as
$t\equiv s_{4}\equiv -(k_{4}+k_{1})\pdot(k_{4}+k_{1})$.

%%%%%%%%%%%%%%%%%%%%%%%%%%%%%%%%%%%%%%%%%%%%%%%%%%%%%%%%%%%%%%%%
%%%%%%%%%%%%%%%%%%%%%%%%%%%%%%%%%%%%%%%%%%%%%%%%%%%%%%%%%%%%%%%%
\section{The Zero Slope Limit and the Low Energy Effective Action}
%%%%%%%%%%%%%%%%%%%%%%%%%%%%%%%%%%%%%%%%%%%%%%%%%%%%%%%%%%%%%%%%
In the last section, we have evaluated the three and four point amplitudes
with the initial and final tachyons and $N-2$
vectors $(N=3,4)$ present.
Let us try to extract physical significance from these.

The three point amplitude eq.~(\ref{eq:A3}) contains the two multiplicative
factors $e^{\frac{i}{2}\theta^{ij}k_{1i}k_{2j}}$ and
$e^{{\cal C}(\nu_{I})}$
both of which are listed in the last section as prominent features.
The first factor represents the noncommutativity of the D$p$-brane
worldvolume.
The second factor will be discussed shortly.
Aside from these factors and $(p+1)$-dimensional delta functions
representing momentum conservation,
eq.~(\ref{eq:A3}) is interpreted as coming from field theory
vertex of
\begin{equation}
\Phi i\stackrel{\leftrightarrow}{\partial} \Phi^{\dag} \pdot A
 +\frac{1}{2}
  \Phi\Phi^{\dag} J^{MN} F_{MN}~,
\end{equation}
where $\Phi$ is a complex tachyon field and
$A_{M}$ and $F_{MN}$ are the gauge field and its field strength
respectively.
The first term is the gauge-scalar derivative interaction
while the second factor is a new interaction coming from our
$p$-$p'$ open string system.

The four point amplitude eq.~(\ref{eq:A4}) is quite complex
but one can still systematically investigate the singular behavior
of the integrand around its end points $x=0$, $1$.
This behavior is sufficient to tell us the zero slope limit
of the amplitude and the content of the low energy field theory.
We will focus upon this in the remainder of this section.
To be more accurate we consider the sum of eq.~(\ref{eq:A4})
and the one obtained from this by $k_3 \leftrightarrow k_4$,
$\zeta_{3}\leftrightarrow \zeta_{4}$
in accordance with the two open string (dual) diagrams.

%%%%%%%%%%%%%%%%%%%%%%%%%%%%%%%%%%%%%%%%%%%%%%%%%%%%%%%%%%%%%%%%%%%
The nontrivial zero slope limit is given by sending $\alpha' \to 0$
while keeping the parameter $\theta^{ij}$ of noncommutativity
and the open string metric fixed.
{}From eqs.~(\ref{eq:g-munu}) and (\ref{eq:B-ij}) this means
that \cite{SW}
\begin{equation}
\begin{array}{rcl}
\alpha' &\sim& \varepsilon^{1/2} \to 0~,  \\
g &\sim& \varepsilon \to 0~, \\
|b_{I}| &\sim& \varepsilon^{-1/2} \to \infty~.
\end{array} \end{equation}
In this limiting procedure,  $\alpha' b_{I}$ becomes finite:
$\alpha' b_{I} \to \beta_{I}$.
Let us first look at the multiplicative factor ${\cal C}(\nu_{I})$
defined in eq.~(\ref{eq:C}).
Using
%\begin{equation}
%\nu_{I} \approx 
%\left\{ 
%\begin{array}{ll}
%   {}-\frac{1}{\pi b_{I}},&b_{I}\to -\infty\\
%     \frac{1}{2}+\frac{b_{I}}{\pi},&b_{I}\approx 0\\
%     1-\frac{1}{\pi b_{I}},&b_{I}\to \infty
%\end{array} 
%\right. ~,
%\end{equation}
$\mbox{\boldmath$\psi$}(1)= -\gamma$ and eq.~(\ref{eq:psib}),
we obtain 
\begin{equation}
{\cal C}(\nu_I)\to -\pi\sum_{I,\bar{J}}\left|\beta_{I}\right|
\kappa_{I}\overline{\kappa}_{\overline{J}}G^{I\overline{J}}
=-\frac{\pi}{2}\sum_{I}\left|\beta_{I}\right|
  \left(k\kuroten k\right)_{I}~.\label{eq:damping2}
\end{equation}
%If $\left|\beta_{I}\right|$ are chosen to be independent of $I$:
%$\left|\beta_{I}\right|=\beta$,
%\begin{equation}
%{\cal C}(\nu_I) \to -\frac{\pi}{2} \beta\, k \ppdot k~.
%\end{equation}
%%%%%%%%%%%%%%%%%%%%%%%%%%%%%%%%%%%%%%%%%%%%%%%%%%%%%%%%
So this exponential multiplicative factor acts as a gaussian damping
factor when vectors propagate into the $x^{p+1},\ldots,x^{p^{\prime}}$
directions.

Let us come back to the four point amplitude (\ref{eq:A4}).
If the integrand is regular, one could take the
$\alpha^{\prime}\rightarrow 0$ limit inside the integral and
this will not give us any nontrivial contribution.
If the integrand is singular at some point,
it will still not give us much as long as one can avoid such
singularity by a contour deformation.
The nontrivial contribution in the $\alpha^{\prime}\rightarrow 0$
limit, therefore, is obtained only when we have end point singularities.

\begin{figure}[htb]
 \epsfxsize=30em
 \centerline{\epsfbox{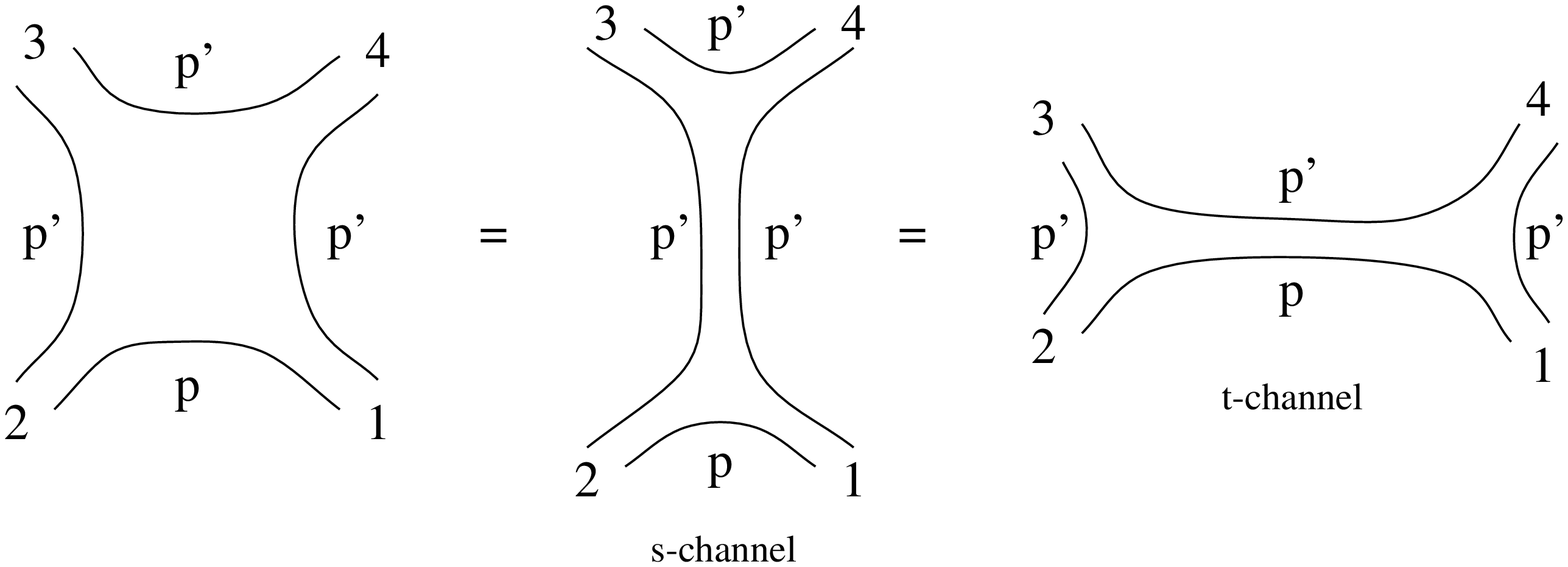}}
 \caption{The string diagram corresponding to
         the $s$-channel and the $t$-channel.}
 \label{fig:st-dual}
\end{figure}

\begin{figure}[htb]
 \epsfxsize=30em
 \centerline{\epsfbox{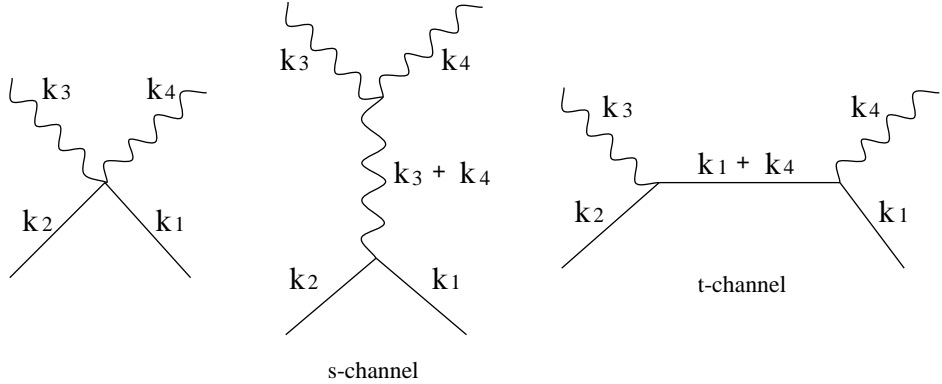}}
 \caption{Field theory diagrams.}
 \label{fig:feynman}
\end{figure}

Let us focus on the behavior of the integrand
near $x=0$ from which we can read off the mass of particles
exchanged in the $t$-channel.
Fig.\ref{fig:st-dual} indicates that the $t$-channel poles originate
in the propagation of the $p$-$p^{\prime}$ open string.
Thus the complicated behavior of the integrand near $x=0$ should
reflect the spectrum of the $p$-$p^{\prime}$ open string \cite{CIMM}.
In order to identify the $t$-channel poles,
we expand the integrand of the amplitude (\ref{eq:A4})
around $x=0$:
\begin{equation}
A_{4}=\int^{1}_{0} dx \sum_{A} f_{A}x^{-\alpha^{\prime}t+K_{A}}~,
\end{equation} 
where the coefficients $f_{A}$ are functions
of momenta and polarization tensors.
The term $f_{A} x^{-\alpha^{\prime}t+K_{A}}$
in the integrand of the above equation
yields the $t$-channel pole at $\alpha^{\prime}t=K_{A}+1$,
when it is integrated
near $x=0$: $\displaystyle\int_{0}^{\delta} dx\ldots$~.
{}From explicit computation we find that the $t$-channel poles
exist at
\begin{eqnarray}
&&\alpha^{\prime}t = \alpha^{\prime}m_{T}^{2}
 + W 
 + \sum_{I^{\prime}}M_{I^{\prime}}(n+1-\nu_{I^{\prime}})
 + \sum_{L^{\prime}}M^{\prime}_{L^{\prime}}(n^{\prime}+\nu_{L^{\prime}})
 + N~,
\label{eq:t-poles}\\
&&\begin{array}{rcl}
\mbox{with}\qquad  W &=&0~,\ 1~,\ 1-\nu_{I}~,\ \nu_{I}~,\ 1+\nu_{I}~,
              \ 2-\nu_{I}~,\\
            &&1-\nu_{I}-\nu_{L}~,\ \nu_{I}+\nu_{L}~,
\ 1+\nu_{I}+\nu_{L}~, \mbox{ or }1-\nu_{I}+\nu_{L}~,
\end{array} \nonumber
\end{eqnarray}
where $n$, $n^{\prime}$, $M_{I^{\prime}}$,
$M^{\prime}_{L^{\prime}}$ and $N$
are non-negative integers.
The terms proportional to $M_{I^{\prime}}$ and
$M^{\prime}_{L^{\prime}}$
come from the exponential of the hypergeometric function
${\cal H}$,
\begin{eqnarray}
&& \exp\left[-\alpha^{\prime}\sum_{I}\left\{
  \left(k_{3}\kuroten k_{4}+k_{3}\peke k_{4}\right)_{I}
       {\cal H}\left(\nu_{I}:\frac{1}{x}\right)
 +\left(k_{3}\kuroten k_{4}-k_{3}\peke k_{4}\right)_{I}
       {\cal H}(\nu_{I}:x)
      \right\}\right]\nonumber\\
&&=\exp\left(\alpha^{\prime}
       \sum_{I}\pi b_{I}\left(k_{3}\peke k_{4}\right)_{I}\right)
\nonumber\\
&&\hspace{2em}\times
\sum_{M=0}^{\infty}\frac{1}{M!}
   \left(-\alpha^{\prime}\sum_{I^{\prime}}
      \left(k_{3}\kuroten k_{4}+k_{3}\peke k_{4}\right)_{I^{\prime}}
      \sum_{n=0}^{\infty}\frac{x^{n+1-\nu_{I^{\prime}}}}
      {n+1-\nu_{I^{\prime}}}
    \right)^{M}\nonumber\\
&& \hspace{2em}\times
  \sum_{M^{\prime}}^{\infty}\frac{1}{M^{\prime}!}
   \left(-\alpha^{\prime}\sum_{L^{\prime}}
      \left(k_{3}\kuroten k_{4}-k_{3}\peke k_{4}\right)_{L^{\prime}}
      \sum_{n^{\prime}=0}^{\infty}
      \frac{x^{n^{\prime}+\nu_{L^{\prime}}}}
           {n^{\prime}+\nu_{L^{\prime}}}\right)^{M^{\prime}}~.
\end{eqnarray}
The $t$-channel poles eq.~(\ref{eq:t-poles})
correspond to the spectrum of the $p$-$p^{\prime}$
open string \cite{CIMM}.
In view of the analysis in \cite{CIMM},
we expect that a large number of light states should be exchanged
in the zero slope limit
in which one of $\nu_{I}$
goes to unity and the others approach zero.
In order to specify the situation,
we assume, without loss of generality, that
$\nu\equiv\nu_{\frac{p+2}{2}}$ goes to $1$
and $\nu_{\widetilde{I}}$
$(\widetilde{I}\neq \frac{p+2}{2})$ go to $0$
in the zero slope limit.
In this zero slope limit  many light states
are realized by the poles in eq.~(\ref{eq:t-poles}) with
$(n,n^{\prime},M_{\widetilde{I}^{\prime}},
  M^{\prime}_{\frac{p+2}{2}},N)=0$,
\begin{equation}
 W=0~,\ 1-\nu~,\ \nu_{\widetilde{I}}~,
\ 1-\nu-\nu_{\widetilde{I}}~,\ 1-\nu+\nu_{\widetilde{I}}~,
\end{equation}
and $M_{\frac{p+2}{2}}$ and $M^{\prime}_{\widetilde{I}}$ being
arbitrary non-negative integers.
Aside from the multiplicative factors and the momentum conserving delta
functions, the massless pole obtained in the zero slope limit
turns out to be
\begin{eqnarray}
&&\sim \left\{ \frac{1}{t-m^2_T}
  \left[\frac{1}{2}\left\{\Big(k_{2}-(k_{1}+k_{4})\Big)\pdot\zeta_{3}
             {}-ik_{3}\ppdot J\zeta_{3} \right\}
        \left\{\Big((k_{2}+k_{3})-k_{1}\Big)\pdot\zeta_{4}
             {}-ik_{4}\ppdot J\zeta_{4} \right\}\right]\right.
\nonumber\\
&&\hspace{2em}+\frac{1}{t-(m^2_T+\frac{1}{\pi |\beta|})}\left[
-(k_3\kuroten k_4+ k_3\peke k_4)_{\frac{p+2}{2}}\zeta_3\pdot\zeta_4 \right.\nonumber \\
&&\hspace{10em}+\Big(\frac{1}{2\pi |\beta|}-k_3\pdot k_4\Big)(\zeta_3\kuroten\zeta_4
+\zeta_3\peke \zeta_4)_{\frac{p+2}{2}} \nonumber \\
&&\hspace{10em} +\frac{1}{2}\left\{(k_2-k_1+k_4)\pdot\zeta_3-ik_3\ppdot J\zeta_3\right\}
(k_3\kuroten\zeta_4
+k_3\peke \zeta_4)_{\frac{p+2}{2}} \nonumber \\
&&\hspace{10em} \left. -\frac{1}{2}\left\{(k_2-k_1-k_3)\pdot\zeta_4-ik_4\ppdot J\zeta_4\right\}
(k_4\kuroten\zeta_3
-k_4\peke \zeta_3)_{\frac{p+2}{2}} \right]\nonumber \\
&&\hspace{2em}+\sum_{\widetilde{I}}\frac{1}{t-(m^2_T+\frac{1}{\pi |\beta_{\widetilde{I}}|})}\left[
\frac{1}{2\pi|\beta_{\widetilde{I}}|}(\zeta_3\kuroten\zeta_4-\zeta_3\peke\zeta_4)_{\widetilde{I}} \right.\nonumber\\
&&\hspace{12em} +\frac{1}{2}
       \left\{\Big(k_{2}-(k_{1}+k_{4})\Big)\pdot\zeta_{3}
             {}-ik_{3}\ppdot J\zeta_{3} \right\}
       \left(k_{3}\kuroten\zeta_{4}-k_{3}\peke\zeta_{4}
       \right)_{\widetilde{I}}\nonumber\\
&&\hspace{12em} \left. -\frac{1}{2}  
   \left\{\Big((k_{2}+k_{3})-k_{1}\Big)\pdot\zeta_{4}
             {}-ik_{4}\ppdot J\zeta_{4} \right\}\left(k_{4}\kuroten\zeta_{3}
         +k_{4}\peke\zeta_{3}\right)_{\widetilde{I}}\right]\nonumber\\
&&\hspace{2em}-\sum_{\widetilde{I},\widetilde{L}}
\frac{1}{t-(m^2_T+\frac{1}{\pi |\beta_{\widetilde{I}}|}
+\frac{1}{\pi |\beta_{\widetilde{L}}|})}\frac{1}{2}
  \left(k_{4}\kuroten\zeta_{3}+k_{4}\peke\zeta_{3}\right)_{\widetilde{I}}
  \left(k_{3}\kuroten\zeta_{4}-k_{3}\peke\zeta_{4}\right)_{\widetilde{L}}
\nonumber\\
&&\hspace{2em}+\sum_{\widetilde{I}}\frac{1}{t-(m^2_T+\frac{1}{\pi |\beta|}
-\frac{1}{\pi |\beta_{\widetilde{I}}|})}\times \left[-\frac{1}{2}
\left(k_{3}\kuroten k_{4}+k_{3}\peke k_{4}\right)_{\frac{p+2}{2}}
\left(\zeta_{3}\kuroten\zeta_{4}+\zeta_{3}\peke\zeta_{4}\right)_{\widetilde{I}}
\right. \nonumber\\
&&\hspace{12em} -\frac{1}{2}
\left(k_{3}\kuroten k_{4}+k_{3}\peke k_{4}\right)_{\widetilde{I}}
\left(\zeta_{3}\kuroten\zeta_{4}+\zeta_{3}\peke\zeta_{4}\right)_{\frac{p+2}{2}}
 \nonumber\\
&&\hspace{12em} +\frac{1}{2}
\left(k_{4}\kuroten\zeta_{3}-k_{4}\peke\zeta_{3}\right)_{\frac{p+2}{2}}
\left(k_{3}\kuroten\zeta_{4}+k_{3}\peke\zeta_{4}\right)_{\widetilde{I}}
 \nonumber\\
&&\hspace{12em}\left. +\frac{1}{2}
\left(k_{4}\kuroten\zeta_{3}-k_{4}\peke\zeta_{3}\right)_{\widetilde{I}}
\left(k_{3}\kuroten\zeta_{4}+k_{3}\peke\zeta_{4}\right)_{\frac{p+2}{2}}
\right] \nonumber\\
&&\hspace{2em} - \sum_{\widetilde{I}}\frac{1}{t-(m^2_T+\frac{1}{\pi |\beta|}
+\frac{1}{\pi |\beta_{\widetilde{I}}|})}\times \left[\frac{1}{2}
\left(k_{3}\kuroten k_{4}+k_{3}\peke k_{4}\right)_{\frac{p+2}{2}}
\left(\zeta_{3}\kuroten\zeta_{4}-\zeta_{3}\peke\zeta_{4}\right)_{\widetilde{I}}
\right. \nonumber\\  
&&\hspace{12em}\left.\left. +\frac{1}{2}
\left(k_{3}\kuroten k_{4}-k_{3}\peke k_{4}\right)_{\widetilde{I}}
\left(\zeta_{3}\kuroten\zeta_{4}+\zeta_{3}\peke\zeta_{4}\right)_{\frac{p+2}{2}}
\right] \right\} \nonumber \\
&&\times \exp\left\{-\pi\sum_{I}\left|\beta_{I}\right|
       \left(k_{3}\kuroten k_{4}\right)_{I}\right\}~,
\label{eq:t-residue}
\end{eqnarray}
where we have used
\begin{equation}
 \left\{
   \begin{array}{lccc}
    \displaystyle \nu\simeq 1-\frac{1}{\pi b_{\frac{p+2}{2}}}
    &\Rightarrow 
    &\displaystyle \frac{\alpha^{\prime}}{1-\nu}
       \rightarrow\pi \beta_{\frac{p+2}{2}}
    &(>0) \\
    \displaystyle \nu_{\widetilde{I}}
        \simeq -\frac{1}{\pi b_{\widetilde{I}}}
    &\Rightarrow
    &\displaystyle \frac{\alpha^{\prime}}{\nu_{\widetilde{I}}}
       \rightarrow -\pi \beta_{\widetilde{I}}
    &(>0)
   \end{array}
\right.~,
\end{equation}
in the zero slope limit.
The first term in eq.~(\ref{eq:t-residue})
comes from the tachyon state exchange.
Here the vertices derived from
three point amplitude emerge.
{}From the $t$-channel diagram in Fig.\ref{fig:feynman},
we find that these vertices depend on momenta in a proper way.
It is worth noting that
combining the exponential factor in eq.~(\ref{eq:t-residue})
with the multiplicative factor
$\exp\Big({\cal C}_{3}(\nu_{I})+{\cal C}_{4}(\nu_{I})\Big)$,
we obtain a gaussian damping factor in the
$x^{p+1},\ldots,x^{p^{\prime}}$ directions
in the zero slope limit,
\begin{equation}
 \exp\left[-\frac{\pi}{2}\sum_{I}\left|\beta_{I}\right|
  \left((k_{3}+k_{4})\kuroten (k_{3}+k_{4})\right)_{I}\right]~.
\label{eq:damping}
\end{equation}

Next we focus on the behavior of the integrand near $x=1$
from which we can read off the $s$-channel poles.
{}From Fig.\ref{fig:st-dual}, the $s$-channel poles
come from the propagation of the $p^{\prime}$-$p^{\prime}$
open string.
In a similar way to the $t$-channel,
by expanding the integrand around $x=1$
and integrating it near $x=1$:
$\displaystyle\int^{1}_{1-\delta}dx\ldots$~,
we find that $s$-channel poles correspond to
the $p^{\prime}$-$p^{\prime}$ open string spectrum.
In particular, aside from the multiplicative factor
and momentum conserving delta functions,
the massless pole turns out to be
\newpage
\begin{eqnarray}
&&\sim
\frac{1}{2s}\left[  \left\{
   (k_{2}-k_{1})\pdot\zeta_{3}-i(k_{3}+k_{4})\ppdot J\zeta_{3}
  \right\} k_{3}\prdot\zeta_{4}\right.\nonumber\\
&&\hspace{4em}-k_{4}\prdot\zeta_{3} \left\{
       (k_{2}-k_{1})\pdot\zeta_{4}
     {}-i(k_{3}+k_{4})\ppdot J\zeta_{4} \right\}\nonumber\\
&&\hspace{4em} -2 \sum_{I}\nu_{I}
    \left(k_{3}\peke k_{4}\right)_{I}\zeta_{3}\prdot\zeta_{4}
   {}-2k_{3}\prdot k_{4} \sum_{I}\nu_{I}
     \left(\zeta_{3}\peke\zeta_{4}\right)_{I}\nonumber\\
&&\hspace{4em}\left.
   + \left\{-t+m_{T}^{2} +k_{3}\ppdot k_{4}
     {}-\sum_{I}(1-2\nu_{I})
       \left(k_{3}\peke k_{4}\right)_{I}
    \right\}    \zeta_{3}\prdot\zeta_{4}
\right]\nonumber\\
&&\hspace{1.5em}\times \exp\left[2\alpha^{\prime}\sum_{I}
     \left(k_{3}\kuroten k_{4}\right)_{I}
     \left\{\gamma+\frac{1}{2}\left(\mbox{\boldmath$\psi$}(\nu_{I})
            +\mbox{\boldmath$\psi$}(1-\nu_{I})\right)\right\}
         \right]\nonumber\\
&&+\left(k_{3}\leftrightarrow k_{4};
         \zeta_{3}\leftrightarrow\zeta_{4}\right)~,
\label{eq:nearx=1}
\end{eqnarray}
where
\begin{equation}
 s\equiv -(k_{3}+k_{4})\prdot(k_{3}+k_{4})
  =-2 k_{3}\prdot k_{4}~.
\end{equation}
Here
by using eqs.~(\ref{eq:power}), (\ref{eq:loghyper}) and (\ref{eq:psib})
we have expanded the hypergeometric function ${\cal H}$
around $x=1$ as
\begin{eqnarray}
 &&{\cal H}\left(\nu;\frac{1}{x}\right)
   =-\frac{\pi}{2}b-\ln (1-x)\nonumber\\
&&\hspace{2em}
    +\sum_{m=0}^{\infty}\frac{(-1+\nu)_{m}}{m!}(1-x)^{m}
     \sum_{n=0}^{\infty}\frac{(1-\nu)_{n}}{n!}
  \left\{\mbox{$\psi$}(n+1)-\mbox{$\psi$}(n+1-\nu)\right\}(1-x)^{n}~,
\nonumber\\
 &&{\cal H}\left(\nu;x\right)
   =\frac{\pi}{2}b -\ln (1-x)\nonumber\\
&&\hspace{2em}+\sum_{m=0}\frac{(-\nu)_{m}}{m!}(1-x)^{m}
    \sum_{n=0}^{\infty}\frac{(\nu)_{n}}{n!}
    \left\{\mbox{$\psi$}(n+1)-\mbox{$\psi$}(n+\nu)\right\}
    (1-x)^{n}~.
\end{eqnarray}
We find that in our result eq.~(\ref{eq:nearx=1})
the first two terms are in accordance with
the vertex seen at the three point amplitude (\ref{eq:A3}).
This vertex shows up again with proper momentum dependence
(see Fig.\ref{fig:feynman}) as well as
in the $t$-channel pole corresponding to the tachyon exchange.
Combined with the multiplicative factor
$\exp\Big({\cal C}_{3}(\nu_{I})+{\cal C}_{4}(\nu)\Big)$,
the exponential factor in eq.~(\ref{eq:nearx=1})
gives us a gaussian damping factor,
\begin{equation}
 \exp\left[\alpha^{\prime}\sum_{I}\left\{\gamma
   +\frac{1}{2}\Big(\mbox{\boldmath$\psi$}(\nu_{I})
                    +\mbox{\boldmath$\psi$}(1-\nu_{I})\Big)\right\}
     \left((k_{3}+k_{4})\kuroten (k_{3}+k_{4})\right)_{I}
 \right]~.
\end{equation}
It is noteworthy that in the zero slope limit
this gaussian damping factor turns out to be the same
as that of $t$-channel (eq.~(\ref{eq:damping})).

While we do not try to derive here the complete action of the low energy
noncommutative field theory in $p'+1$ dimensions, it is still possible to
exhibit the interactions which reproduce the parts of the
amplitudes in the zero slope limit which are expressible in terms
of the inner product with respect to the open string metric.
We find 
that this part of the action is
\begin{eqnarray}
&&S = S_0 + S_1~,\nonumber\\
&&\mbox{with}\quad
S_0 = \frac{1}{g_{YM}^{\ 2}}\int d^{p'+1}x \sqrt{-G}
\left\{ -\left(D_{\mu}\Phi \right)^{\dag}\ast\left(D^{\mu}\Phi\right)
       {}-m^2 \Phi^{\dag} \ast \Phi
       {} -\frac{1}{4} F_{MN} \ast F^{MN}\right\}~,\nonumber\\
&&\hspace{3em}
S_1 = \frac{1}{2g^{\ 2}_{YM}}\int d^{p'+1}x \sqrt{-G}
\Phi^{\dag} \ast F_{MN} J^{MN} \ast \Phi~,
\label{eq:leea}
\end{eqnarray}
where 
\begin{eqnarray}
&&D_{\mu}\Phi =\partial_{\mu}\Phi-iA_{\mu} \ast \Phi~,
\qquad
\left( D_{\mu} \Phi \right)^{\dag}
  =\partial_{\mu}\Phi^{\dag}+i \Phi^{\dag}\ast A_{\mu}~,\nonumber \\
&&F_{MN}=\partial_{M}A_{N}-\partial_{N}A_{M}
    {}-i \left[A_{M},A_{N}\right]_{\ast}~,\qquad
 \left[A_{M}, A_{N}\right]_{\ast}= A_{M} \ast A_{N}
    {}-A_{N} \ast A_{M}~,
%&&\phi=\phi(x_{\sigma}; x_{i}=0)~,\quad
%  (\sigma=0,\ldots,p;\ i=p+1,\ldots,p^{\prime})~,
%\nonumber\\
%&& A_{\mu}= A_{\mu}(x_{\rho})~,
%\quad \left(\mu,\rho=0,\ldots,p^{\prime}\right)~,
\end{eqnarray}
the $\ast$ product of two functions $f$ and $g$ is given by
\begin{equation}
f(x) \ast g(x) = \left.
e^{\frac{i}{2}\theta^{\mu\rho}
\frac{\partial}{\partial y^{\mu}}\frac{\partial}{\partial z^{\rho}}}
f(y)g(z)\, \right|_{y,z\to x}~,
\end{equation}
and $g_{YM}$ is the effective Yang-Mills coupling
defined by using the open sting coupling $G_{s}$
and that of the closed string $g_{s}$ as \cite{SW}
\begin{equation}
 \frac{1}{g_{YM}^{\ 2}}
 =\frac{\left(\alpha^{\prime}\right)^{\frac{3-p^{\prime}}{2}}}
       {\left(2\pi\right)^{p^{\prime}-2} G_{s}}
  =\frac{\left(\alpha^{\prime}\right)^{\frac{3-p^{\prime}}{2}}}
       {\left(2\pi\right)^{p^{\prime}-2}g_{s}}
   \left(\frac{\det (g+2\pi\alpha^{\prime}B)}
              {\det G}\right)^{\frac{1}{2}}~.
\end{equation}
In eq.~(\ref{eq:leea}) we have determined the seagull interaction
corresponding to the non-pole term by invoking the noncommutative
$U(1)$ invariance.

Let us finally discuss the gaussian damping factor
eq.~(\ref{eq:damping}) which have originated from
the exponential multiplicative factor eq.~(\ref{eq:C})
and the lowest modes in the hypergeometric function ${\cal H}$.
Recall that there is no momentum conservation for
the $x^{p+1},\ldots,x^{p^{\prime}}$-directions and
that the tachyon momenta $k_1$ and $k_2$ are constrained to lie on
the $x^{0},\ldots,x^{p}$-directions. Without the gaussian damping
factor, our picture would be that $(N-2)$ incident 
noncommutative $U(1)$ photons travel freely in
the $x^{p+1},\ldots,x^{p^{\prime}}$-directions until they get stopped by
the D$p$-brane. The actual spacetime picture
which we have exhibited here is that the lowest mode of the
$p$-$p^{\prime}$ open string develops a physical scale
$\sqrt{|\beta_{I}|}$ and that this mode creates a cloud
around the D$p$-brane in the zero slope limit.
The noncommutative $U(1)$ photons get decelerated by the presence
of this cloud, which is reflected in our damping factor
eq.~(\ref{eq:damping}).
The mean free paths will be measured by $\sqrt{|\beta_I|}$.

In this situation, the tachyon field in these directions
should be expanded by the coherent states
$\left\{\left\langle x^{p+1},\ldots,x^{p^{\prime}} | \nu_{I}
 \right\rangle \right\}$
associated with the would-be zero modes $\alpha^I_{1-\nu_{I}}$
(or $\overline{\alpha}^{\overline{I}}_{\nu_{I}}$).
On this basis, the complete analysis of low-lying states and
$\nu_I$ dependent interactions obtained from the residual parts of
the amplitudes will lead to the full-fledged form of the
tachyon-vector interactions in these directions. 
The appearance of the coherent states here
suggests that the fields which have originated from
the $p$-$p^{\prime}$ open string should support
noncommutative solitons
on the D$p^{\prime}$-brane worldvolume
which has recently been found in \cite{GMS}.

%%%%%%%%%%%%%%%%%%%%%%%%%%%%%%%%%%%%%%%%%%%%%%%%%%%%%%%%%%%
\section*{Acknowledgements:}
We are grateful to Professor E. Date
for helpful discussions on this subject.

\vspace{2ex}

%%%%%%%%%%%%%%%%%%%%%%%%%%%%%%%%%%%%%%%%%%%%%%%%

\appendix

%%%%%%%%%%%%%%%%%%%%%%%%%%%%%%%%%%%%%%%%%%%%%%%%%%%%%%%%%%%%%%%%

\section{More on the two-point function
         at the worldsheet boundary}\label{sect:hyper}

The radius of convergence of the hypergeometric series
$F(a,b;c;z)$ is unity.
Thus we have evaluated the hypergeometric series
on its convergent circle in eq.~(\ref{eq:hyperseries})
in deriving the noncommutativity term (\ref{eq:samenc}).
In this appendix we will give another derivation of
eq.~(\ref{eq:hyperseries}) to verify
the noncommutativity term (\ref{eq:samenc})
and the two-point function (\ref{eq:propCIMM-3}).

Let us focus on the relation,
\begin{equation}
 {\cal F}(\nu;z)
=-\ln (1-z)+z^{\nu}\sum_{n=0}^{\infty}\frac{(\nu)_{n}}{n!}
   \left\{\mbox{\boldmath$\psi$}(n+1)-\mbox{\boldmath$\psi$}(n+\nu)
   \right\} (1-z)^{n}~.\label{eq:loghyper}
\end{equation}
One can obtain this relation by using eq.~(\ref{eq:power})
and a formula for the hypergeometric function \cite{Erdelyi},
\begin{eqnarray}
\lefteqn{ F(a,b;a+b;z)=\frac{\Gamma(a+b)}{\Gamma(a)\Gamma(b)}
\sum_{n=0}^{\infty}
   \frac{(a)_{n}(b)_{n}}{\left(n!\right)^{2}}
   \Big[2\mbox{\boldmath$\psi$}(n+1)-\mbox{\boldmath$\psi$}(a+n)
   } \nonumber\\
&&\hspace{15em} -\mbox{\boldmath$\psi$}(b+n) -\ln(1-z)\Big]
   (1-z)^{n}~,
\end{eqnarray}
which is derived from the following relation
by putting $c=a+b+\delta$ and by taking the limit of
$\delta\rightarrow 0$:
\begin{eqnarray}
\lefteqn{ F(a,b;c;z)=\frac{\Gamma(c)\Gamma(c-a-b)}{\Gamma(c-a)\Gamma(c-b)}
   F(a,b;a+b-c+1;1-z)}\nonumber\\
&&\hspace{3em}
   +\frac{\Gamma(c)\Gamma(a+b-c)}{\Gamma(a)\Gamma(b)}(1-z)^{c-a-b}
    F(c-a,c-b;c-a-b+1;1-z)~.
\end{eqnarray}
{}From eq.~(\ref{eq:loghyper}), one can find that
$\displaystyle\lim_{z\rightarrow 1}
\left\{ {\cal F}\left(1-\nu_{I};\frac{1}{z}\right)
 {}-{\cal F}\left(\nu_{I};z\right) \right\}$
is sensitive to the way of taking the limit.
In the original expression (\ref{eq:propCIMM-2}) of
the two point function, however,
the way of sending $z\rightarrow 1$ is fixed
in a definite way on the real axis because of the step function
in front of each hypergeometric function.
The constant noncommutativity term (\ref{eq:samenc})
should be more precisely described as
\begin{eqnarray}
\lefteqn{
\frac{4}{\varepsilon} \frac{\delta^{I\overline{J}}}{1+b_{I}^{2}}
\left\{ \lim_{\tau_{1}\rightarrow\tau_{2}+0}
        {\cal F}\left(1-\nu_{I};\frac{e^{\tau_{2}}}{e^{\tau_{1}}}\right)
{}-\lim_{\tau_{1}\rightarrow\tau_{2}-0}
        {\cal F}\left(\nu_{I};\frac{e^{\tau_{1}}}{e^{\tau_{2}}}\right)
     \right\}
}\nonumber\\
&&=\frac{4}{\varepsilon} \frac{\delta^{I\overline{J}}}{1+b_{I}^{2}}
\lim_{\tau_{1}\rightarrow\tau_{2}}\left[
 \left\{ -\ln\left|1-\frac{e^{\tau_{2}}}{e^{\tau_{1}}}\right|+
        \left(\frac{e^{\tau_{2}}}{e^{\tau_{1}}}\right)^{1-\nu_{I}}
        \Big(\mbox{\boldmath$\psi$}(1)-\mbox{\boldmath$\psi$}(1-\nu_{I})
        \Big)   \right\}
\right.\nonumber\\
&&\hspace{8em}
  {}-\left\{-\ln \left|1-\frac{e^{\tau_{1}}}{e^{\tau_{2}}}\right|
            +\left(\frac{e^{\tau_{1}}}{e^{\tau_{2}}}\right)^{\nu_{I}}
         \Big(\mbox{\boldmath$\psi$}(1)-\mbox{\boldmath$\psi$}(\nu_{I})
         \Big)\right\}
\nonumber\\
&&=\frac{4}{\varepsilon}\frac{\delta^{I\overline{J}}}{1+b_{I}^{2}}
  \pi b_{I}~,
\end{eqnarray}
where we have used a relation for the digamma function,
\begin{equation}
 \pi b_{I}=-\pi\cot (\pi\nu_{I})
 =\mbox{\boldmath$\psi$}(\nu_{I})-\mbox{\boldmath$\psi$}(1-\nu_{I})~.
\label{eq:psib}
\end{equation}
Thus we obtained the same noncommutativity term as eq.~(\ref{eq:samenc})
through more careful treatment of the hypergeometric functions
and this provides another verification to 
the two-point function (\ref{eq:propCIMM-3}).

\newpage

%%%%%%%%%%%%%%%%%%%%%%%%%%%%%%%%%%%%%%%%%%%%%%%%%%%%%%%%%%%%%%%%%
%%%  references 
%%%%%%%%%%%%%%%%%%%%%%%%%%%%%%%%%%%%%%%%%%%%%%%%%%%%%%%%%%%%%%%%%

\end{document}